\documentclass[a4paper]{article}
\usepackage[latin1]{inputenc}
\usepackage{graphics}
\usepackage{amssymb}
\usepackage{amsfonts}
\usepackage[thmmarks]{ntheorem}
\usepackage[all]{xy}
\usepackage{longtable}

\newcommand{\eqref}[1]{(\ref{#1})}

\def\A{{\mathcal A}}
\def\H{{\mathcal H}}
\def\K{{\mathcal K}}
\def\D{{\mathcal D}}
\def\C{{\mathcal C}}
\def\B{{\mathcal B}}

\def\SC{{\rm S}{\mathcal C}}

\def\S{{\mathcal S}}

\def\RR{{\mathbb R}}
\def\CC{{\mathbb C}}
\def\NN{{\mathbb N}}
\def\ZZ{{\mathbb Z}}

\def\HH{{\mathbb H}}
\def\ss{\mathfrak{s}}
\def\SS{\mathfrak{S}}

\def\End{\mbox{\rm End}}
\def\Hom{\mbox{\rm Hom}}
\def\Aut{\mbox{\rm Aut}}
\def\Hol{{\rm Hol}}
\def\Vert{{\rm Vert}}
\def\SVert{\Gamma{\rm Sl}}

\def\bra{\langle}
\def\ket{\rangle}
\def\id{{\rm Id}}
\def\tr{\mbox{\rm Tr}}
\def\ker{\mbox{\rm Ker}}
\def\im{\mbox{\rm Im}}

\def\diag{{\rm diag}}
\def\Ad{{\rm Ad}}

\def\bea{\begin{eqnarray}}
\def\eea{\end{eqnarray}}
\def\Span{\mbox{\rm Span}}
\def\Spin{{\rm Spin}}
\def\Diff{{\rm Diff}}
\newcommand\transp[1]{\hspace{0.4mm}{}^t\hspace{-0.6mm}#1}

\def\be{\begin{equation}}
\def\ee{\end{equation}}

\newenvironment{rem}[1][{}]{\smallbreak \noindent  {\bf Remark #1}\small }

 \newenvironment{ex}[1][{}]{\smallbreak \noindent  {\bf Example #1}\small }

\newtheorem{theorem}{Theorem}
\newtheorem{definition}{Definition}
\newtheorem{lemma}{Lemma}
\newtheorem{cor}{Corollary}
\newtheorem{propo}{Proposition}
\theoremstyle{nonumberplain}
\theorembodyfont{\normalfont}
\theoremseparator{:}
\theoremsymbol{$\P$}

\newtheorem{demo}{Proof}

\begin{document}
\title{Algebraic backgrounds: a framework for noncommutative Kaluza-Klein theory}
\author{Fabien Besnard}

\maketitle
\begin{abstract}
We investigate the representation of diffeomorphisms in Connes' Spectral Triples formalism. By encoding the metric and spin structure in a moving frame, it is shown on the paradigmatic example of spin semi-Riemannian manifolds that the bimodule of noncommutative 1-forms $\Omega^1$ is an invariant structure in addition to the chirality, real structure and Krein product. Adding  $\Omega^1$ and removing the Dirac operator from an indefinite  Spectral Triple we obtain a structure which we call an \emph{algebraic background}. All the Dirac operators compatible with this structure then form the configuration space of a noncommutative Kaluza-Klein theory. In the case of the Standard Model, this configuration space is stricty larger than the   one obtained from the fluctuations of the metric, and contains in addition to the usual gauge fields the $Z_{B-L}'$-boson, a complex scalar field $\sigma$, which is known to be required in order to obtain the correct Higgs mass in the Spectral Standard Model, and flavour changing fields. The latter are invariant under automorphisms and can be removed without breaking the symmetries. It is remarkable that, starting from the conventional Standard Model algebra $\CC\oplus \HH\oplus M_3(\CC)$,  the ``accidental'' $B-L$ symmetry is necessarily gauged in this framework.
\end{abstract}
\tableofcontents

\section{Introduction}
Kaluza-Klein theory is probably the best idea to date concerning the unification of the different forces, and it lives up to this day, in one form or another, in many approaches to this question. However, it suffers from the instability of the compact dimensions which tend to collapse into singularities (\cite{penrose}, section 10.3). In String Theory, it is crucial that the extra-dimensions stay small without collapsing: this is known as the ``moduli stabilization'' prolem. Its resolution is believed to lead to the huge Landscape  \cite{KKLT}.

Another take on this issue could be that the compact extra dimensions \emph{did} collapse in the early moments of the universe, though quantum gravity effects prevented the appearance of a singularity (similarly to what happens to the Schwarzschild singularity in Loop Quantum Gravity, \cite{GP}). In this view the extra dimensions are in a quantum regime, and the variables describing them belong to a finite-dimensional algebra\footnote{One can even speculate that the different families of particles are a manifestation of some quantum number comparable to the principal quantum number of the Bohr atom.}. This   is exactly what Connes' Noncommutative Geometry offers with the framework of almost-commutative spacetimes.

In brief, the idea is to generalize GR by replacing a   spin manifold with a so-called spectral triple $(\A,\H,D)$, where $\A$ is an algebra, $\H$ a Hilbert space and $D$ a Dirac operator. When the algebra is commutative, one   exactly recovers the classical geometric data (under several additional natural assumptions, see \cite{reconst}). Another remarkable insight of Connes was to postulate the spectral action, which is of the form $S(D)=\tr(f(D^2/\Lambda^2))$, where $f$ is a cut-off function. It can be shown that it reproduces the Einstein-Hilbert action with a cosmological constant identified with $\Lambda$, plus terms of higher-order in the curvature  \cite{SpecAct}. The most striking application of this action principle is that 
  when the algebra is taken to be  almost-commutative, i.e. when it is the tensor product of a commutative by a finite-dimensional  non-commutative algebra, one can recover, by taking a suitable finite-dimensional part, all the terms of the Standard Model bosonic action, with the correct signs, including the Higgs which pops up naturally. Neutrino mixing and the see-saw mechanism are also natural features of the Spectral Standard Model \cite{SMmix}. There are, nevertheless, a number of problematic issues in this approach. First, the whole theory works only for a Riemannian manifold, i.e. in the Euclidean signature. There has been however some recent progress  the formulation of a semi-Riemannian NCG (\cite{francoeckstein}, \cite{vddgpr}, \cite{vdd},  \cite{part1}, \cite{SST2}, \cite{DFLM}, \cite{bbb2}, \cite{BS}, \cite{thesenadir}).  The only ingredient still missing is the spectral action, which is badly divergent in the non-Riemannian case (see \cite{thesenadir} for a pedagogical explanation). However, if one is more modest and falls back to the older Connes-Lott action (which does not include gravity), it is possible to recover the Standard Model action in the Lorentz signature \cite{thesenadir}. Another problem, possibly the most serious one, is the unimodularity condition, which is equivalent to a restriction of the symmetries of the theory, added by hand. On the side of physical predictions, one can estimate the mass of Higgs boson under the big desert hypothesis \cite{KS}. Unfortunately one obtains a value of around 170 GeV. One can nonetheless correct the theory by introducing a complex scalar field $\sigma$ \cite{resilience}. It has the very welcome features of stabilizing the vacuum in addition to pushing the Higgs mass down to the correct value. However, this field does not come from a fluctuation of the metric like the other bosonic fields in the Spectral Standard Model.  Understanding how the complex scalar could naturally arise has been a major challenge since then, with two  approaches to this question:  Boyle-Farnsworth theory (\cite{BF1}, \cite{BF2}), and  twisted Spectral Triples theory (\cite{DLM}, \cite{DFLM}). Both require  important modifications of the formalism.

Our goal in this paper was not to solve any of these problems, but rather to try to understand some less studied foundational questions. The first is how do we distinguish the dynamical and background structures in NCG ? In GR the answer is easy: the background   is a differentiable manifold, and the dynamical variable is the metric. Clearly, the Dirac operator, which replaces the metric, should be the dynamical variable in NCG. Let us however consider the classical case where  varying $D$ is equivalent to varying the metric on the manifold. In that case it would make no sense to claim that $(\A,\H)$ is the background. For one thing, we need $D$ to recover the differentiable structure, $\A$ alone is not enough. Another problem is that we need the metric to build the scalar product, so that $\H$ contains some information on the  metric. More subtly, we  also need  a spin structure to build the spectral triple (more precisely, $\H$ and another gadget called the real structure), and since it depends on the metric, it is not immediately clear how we can keep the same Hilbert space while varying the metric. The second, and related, problem is the implementation of diffeomorphism invariance. It is generally admitted that diffeomorphism invariance is superseded by unitary invariance in NCG. However, diffeomorphisms change the spinor bundle and the spin structure (and the metric, hence also the scalar product), so it is not \emph{a priori} obvious that we can implement diffeomorphisms inside a fixed background structure like a spectral triple. Finally, in the case of an almost-commutative spectral triple, like that of the Standard Model, the dynamical variable is not an arbitrary Dirac operator, but a so-called fluctuated Dirac. These fluctuations are motivated by Morita self-equivalences, but their physical meaning is not transparent. It would be more in the spirit of Kaluza-Klein theory to consider a general Dirac operator as the variable.

To answer these questions we will consider the example of the moving frame, or tetrad, formulation of GR\footnote{We speak about the ``tetrad-only'' formulation, where the connection is not an independent variable.}, which seems to be the best-suited to introduce a spin structure in the context of a variable metric. We will see that among the structures which are diffeomorphism-invariants, there is the subspace spanned by the gamma matrices in the algebra of endomorphisms of the spinor space. In the NCG pictures this means that the module of noncommutative 1-forms should be invariant under the symmetries, and thus part of the background. We will generalize  this example and define \emph{algebraic backgrounds}. To cut short a long story, an algebraic background is a Spectral Triple plus a bimodule $\Omega^1$ of noncommutative 1-forms minus the Dirac operator.  A Dirac operator $D$ is then defined in the usual way except that we require it in addition to be compatible with $\Omega^1$ in the sense that the 1-forms defined with $D$ belong to $\Omega^1$. Attached to an algebraic background there is thus a configuration space which is the space of all Dirac operators which are compatible with $\Omega^1$, exactly as in GR there is a configuration space which is the space of all metrics which are smooth, i.e. compatible with the differential structure\footnote{We will see that the compatibility with a spin structure complicates the matter a little bit in the case of GR, particularly so in non-Euclidean signature, but not really for algebraic backgrounds}. It is encouraging that the automorphisms of the algebraic background canonically obtained from a spin manifold  exactly correspond to the symmetries of tetradic GR, i.e. diffeomorphisms and ``local Lorentz transformations'', or more precisely smooth maps to the spin group, what we call \emph{spinomorphisms}. By contrast, the automorphism group of the canonical spectral triple over a manifold is somewhat larger if we do not require automorphisms to preserve the Dirac operator, or much smaller if we require them to. However, the configuration space of the algebraic background is slightly larger than that of tetradic GR: in addition to Dirac operators associated with tetrads, it contains so-called \emph{centralizing fields}, which in spacetime dimension 4 consist of a single pseudo-vector field. This can be seen as an unwanted feature. However, these extra fields can be removed without spoiling the theory since they do not mix with the frame field under spino/diffeomorphisms.  It is also possible to define an almost-commutative algebraic background with about the same ingredients as the Spectral Triple of the Standard Model. Remarkably, its automorphism group  turns out to be the symmetry group of the Standard Model coupled with tetradic GR  extended by gauged $B-L$-symmetries\footnote{To prove this result we use many peculiarities of the SM algebraic background, including an interplay between the spacetime and internal parts. We do not whether there exists a more generic proof or if it is an important clue.}. The configuration space is much larger than in the Spectral Standard Model: it contains the usual gauge fields, the $Z_{B-L}'$-boson, the anomalous $X$-field (which can be removed as usual with the unimodularity condition), the complex scalar which is so much sought after, but also many centralizing fields which in this case act on flavours. As is customary with Kaluza-Klein  theories, we obtain more fields than we really want\ldots\ However, much as in the GR case,  it is  possible to get rid of the flavour changing fields without breaking the symmetry. The smallest submodel which is invariant under automorphisms contains just the gauge fields and the $Z_{B-L}'$ (as well as the $X$-field). It can be discarded for physical reasons since it does not include neutrino mixing. The second smallest contains one complex scalar in addition, of $B-L$-charge $2$. Its field content thus exactly coincides with the one found in \cite{BF2} by an entirely different method. It seems to have nice cosmological implications \cite{boylecosmo}. Note that in the twisted spectral triples framework, the $Z_{B-L}'$ does not appear, so that the two approaches lead to different physical predictions.

In summary, the framework of algebraic backgrounds, though it was not its original motivation, makes the $\sigma$-field appear naturally, as well as the $Z_{B-L}'$ boson. Unfortunately, we are only able to discuss the field content, since we do not know any replacement for the spectral action principle at the moment. The unimodularity problem is not solved either. Note however that the problematic $X$-field is centralizing, as well as the $Y$, $Z_{B-L}'$ and flavour fields. It would be possible to project on the non-centralizing fields, but the price would be the suppression of electromagnetism ! However, this strategy might be successful in a model where the  $Y$ is not centralizing, such as the Pati-Salam model. It will be the subject of future research.


The paper is organized as follows. In section 2 we define Clifford, spin-c, and spin structures in the algebraic way, in general semi-Riemannian signature. We prove that this formulation is completely equivalent to the usual one which uses principal bundles. In the non-Euclidean case, an eminent role is played by the so-called \emph{spinor metric}. We also define \emph{tetradic spin structures}, and show how to build a configuration space for tetradic GR in the presence of a spin structure, which is metric-dependent, but whose equivalence class is not. In section 3 we introduce \emph{algebraic backgrounds} and discuss some generalities about them. Section 4 is devoted to the canonical background associated to a semi-Riemannian spin manifold. We compute its automorphism group and configuration space. In section 5 we discuss almost-commutative algebraic backgrounds. In section 6 we apply the previous framework to the algebraic background of the Standard Model. In particular we compute the automorphism group and the configuration space, i.e. the field content.

There are many routine  proofs and calculations in this paper, which could interrupt the fluidity of the reading: we have displayed them in small characters to notify the reader that they can be skipped without much harm. Some of the proofs about spin structures would just be  too long a digression and are relegated to the appendices.

\section{Clifford, spin-c and spin structures}\label{mfss}
Throughout this section $(M,g)$ will be an orientable $n$-dimensional semi-Riemannian manifold with metric $g$ of signature $(p,q)$, with $n=p+q$ even.
\subsection{General notations}
We first fix some notations and conventions. They are the same as in \cite{part1}, to which we refer for more details.

We define the Clifford algebra $Cl(V,g)$ of the real  vector space $V$ with non-degenerate metric $g$ to be generated by vectors of $V$ subject to the relations $uv+vu=2g(u,v)$.  The complexified Clifford algebra $Cl(V,g)\otimes \CC$ will be denoted $\CC l(V,g)$. It is equipped with a canonical real structure $c$, and a $\CC$-linear anti-involution $a\mapsto a^T$ which consists in reversing all products and is sometimes called the transpose or main anti-involution. If we compose the two we obtain an anti-linear anti-involution $a\mapsto a^\times=c(a^T)$. It is the adjoint operation with respect to a canonical non-degenerate sesquilinear form on $\CC l(V,g)$ which extends $g$ (see \cite{part1}).  We  also recall that the chirality element of $\CC l(V,g)$ is $\chi=i^{{n\over 2}+q}e_1\ldots e_n$, where $(e_1,\ldots,e_n)$ is any positively oriented pseudo-orthonormal basis. We will often use the vector space isomorphism $\Theta : \Lambda V\otimes\CC\simeq \CC l(V,g)$ which sends $v_1\wedge \ldots \wedge v_k$ to the antisymmetrization of $v_1\ldots v_k$. The elements of  $\Theta(\Lambda^k V\otimes \CC)$    will be called \emph{$k$-vectors}. A basis of the space of $k$-vectors is given by the ordered products of $k$ elements of a given pseudo-orthonormal basis of $V$.

The (complex) Clifford group, spin-c group and spin group are respectively defined to be (writing $V^\CC=V\otimes\CC$):

\begin{enumerate}
\item $\Gamma_\CC(V,g)=\{u\in\CC l(V,g)|,\Ad_u(V^\CC)\subset V^\CC\}$,
\item $\Spin^c(V,g)=\{u\in \CC l(V,g)|\Ad_u(V^\CC)\subset V^\CC,\chi u=u\chi, uu^\times=\pm 1\}$
\item $\Spin(V,g)=\{u\in \CC l(V,g)|\Ad_u(V^\CC)\subset V^\CC,\chi u=u\chi, uu^\times=\pm 1,c(u)=u\}$
\end{enumerate}

Remember that the Clifford group is generated by non-isotropic vectors of $V^\CC$, and that its elements satisfy $uu^\times\in\RR^*$. The elements of the spin group are products of an even number of normalized elements of $V$. The elements of the spin-c group are obtained by multiplying spin group elements by unimodular complex numbers. For all this, see \cite{crum}, chap 40. The spin group has two connected components, corresponding to the sign of the product $uu^\times$, except when $(p,q)=(1,1)$. We write $\Spin(V,g)^0$, and more generally for any topological group $G$ we write $G^0$, for the component of the identity. In the whole text, when we speak of the ``spin group'', we always have the neutral component in mind. If $(p,q)=(1,1)$, $\Spin(V,g)$ has $4$ connected components. We should take two of these components to obtain a (trivial) double cover of $SO(V,g)^0$. We will not consider this exceptional case again in this paper.

When $(V,g)$ is $\RR^{p,q}$, that is $\RR^{p+q}$ equipped with the standard metric of signature $(p,q)$, we write the above groups $\Gamma_\CC(p,q)$, $\Spin^c(p,q)$, $\Spin(p,q)$. We also use the notations $Cl(p,q)$ and $\CC l(p,q)$.

An element $A$ of $\End(V)$ extends to an automorphism of $Cl(V,g)$ iff it is an isometry, and when it is the case, the extension is unique and will be denoted $\tilde A$ in the sequel. On the other hand, any element $u$ of $\Spin(V,g)$ defines a direct isometry of $V$ by its adjoint action, and the map $\Spin(V,g)\rightarrow SO(V,g)$ sending $a$ to $\Ad_a$ is a 2:1 covering map. The spin group is simply connected when $g$ is Euclidean (i.e. $q=0$),  Lorentzian ($p=n-1,q=1$),  anti-Euclidean ($p=0$) or anti-Lorentzian ($p=1$, $q=n-1$), except when $(p,q)=(2,0)$ or $(0,2)$ (there are also the exceptions $(1,2)$ and $(2,1)$ in odd dimensions) (\cite{Var}, chap. 5).

\subsection{Algebraic spin structures}\label{algspinstru}
Spin structures have been introduced in the 1950's (in \cite{hael}) through the ``topological definition'' to be recalled below, which uses principal bundles, though spinor fields had already been in use in physics for more than two decades. The definition of spin structures which is the most natural for both Noncommutative Geometry and particle physics is the one we call ``algebraic'' and uses vector bundles. It is known for a long time in the Euclidean case (see \cite{plymen} and also the detailed account in \cite{gracia}). In the general case, one has to make the role played by the spinor metric more explicit, so we prefer to present the theory from scratch. We start with a weaker structure which will help clarifying the discussion.

\begin{definition} Let $or$ be an orientation on $M$. A $(g,or)$-Clifford structure  is a triple $(\S,\rho,\chi)$ where
\begin{enumerate}
\item $\S$ is a complex vector bundle over $M$,
\item\label{c2} $\rho : \CC l(TM,g)\rightarrow \End(\S)$ is a bundle isomorphism,
\item\label{c3} $\chi$, the chirality element, is the image under $\rho$ of the chirality element of the Clifford bundle corresponding to $or$.
\end{enumerate}
\end{definition}
Note that \ref{c2} means in particular that for each $x$, the fibre $ {\cal S}_x$ is an irreducible spinor module with the action of $\CC l(T_xM,g_x)$ given by $\rho_x$. According to the general definitions of the previous section, point \ref{c3} means that  at each $x\in M$, $\chi(x)=(i)^{{n\over 2}+q}\rho(e_1)\ldots \rho(e_n)$ where $(e_1,\ldots,e_n)$ is a positively oriented pseudo-orthonormal basis of $T_xM$. There is a slight abuse of notation in using the same letter to denote the chirality element in the Clifford algebra and the bundle $\End(\S)$. 


\begin{rem} 
 Clifford structures were introduced in \cite{friedtraut}, in the Riemannian signature and with a ``topological definition'', and rediscovered in \cite{thesenadir} in the above form.
\end{rem}

Next, we want to define spin-c structures. For this we first need the notion of spinor metrics.

\begin{definition} A  \emph{spinor metric} for a Clifford structure $(\S,\rho,\chi)$ on  $(M,g,or)$  is a smooth non-degenerate hermitian form $H$ on $\S$ such that for every tangent vector $v\in T_xM$, $\rho_x(v)$ is self-adjoint.
\end{definition}

As we are going to see, the existence of spinor metrics is tied to the space and time orientability of the manifold. Remember that $(M,g)$ is called time-orientable iff there exists a $p$-form $\omega$ such that $\omega_x(v_1,\ldots,v_p)\not=0$ for any linearly independent timelike vectors $(v_1,\ldots,v_p)$ at $x$. The sign of $\omega_x(v_1,\ldots,v_p)$ defines the time-orientation of $(v_1,\ldots,v_p)$, and $p$-forms as above fall into two classes according to the sign they assign to $(v_1,\ldots,v_p)$. These classes are called \emph{time orientations}. In the Lorentzian/anti-Lorentzian cases, they bijectively correspond to smooth choices of future cones all over the manifold. Space orientations are defined \emph{mutatis mutandis}. For more details, see  \cite{part1}, section 3, to which we also refer for the proof of the following theorem:

\begin{theorem}\label{existsspinormetric}
Let $M$ be an orientable semi-Riemannian manifold with a Clifford structure. A spinor metric exists on $M$ iff it is space \emph{and} time orientable.
\end{theorem}

Spinor metrics are unique up to multiplication by a positive scalar function \cite{part1}.

\begin{rem}[1] A manifold is orientable iff its first Stiefel-Whitney class vanishes. Similarly, it is time (resp. space)-orientable iff, given a decomposition $TM=TM^+\oplus TM^-$ of the tangent bundle with $TM^+$ timelike and $TM^-$ spacelike, then $w_1(TM^+)=0$ (resp. $w_1(TM^-)=0$). This condition does not depend on the decomposition. For more details, see \cite{baum} p 45. 
\end{rem} 
\begin{rem}[2]
If $M$ is not supposed to be orientable but only equipped with a spinor bundle, then a spinor metric exists iff a) $M$ is space\footnote{Remember we use the West-coast convention, hence spacelike vectors satisfy $g(v,v)<0$.} orientable if $p$ and $q$ are even, b) $M$ is time-orientable if $p$ and $q$ are odd. In particular, spinor metrics always exist on Riemannian manifolds. This can be proven directly using partitions of unity. This technique does not work for non-Riemannian manifolds because a sum of non-positive non-degenerate metrics can be degenerate.
\end{rem}

\begin{definition}\label{posspinmet} A positive \emph{spinor metric} for a Clifford structure $({\cal S},\rho,\chi)$ on a space and time oriented  manifold $(M,g,or,t-or)$  is a spinor metric such that for all $x\in M$:
\begin{enumerate}
\item For every positively oriented pseudo-orthonormal family of timelike vectors $e_1,\ldots,e_p\in T_xM$, $H_x(.,i^{p-1\over 2}\rho(e_1)\ldots\rho(e_p).)$ is positive-definite if $p,q$ are odd.
\item  For every positively oriented  pseudo-orthonormal family of spacelike vectors $e_1,\ldots,e_q\in T_xM$, $H_x(.,i^{q\over 2}\rho(e_1)\ldots\rho(e_q).)$ is positive-definite if $p,q$ are even.
\end{enumerate}  
\end{definition}
\begin{rem} Call $\eta$ the element $i^{p-1\over 2}\rho(e_1)\ldots\rho(e_p)$ or $i^{q\over 2}\rho(e_1)\ldots\rho(e_q)$ according to the case. Then $\eta$ is self-adjoint with respect to $H$  and $\eta^2=1$ in both cases. It is called a fundamental symmetry.
\end{rem}

This definition is really just a sign fixing. Indeed, it is proven in \cite{part1} (theorem 2) that  there always exists  pseudo-orthonormal families of timelike/spacelike vectors, according to the case, such that the above hermitian forms are \emph{definite}. It is then easy to show that they must be definite for \emph{all} such families, using the fact that the spin group acts transitively on them. For instance, in the odd case, if $(e_1',\ldots,e_p')$ and $(e_1,\ldots,e_p)$ are positively oriented pseudo-orthonormal families of timelike vectors then there exists an element $u\in\Spin(p,q)$ such that $e_i'=\Lambda e_i\Lambda^{-1}$, where $\Spin(p,q)$ is seen as a subgroup of $\CC l(T_xM,g_x)$. Then $H_x(\psi,i^{p-1\over 2}\rho(e_1')\ldots\rho(e_p')\psi)=H_x(\rho( \psi'),i^{p-1\over 2}\rho(e_1)\ldots\rho(e_p)\psi')$, with $\psi'=\rho(\Lambda^{-1}\psi)$.

We can notice that the definition of positive spinor metrics takes on a particularly nice form in the anti-Lorentzian case. Since it is the case which concerns us the most, it is worth writing it down explicitly.

\begin{cor} A positive \emph{spinor metric} for a Clifford structure $({\cal S},\rho,\chi)$ on a space and time oriented  anti-Lorentzian manifold $(M,g,or,t-or)$  is a spinor metric such that  the form $H_x(.,\rho(v).)$ is positive-definite for every future-directed timelike tangent vector $v$. 
\end{cor}
Conversely, the above property characterizes the open future half-cones. All our spinor metrics will be positive from now on.

\begin{definition}\label{defspincstru}
A $(g,or,t-or)$-spin-c structure on a space and time oriented manifold is a tuple $(\S,\rho,\chi,H)$ where
\begin{enumerate}
\item $(\S,\rho,\chi)$ is a Clifford structure,
\item $H$ is positive spinor metric.
\end{enumerate}
\end{definition}

Thanks to theorem \ref{existsspinormetric}, we see that on a space and time oriented manifold, a spin-c structure exists iff a Clifford structure exists. This is probably the reason why Clifford structures are often overlooked. For instance, in \cite{gracia} the definition of (Riemannian) spin-c structures (definition 9.7) is exactly our definition of Clifford structures. However, the distinction between the two structures is important   because: 1) the morphisms are different, 2) a time-orientation plays a role in a spin-c structure in the non-Riemannian case.

We now come to the algebraic definition of spin structures\footnote{Note that this definition is  equivalent to the standard one and \emph{not} to the one  studied in \cite{bdt} and \cite{thelen} also under the name ``algebraic spin structure''.}. The adjoint of an operator $A$ with respect to $H$, and more generally with respect to an indefinite non-degenerate product, is denoted by $A^\times$.

\begin{definition}\label{defspinstru} A $(g,or,t-or)$-spin structure  is a tuple $( {\cal S},\rho,\chi,H,C)$ where $( {\cal S},\rho, \chi,H)$ is a $(g,or,t-or)$-spin-c structure and $C :  {\cal S}\rightarrow  {\cal S}$, is a bundle map which  is antilinear in the fibres, which defines a real structure\footnote{We use the convention  of \cite{gracia} here. The usual charge conjugation operator $J$ of noncommutative geometry is just $\chi C$ where $\chi$ is the chirality operator.} such that $Cl(T_xM,g_x)$ is the real part of $\CC l(T_xM,g_x)$ for $C_x$, and satisfies $CC^\times=\pm 1:= \kappa$.
\end{definition}

Note that in those circumstances $C^2$ is also necessarily equal to a sign $ \epsilon$. The charge conjugation operator is unique up to multiplication by a phase function $\zeta\in{\cal C}^\infty(M,S^1)$. For an anti-Lorentzian metric,  $ \kappa=1$ in all dimensions. More details on these signs will be given is section \ref{AB} below.


The definitions of isomorphisms of Clifford/spin-c/spin structures are the natural ones. For any operator $T: V\rightarrow V'$ we will (abusively) write $\Ad_T$ for the map $A\mapsto TAT^{-1}$.
\begin{definition}
\begin{enumerate}
\item Two  $(g,or)$-Clifford structures $( {\cal S},\rho,\chi)$ and $( {\cal S}',\rho',\chi')$ are called \emph{isomorphic} iff there exists a bundle isomorphism $\Sigma :  {\cal S}\rightarrow  {\cal S}'$ such that the following diagram commutes for all $x\in M$:
\be
\xymatrix{\CC l(T_xM,g)\ar[rr]^{\rho_x}\ar[d]^{\rho_x'}&&\End( {\cal S}_x)\ar[d]^{\Ad_{\Sigma_x}}\cr
\End( {\cal S}_x')\ar[rr]^{\id}&&\End( {\cal S}_x')
}\label{ssequiv}
\ee
(In particular $\chi'=\Ad_\Sigma(\chi)$.)
\item Two  $(g,or,t-or)$-spin-c structures $({\cal S},\rho,\chi,H)$ and $( {\cal S}',\rho',\chi',H')$ are isomorphic if there exists a bundle isomorphism $\Sigma$ such that
	\begin{enumerate}
\item $\Sigma$ defines an isomorphism of Clifford structures,
\item $\Sigma_x : ( {\cal S}_x,H_x)\rightarrow ( {\cal S}_x',H_x')$ is a Krein unitary transformation for all $x$,
	\end{enumerate}
\item Two  $(g,or,t-or)$-spin structures $( {\cal S},\rho,\chi,H,C)$ and $( {\cal S}',\rho',\chi',H',C')$ are isomorphic if there exists a bundle isomorphism $\Sigma$ such that
	\begin{enumerate}
\item $\Sigma$ defines an isomorphism of spin-c structures,
\item $C'= \Ad_\Sigma(C)$.
	\end{enumerate}
\end{enumerate}
\end{definition}
\begin{rem}  The automorphism group of a Clifford (resp. spin-c, resp. spin) structure is $\CC^*$ (resp. $S^1$, resp. $\ZZ_2$).
\end{rem}
Here is a first example of isomorphism of spin structures. As we have explained above, given a spin  structure $({\cal S},\rho,\chi,H,C)$, the set of all possible spin structures restricting to the same Clifford structure $( {\cal S},\rho,\chi)$ is $\{( {\cal S},\rho,\chi,\tau H,\zeta C)|\zeta\in{\cal C}^\infty(M,S^1),\tau\in{\cal C}^\infty(M,]0;+\infty[\}$.

\begin{propo}\label{equivmult}
The spin structures $( \S,\rho,\chi,\tau H, \zeta C)$  and  $(\S,\rho,\chi,\tau' H,\zeta' C)$ are isomorphic iff $\zeta'\zeta^{-1}$ has a square root, and when it is the case there exist  exactly two isomorphisms which are given by $(\Sigma\Psi)_x=  f(x)\Psi_x$ where $f$ is a square root of $\tau'\zeta'\tau^{-1}\zeta^{-1}$.
\end{propo}
\begin{demo}
{\small First let $f$ be as above. We have
\begin{eqnarray*}
\tau(x)H_x(\Sigma_x\Psi_x,\Sigma_x\Psi_x)&=&\tau(x)|f(x)|^2H_x(\Psi_x,\Psi_x)\\
&=&\tau'(x)H_x(\Psi_x,\Psi_x)
\end{eqnarray*}
and then we have $(\Sigma \zeta C\Sigma^{-1}\Psi)_x=f(x) \bar  f(x)^{-1}\zeta(x)C\Psi_x=\zeta(x)'C\Psi_x$. Moreover $\Sigma$ and $\rho$ clearly commute. This shows the existence part. For the uniqueness, notice that $\Sigma$ is required to commute with $\rho$, which shows that $(\Sigma\Psi)_x=f(x)\Psi_x$, with $f$ some function. The result follows easily.}
\end{demo}

\begin{rem}     Using the above proposition, we see that two spin-c structures reducing to the same Clifford structure are isomorphic.
\end{rem}

\begin{ex} Let us give an application of proposition \ref{equivmult}. Consider a frame $(e_1,e_2)$ on the  $2$-torus $T=(\RR/2\pi\ZZ)^2$, and define the Euclidean metric $g$ and the orientation $or$ such that $(e_1,e_2)$ is orthonormal and positive. Let $S_0=\CC^2$, $\gamma_1=\pmatrix{0&1\cr 1&0}$, $\gamma_2=\pmatrix{0&i\cr -i&0}$, $\chi_0=i\gamma_1\gamma_2=\pmatrix{1&0\cr 0&-1}$, $H_0$ is the canonical hermitian form on $\CC^2$ and $C_0\psi=\gamma_1 \psi^*$, where $\psi^*$ is the complex conjugate of $\psi\in\CC^2$. Let also $\rho_0$ be defined by $\rho_0(e_i)=\gamma_i$, $i=1,2$. Then $(T\times S_0,\rho_0,\chi_0,H_0,C_0)$ is a spin structure on $(T,g,or)$.  Now let $a_1,a_2\in\{0,1\}$ and define $\zeta_{a_1,a_2}(\theta_1,\theta_2)=e^{i(a_1\theta_1+a_2\theta_2)}$, where $\theta_{1,2}\in \RR/2\pi\ZZ$. Then it is immediate to see that $\zeta_{a_i,a_j}/\zeta_{a_k,a_l}$ has no square root for $(i,j)\not=(k,l)$. Thus $\ss_{a_1,a_2}:=(T\times S_0,\rho_0,\chi_0,H_0,\zeta_{a_1,a_2}C_0)$ gives four non-isomorphic spin structures on the torus. This exhausts all the isomorphism classes (see below).
\end{ex}
\subsection{Topological definitions}

We now come to the more usual ``topological'' definition of spin structures, via lifts of the frame bundle to a spin-principal bundle. We call $\lambda$ the covering map ${\rm Spin}(p,q)^0\rightarrow SO(p,q)^0$. We denote by ${\rm Fr}(M)$ the $SO(p,q)^0$-principal bundle of space and time oriented pseudo-orthonormal frames. This bundle can be constructed in the following way.  The frame bundle  ${\rm Fr}(M)$ can then be seen as the sub-bundle of ${\rm Hom}(M\times \RR^{p,q},TM)$   whose fibre at $x$ is the space of isometries $f : \RR^{p,q}\rightarrow T_xM$ sending the space and time orientations of $\RR^{p,q}$ to that of $T_xM$. The group $SO(p,q)^0$ acts on the right on ${\rm Fr}(M)$ by $f\cdot r:=f\circ r$.

\begin{definition}\label{topospin}
A \emph{topological spin structure} on $(M,g,or,t-or)$ is a pair $(P,\Lambda)$ where
\begin{enumerate}
\item $P$ is a ${\rm Spin}(p,q)^0$-principal bundle over $M$,
\item $\Lambda : P\rightarrow {\rm Fr}(M)$ is a $2$-fold covering bundle map such that
\begin{equation}
\Lambda(p\cdot a)=\Lambda(p)\cdot \lambda(a)
\end{equation}
for all $p\in P$ and $a\in {\rm Spin}(p,q)^0$.
\end{enumerate}
\end{definition}
This definition is adapted from \cite{fried}, p 35, to the semi-Riemannian space and time oriented context. We also take from this reference the definition of isomorphic spin structures.

\begin{definition}\label{topisom} Two topological spin structures $(P,\Lambda)$ and $(P',\Lambda')$ are called isomorphic if there exists a bundle map $f : P\rightarrow P'$ such that:
\begin{enumerate}
\item For all $u\in \Spin(p,q)^0$, $\forall p\in P$, $f(p\cdot u)=f(p)\cdot u$ (spin-equivariance), and
\item the diagram
$$\xymatrix{
P\ar[rr]^f\ar[dr]_{\Lambda}& & P'\ar[dl]^{\Lambda'}\\
& Q &\\
}$$
commutes.
\end{enumerate}
\end{definition} 

In appendix \ref{equivss}, we show\footnote{It had already been done in \cite{schroder} in the Euclidean case.} how to associate an algebraic spin structure to a topological one, and \emph{vice versa}, and that this association, which sends isomorphisms to isomorphisms, is a bijection at the level of equivalence classes. This enables us to import the known results on the obstruction to the existence of topological spin structures. Namely (\cite{Karoubi}, Prop. 1.1.26, \cite{baum}, p. 78):

\begin{theorem}
A spin structure exists on $(M,g,or,t-or)$ iff the second Stiefel-Whitney class of $M$ vanishes.
\end{theorem}

Note that a spin-c structure exists on  $(M,g,or,t-or)$  iff the third integral Stiefel-Whitney class of $M$ vanishes. As we already said, the obstruction of the existence of a Clifford structure is the same as that of a spin-c structure, by theorem \ref{existsspinormetric}.  Note that a four-dimensional non-compact anti-Lorentzian manifold admits a spin structure if and only if there exists a global section of the frame bundle \cite{geroch1}. If there exists such a globally defined $g$-orthonormal frame on $(M,g)$, it is said to be \emph{metric-parallelizable}. If $g$ is definite, metric-parallelizability is equivalent to parallelizability thanks to the Gram-Schmidt algorithm. In general, it is a stronger condition: see \cite{mounoud} for counter-examples.

We can also import the known results on the isomorphism classes of spin structures. Namely, that $H^1(M,\ZZ_2)\simeq \Hom(\pi_1(M),\ZZ_2)$ acts freely and transitively on the isomorphism classes of spin structures, which is thus a $\ZZ_2$-affine space (\cite{baum}, Satz 2.5). In particular if $\pi_1(M)$ is finitely generated, which is always the case if $M$ is compact, then there is a finite number of equivalence classes of spin structures, and this number is a power of $2$.

 

\subsection{Clifford, spin-c and spin connections}
Let us now define for future use three types of connections on the spinor bundle, each one being adapted to one of the structures introduced above. We write $\Hol_\lambda(\nabla)$ for the parallel transport operator of the connection $\nabla$ along the curve $\lambda$. We let $\nabla^{LC}$ be the Levi-Civita connection on $TM$ for the metric $g$. The parallel transport operator $\Hol_\lambda(\nabla^{LC})$ being an isometry from the tangent spaces at the extremities of $\lambda$, it can be extended to an operator between the corresponding Clifford algebras, which we denote by $\widetilde\Hol_\lambda(\nabla^{LC})$.

\begin{definition}\label{defcon}
\begin{enumerate}
\item Let $(\S,\rho,\chi)$ be an $(g,or)$-Clifford structure on $M$. Then a connection $\nabla$ on $\S$ is a \emph{Clifford connection}  iff for all $x,y\in M, a\in \CC l(T_xM)$, and $\lambda$ a curve from $x$ to $y$, the following diagram commutes:
$$\xymatrix{S_x\ar[rr]^a\ar[d]_{\Hol_\lambda(\nabla)}&& S_x\ar[d]^{\Hol_\lambda(\nabla)}\cr S_y\ar[rr]^{\widetilde{\Hol}_\lambda(\nabla^{LC}(a))}&& S_y}$$
\item Let $(\S,\rho,\chi, H)$ be a $(g,or,t-or)$-spin-c structure on $M$. Then a connection $\nabla$ on $\S$ is a \emph{spin-c connection}   iff it is a Clifford connection and the parallel transport operator $\Hol_\lambda$ is an isometry from $(S_x,H_x)$ onto $(S_y,H_y)$ for every $x,y\in M$ and every curve $\lambda$ joining $x$ to $y$.
\item Let $(\S,\rho,\chi, H,C)$ be a $(g,or,t-or)$ spin structure on $M$. Then a connection $\nabla$ on $S$ is a \emph{spin connection} iff it is a spin-c connection and the following diagram commutes for all $x,y,\lambda$:
$$\xymatrix{S_x\ar[r]^{C_x}\ar[d]_{\Hol_\lambda(\nabla)}& S_x\ar[d]^{\Hol_\lambda(\nabla)}\cr S_y\ar[r]^{C_y}& S_y}$$
\end{enumerate}
\end{definition}

\begin{rem}[1]
It is immediate that the holonomy operator ${\rm Hol}_\lambda(\nabla)$ for $\lambda$ a closed loop at $x$ and $\nabla$ a Clifford (resp. spin-c, resp. spin) connection    belongs to the Clifford group $\Gamma_\CC(T_xM,g_x)$ (resp. spin-c, resp. spin group).
\end{rem}
\begin{rem}[2]
The definitions above are given in the geometric (integrated) point of view on connections. To obtain the differential (infinitesimal) point of view one just has to use the formula
$$(\nabla_X \Psi)(x)=\lim_{t\rightarrow 0}{h_\lambda^{-1}\Psi(\lambda(t))-\Psi(x)\over t}$$
where $\lambda : [0;1]\rightarrow M$ is a curve such that $\lambda(0)=x$ and $({d\over dt}\lambda)(0)=X$, and the similar formula for $\nabla^{LC}$. The infinitesimal version of the defining property of Clifford connections is the ``Leibniz rule"
\be 
\nabla_X(a\cdot\Psi)=\tilde\nabla^{LC}_X(a)\cdot\Psi+a\cdot\nabla_X\Psi\label{Leibnizrule}
\ee
where $\tilde\nabla^{LC}$ is the canonical extension of the Levi-Civita connection to the Clifford bundle, $a$ is a Clifford field, $X$ a vector field and $\Psi$ a spinor field. The infinitesimal version of the isometric property of spin-c connection is the metricity property:
$$X\cdot H(\Psi,\Phi)=H(\nabla_X\Psi,\Phi)+H(\Psi,\nabla_X\Psi)$$
Finally the infinitesimal version of the last property of spin connection is that $\nabla_X$ commutes with $C$.
\end{rem}
\begin{rem}[3] Spin-c connections  appear in the literature on Seiberg-Witten invariants. The only appearance of Clifford connections we could locate is  \cite{bgv}.
\end{rem}

Two connections on $\S$ differ by an $\End(\S)$-valued 1-form, and it is easy to see that this 1-form must be scalar-valued in the case of Clifford connections. If the connections are also metric, the 1-form has values in $i\RR$, and if they commute with $C$ it has values in $\RR$, so we obtain the uniqueness of the spin connection. Moreover the spin connection always exists (see theorem 9.8 in \cite{gracia} for the Riemannian case, \cite{nadir}, chap. 4, for the general case). We will denote the spin connection $\nabla^\ss$, or $\nabla^\ss(g)$ if we want to emphasize the metric.

Finally, let us mention that given a spin structure $\ss=(\S,\rho,\chi, H,C)$, one can define the \emph{canonical Dirac operator} $D^\ss$ by the formula
\begin{equation}
D^\ss=-i\sum_\mu \rho((dx^\mu)^\sharp)\nabla_\mu^\ss=-i\sum_ag(e_a,e_a)\rho(e_a)\nabla_{e_a}^\ss
\end{equation}
where $\nabla^\ss$ is the spin connection, $\sharp$ is the musical isomorphism defined by $g$, and $(e_a)$ is a $g$-frame.

We end with a routine but important proposition.
\begin{propo}\label{spinequiv} Let $\ss$ and $\ss'$ be two $(g,or,t-or)$-spin structures, and $\Sigma$ an isomorphism from $\ss$ to $\ss'$. Then $\nabla^{\ss'}=\Sigma \nabla^\ss\Sigma^{-1}$ and $D^{\ss'}=\Sigma D^\ss\Sigma^{-1}$.
\end{propo}
\begin{demo}
{\small 
We have
\bea
\Sigma D^{\ss}\Sigma^{-1}&=&i\sum_\mu \Sigma\rho((dx^\mu)^\sharp)\nabla^{\ss}_\mu \Sigma^{-1}\cr
&=&i\sum_\mu \rho'((dx^\mu)^\sharp)\Sigma\nabla^{\ss}_\mu \Sigma^{-1}\nonumber
\eea
There thus only remains to show that $\Sigma\nabla^{\ss}  \Sigma^{-1}=\nabla^{\ss'}$. To do this we check that 
$\Sigma\nabla^{\ss}  \Sigma^{-1}$ has the three properties of definition \ref{defcon} and we conclude by the uniqueness of $\nabla^\ss$ as a connection with these properties.  Now   thanks to the following commutative cube:

$$\xymatrix{
& \S_x'\ar[rr]^{\rho'(v)}\ar@{-->}[dd]
& & \S_x'\ar[dd] 
\\
\S_x\ar[dd]_(0.3){{\rm Hol}_\lambda(\nabla^\ss)} \ar[ur]^{\Sigma_x} \ar[rr]^(0.6){\rho(v)}
& & \S_x \ar[ur]_{\Sigma_x} \ar[dd]^(0.3){{\rm Hol}_\lambda(\nabla^\ss)} 
\\
& \S_y'\ar@{-->}[rr]
& & \S_y'
\\
\S_y \ar[rr]^{\rho({\rm Hol}_\lambda(\nabla^{\rm LC})(v))}\ar@{-->}[ur]^{\Sigma_{y}}& & \S_y \ar[ur]^{\Sigma_{y}}
}
$$

we immediately obtain that $\Sigma {\rm Hol}_\lambda(\nabla^\ss)\Sigma^{-1}={\rm Hol}_\lambda(\Sigma \nabla^\ss \Sigma^{-1})$ has the intertwining property. Moreover:
\bea
H'({\rm Hol}_\lambda (\Sigma\nabla^\ss \Sigma^{-1})\Psi_x,{\rm Hol}_\lambda(\Sigma \nabla^\ss \Sigma^{-1})\Psi_x)&=&H(\Sigma_y{\rm Hol}_\lambda(\nabla^\ss)\Sigma_x^{-1}\Psi_x,\Sigma_y{\rm Hol}_\lambda(\nabla^\ss)\Sigma_x^{-1}\Psi_x)\cr
&=&H( {\rm Hol}_\lambda(\nabla^\ss)\Sigma_x^{-1}\Psi_x,   {\rm Hol}_\lambda(\nabla^\ss)\Sigma_x^{-1}\Psi_x)\cr
&=&H(\Sigma_x^{-1}\Psi_x,\Sigma_x^{-1}\Psi_x)\cr
&=&H'(\Psi_x,\Psi_x)\nonumber
\eea
And finally we have $[\Sigma \nabla^\ss\Sigma^{-1},C']=[\Sigma \nabla^\ss \Sigma^{-1},\Sigma C\Sigma^{-1}]=\Sigma [\nabla^\ss,C]\Sigma ^{-1}=0$.
}
\end{demo}

\subsection{Spin structures associated with moving frames}\label{sec25}
A moving frame $e$ is a field of bases of the tangent space.  In this section we suppose that $M$ admits globally defined moving frames, i.e. that $TM$ is trivial. This might appear to be a very restrictive condition, but we have seen that in the case which is the most interesting physically, that of four-dimensional anti-Lorentzian non-compact manifolds, it is not a restriction at all, since this is a necessary condition for the existence of spin structures. Moreover, and possibly more importantly, this condition allows us to encode both the metric and spin structure in a single dynamical object.

Let $e=(e_t,e_s)$ be a moving frame,  where $e_t$ gathers the first $p$ vectors and $e_s$ the $q$ last ones. Such a frame  defines a metric $g_e$ which is the unique metric of signature $(p,q)$ for which $e$ is a pseudo-orthonormal basis with vectors of $e_t$ (resp. $e_s$) of positive (resp. negative) squared length. It also defines an orientation $or_e$ and a time-orientation $t-or_e$, such that $e_s$ and $e_t$ are positively oriented. Let us now show how a moving frame  also defines a spin structure on $M$. 

\begin{rem} In this section the metric is determined by the frame, but at times a metric $g$ will be fixed in advance. In this circumstance, by ``frame'' we will always mean a pseudo-orthonormal frame with respect to $g$, or $g$-frame. We can also speak  of $(g,or)$-frame and $(g,or,t-or)$-frames, with obvious meanings. We will also sometimes call frames \emph{tetrads}, even though the space+time dimension is not retricted to $4$.
\end{rem}

We first need to define a standard spinor space $S_0=\CC^{2^{n/2}}$, which is a an irreducible representation  $\rho_0: Cl(p,q)\rightarrow \End(S_0)$. We let $\chi_0, H_0, C_0$ be the chirality, spinor metric and real structure defined by this representation. This involves a choice of positive scalar factor for $H_0$ and phase for $C_0$. We also define the gamma matrices (with down indices) to be $\gamma_a=\rho_0(\epsilon_a)$, where $(\epsilon_a)_a$ is the canonical basis of $\RR^{p,q}$. We also write $V_0=\rho_0(\RR^{p,q})=\Span_\RR\{\gamma_a|a=1,\ldots n\}$ and $V_0^\CC=\rho_0(\RR^{p,q}\otimes\CC)=\Span_\CC\{\gamma_a|a=1,\ldots,n\}$.

\begin{ex}
In dimension $1+3$, we can choose\footnote{We let dimension indices run from $1$ to $n$ in all the text, except in the Lorentz/anti-Lorentz cases, where we return to the more usual $0,\ldots,n-1$ convention.}
\be
\gamma_0=\pmatrix{0&I_2\cr I_2&0};\gamma_k=\pmatrix{0&\sigma_k\cr -\sigma_k&0},k=1,2,3\label{chiralbasis}
\ee
\be
H_0(\psi,\psi'):=\psi^\dagger\gamma_0\psi'\label{spinmet}
\ee
\be
\chi_0=\gamma_5=i\gamma_0\ldots\gamma_3=\pmatrix{-I_2&0\cr 0&I_2}
\ee
and
\be
C_0\psi=\gamma_5\gamma_2\psi^*\label{conj}
\ee
where $\psi^*$ is the complex conjugate of $\psi$. Note that $H_0$ is a spinor metric (or more precisely a Robinson form) thanks to the fact that the basis is chiral, i.e. that $\gamma_0^\dagger=\gamma_0$ and $\gamma_k^\dagger=-\gamma_k$, $k=1,2,3$. Concrete representations of gamma matrices, $H_0$, $\chi_0$ and $C_0$ in all dimensions and for all signatures can be found in \cite{bbb2}.
\end{ex}

\begin{propo}\label{tetradss} Let $e=(e_a)_{a=1,\ldots,n}$ be a moving frame. We define $\ss_e$ to be $({\cal S}_0,\rho_e,\chi_0,H_0,C_0)$, where $ {\cal S}_0=M\times S_0$, $\rho_e$ is the unique representation of the Clifford bundle such that
\be
\forall x\in M, (\rho_e)_x(e_a(x))=\gamma_a,\label{rep}
\ee
$\chi_0,H_0$ and $C_0$ are constant and defined (slightly abusively) to be the objects bearing the same name on each fibre. Then $\ss_e$ is a spin structure for the metric, orientation and time-orientation defined by $e$.
\end{propo}
The proof is immediate. We call a $\ss_e$ a \emph{tetradic} spin structure.

\begin{propo}\label{tetradsurj} A $(g,or,t-or)$-spin structure $(\S,\rho,\chi,H,C)$ is tetradic iff 
\begin{enumerate}
\item $\S=\S_0$, $\chi=\chi_0$, $H=H_0$, $C=C_0$,
\item $\rho_x(T_xM\otimes\CC)=V_0^\CC$ for all $x\in M$.
\end{enumerate}
\end{propo} 
\begin{demo}
{\small 
The necessity of these conditions is obvious. Let us show that they are sufficient. Let  us define  $e_a(x):=\rho^{-1}(x)(\gamma_a)$. Since $\rho$ is smooth, $e_a$ is a smooth section of $\CC l(TM)$. By the second condition, $e_a$ is a complex vector field, and   since $\gamma_a$ commutes with $C_0$, $e_a$ is a real vector field.  The moving frame $e=(e_a)_a$ is positively oriented since $\chi(x)=\chi_0$ and it is also time-oriented because the spinor metric $H_x=H_0$ is positive.}
\end{demo}
\begin{rem}
In dimensions $\le 4$, the second condition is not needed. Indeed, in these dimensions the facts that $e_a(x)$ commutes with $C$, anti-commutes with $\chi$ and satisfy $e_a(x)^\times=e_a(x)$ suffice to prove that $e_a(x)$ is a (real) vector.
\end{rem}

In the sequel we will only consider tetradic spin structures. We do not necessarily lose anything in doing so, as the next proposition shows.

\begin{theorem}\label{tetradeq} Let $(M,g,or,t-or)$ be   metric-parallelizable. If $H^1(M,\ZZ_2)$ is finite  and\footnote{Remember that the case $(p,q)=(1,1)$ has been excluded at the beginning.} $(p,q)\not=(2,2)$, then every spin structure on $(M,g,or,t-or)$ is isomorphic to a tetradic one. 
\end{theorem}

For a proof, see appendix \ref{preuvetetradss}. We show there by elementary means that the sets of equivalence of spin structures and tetradic spin structures are in bijection. It is quite possible that a more sophisticated approach would yield the result without the finiteness hypothesis, in the non-exceptional case. In the latter case, we show in the appendix that   there are counter-examples.  

\begin{ex} We gave as an example the four non-isomorphic spin structures $\ss_{0,0}$, $\ss_{1,0}$, $\ss_{0,1}$ and $\ss_{1,1}$ on the Euclidean torus  at the end of section \ref{algspinstru}. If $(a_1,a_2)\not=(0,0)$, $\ss_{a_1,a_2}$ is not tetradic since the real structure $\zeta_{a_1,a_2}C_0$ is not constant. According to the previous theorem, there must be an isomorphism from these spin structures to tetradic ones. It is explictitly given by  $\Sigma(\theta_1,\theta_2):=\pmatrix{e^{ia_1\theta_1}&0\cr 0&e^{ia_2\theta_2}}$. It  is an isomorphism from $\ss_{a_1,a_2}$ to the tetradic spin structure $\ss_{a_1,a_2}'=(T\times S_0,\rho_{a_1,a_2}',\chi_0,H_0,C_0)$,  where $\rho_{a_1,a_2}'$ is just $\rho$ pre-composed with the rotation of the tangent plane about an angle $a_2\theta_2-a_1\theta_1$.  
\end{ex}

Consider two tetradic spin structures $\ss=(\S_0,\rho,\ldots,C_0)$ and $\ss'=(\S_0,\rho',\ldots,C_0)$.  They are isomorphic exactly when there a exists a smooth $\Sigma : M\rightarrow \Spin(p,q)^0$ such that $\rho'=\Ad_\Sigma\circ \rho$.

\begin{definition} A \emph{spinomorphism} (of $(S_0,\chi_0,H_0,C_0,V_0)$) is a smooth map  $\Sigma : M\rightarrow \Spin(p,q)^0\subset \End(S_0)$. Spinomorphisms form a group, called the spinomorphism group and denoted by $\Gamma(\Spin(p,q)^0)$. 
\end{definition}

\subsection{Action of diffeomorphisms on spinor fields}

Let $\theta : M\rightarrow M$ be a diffeomorphism. The natural action of $\theta$ on spinor fields is by pushforward:
\be
\theta_*\Psi: y\mapsto \Psi(\theta^{-1}(y))\label{pfspinfield}
\ee
which sends a section of the spinor bundle $\S$ to a section of the pushed forward bundle $\theta_*\S$, whose fibre at $y$ is $\S_{\theta^{-1}(y)}$. Still, in order to interpret $\theta_*\Psi$ as a spinor field, we must provide the appropriate metric and   spin structure. The metric $g$ on $M$ is pushed forward by $\theta$ according to
\be
\theta_*g_{y}(T_{x}\theta(u),T_{x}\theta(v))=g_x(u,v)\label{pfmetric}
\ee
where $y=\theta(x)$. This formula can be neatly restated by   saying  that the tangent map $T_x \theta$ is an isometry from $(T_x M,g_x)$ to $(T_{y}M,\theta_*g_{y})$. This isometry can be extended to a Clifford algebra isomorphism $\widetilde{T_x\theta}$. By composition, the Clifford bundle $\CC l(M,\theta_*g)$ naturally acts on $ {\cal S}_0$ according to the following commutative diagram which defines $\theta_*\rho$:
\be
\xymatrix{
\CC l(T_x M,g_x)\ar[rr]^{\widetilde{Tx\theta}}\ar[d]_{\rho_x}&&\CC l(T_{y}M,\theta_*g_{y})\ar@{-->}[d]^{\theta_*\rho_{y}}\cr
\End(\S_x)\ar[rr]^\id&&\End(\theta_*\S_y)
}\label{pfspinc}
\ee

In other words we haved pushed by $\theta$ the representation of the Clifford bundle. Let us consider in particular what happens to the chirality. If $(e_1,\ldots,e_n)$ is a $(g,or)$-basis at $x$, then $(T_x\theta(e_1),\ldots,T_x\theta(e_n))$ is a $(\theta_*or,\theta_*g)$-basis at $y=\theta(x)$. Thus the chirality element $\theta_*\chi_y$ is natural defined to be
\bea
\theta_*\chi_y&=&\theta_*\rho_y(\widetilde{T_{x}\theta}(i^{{n\over 2}+q}e_1\ldots e_n))\cr
&=&\rho_x(i^{{n\over 2}+q}e_1\ldots e_n))\cr
&=&\chi_x
\eea
Now if we define similarly   $\theta_*H$ and $\theta_*C$ by
\be
\theta_*H_y=H_{x},\qquad \theta_*C_y=C_{x},
\ee
we obtain a push-forward of the complete spin structure.

\begin{propo}\label{pfss}  Let $\ss=(\S,\rho,\chi,H,C)$ be a $(g,or,t-or)$-spin structure. Then $\theta_*\ss:=(\theta_*\S,\theta_*\rho,\theta_*\chi,\theta_*H,\theta_*C)$ is a $(\theta_*g,\theta_*or,\theta_*t-or)$-spin structure.
\end{propo}
The proof is immediate and left to the reader.

However natural this action may be, it poses a serious problem in Noncommutative Geometry. To be specific, in NCG we want to encode the geometry in an algebraic structure where a   space   of sections of a \emph{fixed} spinor bundle has the prime role. Thus we will need to associate diffeomorphisms with   some particular operators on a \emph{fixed} space. Hence we do not want diffeomorphisms to change the  space of sections of the spinor bundle ! This difficulty disappears when the spinor bundle is trivial, since in this case $\theta_*\S=\theta$. In particular, on a metric-parallelizable manifold and with a tetradic spin structure, proposition \ref{pfss} yields 
\be
\theta_*\ss_e=(\S_0,\theta_*\rho_e,\chi_0,H_0,C_0)
\ee
Now, since any vector field can be pushed forward by $\theta$, we can push forward a moving frame $e$ to $\theta_*e$, defined by
\be 
\theta_*e_a(\theta(x))=T_x\theta(e_a(x))
\ee
We then have two natural actions of diffeomorphism on (tetradic) spin structures: $\ss_e\mapsto\theta_*\ss_e$ defined as above, and $\ss_e\mapsto\ss_{\theta_*e}$. Fortunately the two actions coincide.

\begin{propo} Let $e$ be a moving frame, $g_e$, $or_e$ and $t-or_e$ be the metric, orientation and time-orientation it defines. Then the two $(g_e,or_e,t-or_e)$-spin structures $\ss_{\theta_*e}$ and $\theta_*\ss_e$ coincide.
\end{propo}
\begin{demo}
{\small 
We only need to check the coincidence of the representations. By definition, $\rho_{\theta_*e}$ sends $\theta_*e_a$ to $\gamma_a$. On the other hand, $\theta_*\rho_e(\theta_*e_a)=\rho_e(e_a)=\gamma_a$ using (\ref{pfspinc}).}
\end{demo}
\begin{rem} Note that the chirality element stays the same even if $\theta$ reverses the orientation. The reason is that a change of sign in the volume form is compensated by   changes of signs in the representation of some basis vectors. 
\end{rem}
 

\subsection{Local transformation of spin structures and metrics}
Let us introduce one more operation on spin structures. We define the group 
$${\Gamma({\rm Gl}}(M))=\{r\in \Gamma(\End(TM))|r_x\mbox{ is invertible for all }x\in M\}$$
Hence the elements of ${\Gamma({\rm Gl}}(M))$ are fields of invertible endomorphisms of the tangent space, i.e. vertical automorphisms of the tangent bundle\footnote{In \cite{doss} ${\Gamma({\rm Gl}}(M))$ is called the ``gauge group'', but we will refrain doing so  to avoid any confusion.}. Let ${\rm Met}_{p,q}(M)$ be the space of  metrics on $M$ of signature $(p,q)$. If $g\in {\rm Met}_{p,q}(M)$, then we can define the metric $r\cdot g$ by
\be
r\cdot g_x(v,w)=g_x(r^{-1}(v),r^{-1}(w))
\ee
Hence, by construction, $r_x$ is an isometry from $(T_xM,g_x)$ to $(T_xM,r\cdot g_x)$. It is immediate to check that $g\mapsto r\cdot g$ defines a left action of ${\Gamma({\rm Gl}}(M))$ on ${\rm Met}_{p,q}(M)$. It is also immediate to define an action of ${\Gamma({\rm Gl}}(M))$ on orientations. Finally, as we prove in the lemma below, if $t$ is a time-orientation for $g$ then the   image  of $t$ under $r$ defines a time-orientation for $r\cdot g$.

\begin{lemma}\label{lemachin}  Let $\omega$ be a time-orientation $p$-form for $g$. Then $r_*\omega: (v_1,\ldots,v_p)\mapsto \omega(r^{-1}(v_1),\ldots,r^{-1}(v_p))$ is a time-orientation form for $r\cdot g$. Moreover, $\omega_1,\omega_2$ are in the same time-orientation class for $g$ iff $r_*\omega_1$ and $r_*\omega_2$ are in the same time-orientation class for $r\cdot g$.
\end{lemma}
\begin{demo}{\small 
Let $(v_1',\ldots,v_p')$ be a familly of linearly independent vectors at some point $x\in M$ and which are timelike for $r\cdot g$. Then $r^{-1}(v_1'),\ldots,r^{-1}(v_p')$ are linearly independent and timelike for $g$. Thus $\omega_x(r^{-1}(v_1'),\ldots,r^{-1}(v_p'))\not=0$ and this proves that $r_*\omega$ is a time-orientation form for $r\cdot g$. Now it is obvious that $r_*\omega_1$ and $r_*\omega_2$ take on the same sign on a familly $(v_1',\ldots,v_p')$ iff $\omega_1$ and $\omega_2$ do on the inverse image of this familly by $r$. Since $r$ bijectively sends linearly independent families of $g$-timelike to linearly independent families of $r\cdot g$-timelike vectors, the second part of the lemma is clear.  }
\end{demo}

\begin{propo}\label{vertactss}
Let $r\in {\Gamma({\rm Gl}}(M))$ and let $\ss=( {\cal S},\rho,\chi,H,C)$ be an $(g,or,t-or)$-spin structure. Then $r\cdot\ss:=( {\cal S},\rho\circ r^{-1},\chi,H,C)$ is a $(r\cdot g,r\cdot or,r\cdot t-or)$-spin structure. This defines a left action of ${\Gamma({\rm Gl}}(M))$ on the collection of spin structures. Moreover, if $\ss\simeq \ss'$ then $r\cdot\ss\simeq r\cdot \ss'$.
\end{propo}
\begin{demo}
{\small 
The proof that $({\cal S},\rho\circ r^{-1},r\cdot \chi)$ is a Clifford structure is immediate. Let us check that the orientation and chirality are consistent. Let $(e_1',\ldots,e_n')$ be a $(r\cdot g,r\cdot or)$-basis. By definition this implies that $(r^{-1}(e_1'),\ldots,r^{-1}(e_n'))$ is positive for $or$. Thus $\chi=i^{{n\over 2}+q}\rho\circ r^{-1}(e_1')\ldots \rho\circ r^{-1}(e_n')=\rho\circ r^{-1}(i^{{n\over 2}+q}e_1'\ldots e_n')$, which is precisely what we need. Then $H$ and $C$ remain adapted to this new Clifford structure thanks to the fact that $\rho\circ r^{-1}(T_xM)=\rho(T_xM)$ for all $x$. Finally let us prove that $H$ is still a positive spinor metric in the case $p/q$ odd, leaving the other case to the reader. Let $(e_1',\ldots,e_p')$ be a $r\cdot g$-orthonormal and $r\cdot t-or$-positively oriented family of timelike vectors.  Then by lemma \ref{lemachin},   $H(\Psi,i^{p-1\over 2}\rho\circ r^{-1}(e_1')\ldots\rho\circ r^{-1}(e_p') \cdot \Psi)>0$.

Let us now show that $r$ carries isomorphic spin structures to isomorphic spin structures. To do this, we look at the diagram below:
$$\xymatrix{
& \CC l(TM,r\cdot g) \ar[rr]^{\id}\ar@{-->}'[d][dd]
& &  \CC l(TM,r\cdot g) \ar[dd]
\\
\CC l(TM,g) \ar[ur]^{\tilde r} \ar[rr]^>>>>>>>>>>>{\id}\ar[dd]^\rho
& & \CC l(TM,g) \ar[dd]^{\rho'}\ar[ur]^{\tilde r}
\\
& \End(\S)\ar@{-->}[rr]^(0.4){{\rm Ad}_\Sigma}
& & \End(S)
\\
\End(\S) \ar[rr]^{{\rm Ad}_\Sigma}\ar@{-->}[ur]^\id
& & \End(\S) \ar[ur]^{\id}
}$$
The vertical  arrows at the back define the action of $\CC l(TM,r\cdot g)$ on $\S$  which is transported thanks to $\tilde r$. The front face is the isomorphism of $g$-Clifford structures. We need to check that that the face at the back also commutes, which will tell us that the $\tilde r$-transported $g'$-Clifford structures are also isomorphic. Since every other faces of the cube commute and every arrow is an isomorphism, the face at the back also commutes by the cube lemma (see \cite{mitchell}, p. 43). Now the isomorphism of the $\tilde r$-transported $g'$-spin structures is immediate since they are the same as the original $g$-spin structure and the intertwiner $Ad_\Sigma$ is the same for the action of both Clifford bundles.
}
\end{demo}


There is a simple characterization of the orbits of spin structures under ${\Gamma({\rm Gl}}(M))$ in the (anti-)Lorentzian case.

\begin{propo}\label{prop6}
Let  $\ss=( {\cal S},\rho,\chi,H,C)$ be an $(g,or,t-or)$-spin structure and $\ss'=( {\cal S},\rho',\chi,H,C)$ be a $(g',or',t-or')$-spin structure where $g$ and $g'$ are anti-Lorentzian (or Lorentzian) metrics on $M$. The following claims are equivalent.
\begin{enumerate}
\item There exists $r\in {\Gamma({\rm Gl}}(M))$ such that $g'=r\cdot g$, $or'=r\cdot or$, $t-or'=r\cdot t-or$ and $\ss'=r\cdot \ss$.
\item For all $x\in M$, $\rho_x'({\rm Fut}_x')=\rho_x({\rm Fut}_x)$, where ${\rm Fut}_x'$ and ${\rm Fut}_x$ are the future half-cones for $t-or'$ and $t-or$, respectively.
\end{enumerate}
\end{propo}
\begin{demo}
{\small 
$1\Rightarrow 2$: At every $x\in M$ and for any $v\in T_xM$ and any non-zero $\psi\in S$, one has $H(\psi,\rho'(v)\psi)>0$ iff $v\in {\rm Fut}_x'$ since $\ss'$ is a spin structure, and $H(\psi,\rho(v)\psi)>0$ iff $v\in {\rm Fut}_x$ since $\ss$ is a spin structure. Now since $\ss'=r\cdot \ss$, we have $\rho'=\rho\circ r^{-1}$ and it yields 2.

Conversely,  $(\rho')^{-1}\circ \rho : Cl(T_xM,g_x)\rightarrow Cl(T_xM,g_x')$ is an algebra  isomorphism, and by 2 it  sends $T_xM$ to $T_xM$. Hence it is the extension of an isometry $r_x : (T_xM,g_x)\rightarrow (T_xM,g_x')$, which depends smoothly on $x$ since it is the restriction to $TM$ of $(\rho')^{-1}\circ\rho$. To finish checking that $\ss'=r\cdot \ss$ we just need to prove that $or'=r\cdot or$ and $t-or'=r\cdot t-or$. The second part is just what 2 says. For the first part, consider a $g$-pseudo-orthonormal basis $(e_1,\ldots,e_n)$ which is positive for $or$. Then $\rho'(-i)^{{n\over 2}+n-1}r(e_1)\ldots r(e_n))=\rho(-i)^{{n\over 2}+n-1}e_1\ldots e_n)=\chi$. Since $\ss'$ is a spin structure, this proves that $(r(e_1),\ldots,r(e_n))$ is positive for $or'$. Thus $or'=r\cdot or$.}
\end{demo}

\subsection{A configuration space for GR}
We now want to define configuration space ${\cal C}$ for GR. Let us first consider the usual formulation, where the metric is the variable. The space ${\rm Met}_{p,q}$ can be given a topology by considering the quadratic form associated to a metric as a function $TM\rightarrow \RR$ and equipping the continuous functions on $TM$ with the compact-open topology \cite{mounoud,doss}. Recall that since $\RR$ is a uniform space, this topology is the topology of uniform convergence on compact subsets of $TM$ (\cite{kelley}, chap. 7).  A reasonable configuration space would be a path-connected component of ${\rm Met}_{p,q}$. In \cite{doss} there is a nice description of these components:

\begin{theorem}\label{metcomp}  Let $g_1,g_2\in {\rm Met}_{p,q}(M)$. Then the following are equivalent.
\begin{enumerate}
\item $g_1,g_2$ are in the same path-component,
\item $g_1,g_2$ are in the same ${\Gamma({\rm Gl}}(M))^0$-orbit,
\item each timelike subbundle for $g_1$ is homotopic to each timelike subbundle for $g_2$,
\item there exist a timelike subbundle for $g_1$ and a timelike subbundle for $g_2$ which are homotopic,
\item For each splitting of $TM=TM_+^1\oplus TM_-^1$ into timelike and spacelike subbundles for $g_1$, and each similar splitting  $TM=TM_+^2\oplus TM_-^2$    for $g_2$, there exists $r\in {\Gamma({\rm Gl}}(M))^0$ such that $TM_\pm^{2}=r(TM_\pm^1)$.
\end{enumerate}
\end{theorem}

\begin{rem} In particular ${\rm Met}_{n,0}$ and  ${\rm Met}_{0,n}$ are path-connected, but we do not need this result in these cases since these are convex cones.
\end{rem}

For the proof and the formal definition of homotopy of timelike bundles see 3.1, 3.11, 3.12 and 3.14 in \cite{doss}.   Here we are particularly interested with the second point. In it, ${\Gamma({\rm Gl}}(M))$ is also given the compact-open topology, considered as a subspace of ${\cal C}(TM,TM)$, and ${\Gamma({\rm Gl}}(M))^0$ is the path-component of the identity. Note that in our case, $M$ is parallelizable and the topologies on ${\rm Met}_{p,q}$ and ${\Gamma({\rm Gl}}(M))$ can be described in simpler terms. Indeed, fixing a frame, ${\rm Met}_{p,q}$ and ${\Gamma({\rm Gl}}(M))$ can be identified with  subsets of ${\cal C}(M,M_n(\RR))$ and it is easy to see\footnote{This can be proven as follows. First, let a metric $g$  be given by a map $G : M\rightarrow M_n(\RR)$. It also defines a function $TM\rightarrow \RR$ by $(x,v)\mapsto\bra v,G(x)v\ket$. Now  let $(g_k)$ be a sequence of metrics. It converges to $g$ for the compact-open topology on ${\cal C}(TM,\RR)$ if for all compact ${\cal K}\subset TM$, one has $\sup_{(x,v)\in {\cal K}}|\bra v,(G_k(x)-G(x))v\ket|\rightarrow 0$. Let us consider a compact $K$ of $M$ and let $S$ be the unit sphere in $\RR^n$. Then one has $\sup_{(x,v)\in K\times S}|\bra v,(G_k(x)-G(x))v\ket|=\sup_{x\in K}\|G_k(x)-G(x)\|$ where we use the fact that $G_k$ and $G$ are symmetric matrices and the norm we take on matrices is the operator norm. We thus immediately obtain that $G_k\rightarrow G$ uniformly on $K$. Conversely, it is immediate to prove that if $G_k\rightarrow G$ uniformly on $K$, then $g_k\rightarrow g$ uniformly on every compact of the form $K\times B(0,\rho)$, where $B(0,\rho)$ is the closed ball of radius $\rho$ in $M_n(\RR)$. This is sufficient to prove the uniform convergence on every compact of $TM=M\times \RR^n$. Let us now prove that the two topologies on ${\Gamma({\rm Gl}}(M))$ agree. First, when we see this as a subset of ${\cal C}(TM,TM)$ we must choose a uniform structure on $TM=M\times \RR^n$. We define it by choosing a distance function $d$ defining the topology of $M$ and the canonical norm on $\RR^n$. Thus $M\times V$ becomes a metric space under $\delta((x,v),(x',v'))=\max(d(x,x'),\|v-v'\|)$. Let $x\mapsto r(x)$ be an element of ${\Gamma({\rm Gl}}(M))$. It is identified with the function $\tilde r : TM\rightarrow TM$ by $\tilde r(x,v)=r(x)v$.  It is then easy to check that if $\tilde r_n$ converges uniformly to $\tilde r$ on $K\times S$, then $r_n$ converges uniformly to $r$ on $K$. Conversely, if  $r_n$ converges uniformly to $r$ on $K$, then $\tilde r_n$  converges uniformly to $\tilde r$ on $K\times B(0,\rho)$ for any $\rho>0$.  } that under this identification the compact-open topology described above becomes the topology of uniform convergence on compact subsets of $M$ of functions from $M$ to $M_n(\RR)$.

Hence, in view of   theorem \ref{metcomp}, the configuration space of GR could be taken to be ${\cal C}={\Gamma({\rm Gl}}(M))^0\cdot g$, where $g$ is some metric taken to fix the component. However we also need a spin structure to be able to consider fermion fields. More precisely, since the spin structure depends on the metric, we must assign to each metric in ${\cal C}$ a corresponding spin structure. Let us consider an origin $g$-spin structure $\ss$. Since every element of ${\cal C}$ is of the form $r\cdot g$, for $r\in{\Gamma({\rm Gl}}(M))^0$, we can assign the spin structure $r\cdot\ss$ to $r\cdot g$. However a problem arises, since if $r\cdot g=r'\cdot g$ with $r\not=r'$ , the transported spin structures will be different, and in general not even isomorphic.

The tetradic formulation of GR seems to fare better at first, since ${\Gamma({\rm Gl}}(M))^0$ acts freely on tetrads. Thus, if we fix an origin tetrad $e=(e_a)$, and set ${\cal C}={\Gamma({\rm Gl}}(M))^0\cdot e$, then the spin structure $r\cdot \ss$ is assigned in a unique way to $r\cdot e$. However, we now have an additional gauge symmetry, since two tetrads defining the same metric are physically equivalent. Thus if $r\cdot e$ and $r'\cdot e$ define the same metric, the spin structures $r\cdot\ss$ and $r'\cdot \ss'$ must be isomorphic\footnote{Another to put it is that the isometry sending one metric to the other must lift to an automorphism of the spinor bundle, that is, to a symmetry of the fermionic fields.}, which is not necessarily the case.  

We thus see that we must put some restriction on ${\cal C}$: it must be a ``small enough'' neighbourhood of a given  tetrad, so that the isomorphism class of the transported spin structure does not change. The existence of such a neighbourhood seems intuitively obvious, particularly when there is only a finite number of isomorphism classes of spin structure, but since these depend on the metric, it is not so easy to even give a meaning to this intuition. Appendix \ref{topospin} is devoted to this question. It is proved there that the isomorphism classes of spin structures are clopen sets  for the compact-open topology, at least when $H^1(M,\ZZ_2)$ is finite. It then easily results that there exists an open neighbourhood ${\cal U}$ of the unit   in ${\Gamma({\rm Gl}}(M))$ such that the following holds : 
\be 
\forall r,r'\in {\cal U},  (r\cdot g=r'\cdot g, r\cdot or=r'\cdot or, r\cdot t-or=r'\cdot t-or)\Rightarrow r\cdot\ss\simeq r'\cdot\ss\label{ppu}
\ee
Let us call ${\cal U}^0={\cal U}\cap {\Gamma({\rm Gl}}(M))^0$, where ${\cal U}$ is a maximal open neighbourhood of the unit with   property \eqref{ppu}. The configuration space we take for GR will be ${\cal C}(g,\ss):={\cal U}^0\cdot g$ for the metric formulation, and ${\cal C}(e):={\cal U}^0\cdot e$ for the tetrad formulation. Note that there are many cases where ${\cal U}^0$ is quite large. In particular ${\cal U}^0={\Gamma({\rm Gl}}(M))^0$ when $\Hom(\pi_1(M),\ZZ_2)=0$ or $pq=0$.

\begin{rem} Since all the trouble comes from the dependence of the spin structure on the metric, it would seem wise to use a metric-independent definition of the spin structure. Such a definition exists (\cite{swift}, \cite{dabdoss}, \cite{dabperc}), however, to the author's knowledge, only in Euclidean signature. The fact that $SO(n)$ is a retraction of $Gl(n)$ (via polar decomposition) is used crucially in these approaches, and for this reason one can doubt that they can be generalized to other signatures. 
\end{rem}

\section{Algebraic backgrounds, indefinite spectral triples, and spectral spacetimes}\label{AB}
\subsection{Motivating example: the canonical background of a spin manifold}

Let us choose an origin frame $e_0$, defining an origin metric $g_0$, and a corresponding configuration space ${\cal C}(e_0)$, as constructed in the previous section. Though the frame itself is not in the background, it does define some background structures, namely the orientation, time-orientation and spin structure defined by $e_0$, which we call $or_0, t-or_0$ and $\ss_0$. Note that this last object includes the standard spinor space $S_0$. We can add to the list $\Omega^1:=i\rho_0(TM)$. Indeed, for every frame $e$ in ${\cal C}(e_0)$, $i  \rho_e(TM)=\Omega^1$. Thus, while the frame $e$ is a dynamical object, $\Omega^1$ is not.

What we need to do now is to define an algebraic structure equivalent to the background structure we have and suitable for doing Noncommutative Geometry. The first step is to replace the manifold with an algebra. The one we choose is
\begin{equation}
\A=\tilde {\cal C}^\infty_c(M), \label{alg}
\end{equation} 
the algebra generated by  smooth functions\footnote{We consider real functions, since it will be important   to use real algebras when we come to the Standard Model. However the construction can be carried out without change with complex functions in the case of a manifold.} with compact support and the constant function $1$. In the Riemannian case, the second step is to replace the spinor bundle with the Hilbert space of square integrable sections of it. In the non-Riemannian case, there is a difficulty with the completion pointed out in \cite{doppler}. This is why we will only use sections with compact support. That is, we define 
\begin{equation}
\K=\Gamma^\infty_c({\cal S}),\label{KS}
\end{equation}
the   space of smooth spinor fields with compact support. We need a non-degenerate product on it, that we will call the Krein product, and there is no choice but to define
\be
(\Psi,\Phi)=\int_MH_x(\Psi_x,\Phi_x){\rm vol}_{g_0},\label{kps}
\ee 
where ${\rm vol}_{g_0}$, the semi-Riemannian volume element, is $|\det G_0|^{1\over 2}dx^1\wedge \ldots\wedge dx^n$, with $G_0$ the matrix of $g_0$ in the chosen local coordinates. The problem with \eqref{kps} is that it depends on $g_0$. More precisely, let us introduce the group of \emph{special vertical automorphisms}
\begin{equation}
\SVert(M)^0=\{r\in {\Gamma({\rm Gl}}(M))^0|\forall x\in M, \det(r_x)=1\}
\end{equation}
The action of $\SVert(M)^0$ on $\C(e_0)$ splits this space into orbits: let us call them  \emph{special classes}. In particular, we write $\SC(e_0)$ for the special class of $g_0$. We see that \eqref{kps} depends on $\SC(e_0)$. Now if we really want to define an algebraic structure which exactly encodes the background of GR, we should equip $\K$ not only with \eqref{kps}, but with the whole family of similarly defined Krein products for every special class in $\C(g_0)$. We will nonetheless proceed with the single Krein product \eqref{kps}. It would seem that doing so should yield a theory equivalent to unimodular gravity, but we will see that this is not the case: the symmetry group and configuration space will still be those of GR.


\begin{rem} It could be thought that instead of the single Krein product \eqref{kps} we should use the whole family of products $(.,.)_{[g]}$, where ${\rm vol}_{g}$ replaces ${\rm vol}_{g_0}$ and $[g]$ varies in all the possible special classes. This would lead to an alternative version of Algebraic Backgrounds in which ${\cal K}$ is endowed with a family of products $(.,.)_s$. Note that $\chi^\times$, $J^\times$ and $\pi(a)^\times$ could be taken to be independent of $s$, since it is the case for a canonical background. However an automorphism of such a structure would be a linear operator $U$ commuting with $\chi$ and $J$, leaving $\pi(\A)$ and $\Omega^1$ globally invariant  and such that the action of $U$ on the products $(.,.)_s$ is that of a permutation $f_U$ acting on $s$. Then, in the case of a canonical background, a scale transformation $S_r$ would be an automorphism. Moreover, the group of scale transformation would act transitively on the family $(.,.)_s$. Thus, if one considers an automorphism $U$, we can compose with a scale transformation in order that $S_rU$ leaves at least one product $(.,.)_{s_0}$ invariant. It entails that the group of automorphisms in this sense is generated by scale transformations, diffeomorphisms and spinomorphisms. Since such a group is larger that the one of tetradic GR, we discard this theory. However, it could be interesting to pursue it as a noncommutative generalization of scale invariant gravity theories  \cite{bsz}, \cite{bock}.
\end{rem}


\begin{rem} It is also important to note that in the Euclidean case, the operator norm of multiplication operators does not depend on $g_0$. Hence, in Connes' distance formula, using the Dirac operator associated to a metric $g$, it is the geodesic distance for $g$ which comes out and not the one for $g_0$ ! Note also that in this formula the metric $g$ enters through the definition of the gradient only.
\end{rem}

Let us go on with the list of elements of our structure. There is the  representation $\pi$ of the algebra on spinor fields   by pointwise multiplication:
\begin{equation}
(\pi(f)\Psi)_x:=f(x)\Psi_x,\label{ptmult}
\end{equation}
and also the chirality $\chi$ and real structure $C$, already present in $\ss_0$. Finally, there is the ``module of 1-forms''
\begin{equation}
\Omega^1:=i\rho(\Gamma^\infty_c(TM)).\label{defomega1}
\end{equation}
An element of $\Omega^1$ is thus the operator of Clifford multiplication by a compactly supported vector field, or equivalently, by $1$-form. Considering $\Omega^1$ as part of the background structure is the main point where our construction deviates from usual NCG. We have seen above the motivations for doing so. Let us now generalize the construction and give it a name.

\begin{definition} Let $(M,g_0,or_0,t-or_0)$ be a manifold, semi-Riemannian metric, orientation and time-orientation, and $\ss_0=(\S,\rho,\chi,H,C)$ an adapted spin structure. The tuple $(\A, \K, (.,.),$ $\pi,\chi,C,\Omega^1)$  defined by \eqref{alg}, \eqref{KS}, \eqref{kps}, \eqref{ptmult}, \eqref{defomega1} above is called \emph{the canonical background} over $(M,g_0,or_0,t-or_0,\ss_0)$. 
\end{definition}

Note that this definition makes sense even if $g_0,\ldots,\ss_0$ do not come from a moving frame. However, we will see below in theorem \ref{lemdiff} that we will eventually have to restrict to the tetradic case in order to deal with diffeomorphisms. 

The full notation for a canonical background will be $\B_{can}(M,g,or,t-or,\ss)$, but when we need to display only the dependence on the spin structure, we will write simply $\B_{can}(\ss)$, the other data being understood. As we have seen, a canonical background does not really depend on a metric, but only on its special class. In case $g,or,t-or$ and $\ss$ are given by a frame $e$, we will write $\B_{can}(e)$, even though the construction depends as well on the standard spinor  space $S_0$,  a particular choice of gamma matrices, $H_0$, $C_0$ and $\chi_0$. We will consider that $S_0,\gamma_a$, etc. are fixed once and for all.

As propositions \ref{tetradss} and  \ref{tetradsurj} show, the canonical background $\B_{can}(e)$ just defined comprises all the fixed objects when the moving frame takes all possible values. It is thus putatively the correct arena for a gravity theory, since it involves the differentiable structure\footnote{Through $\A$ and $\Omega^1$. Note that in finite dimension, $*$-algebras are automatically $C^*$-algebras, hence the algebra alone gives access only to the topological structure. In this case we can consider the differentiable structure to be entirely encoded in $\Omega^1$.}, but leaves the metric undetermined. In order to encode the metric, as we will see in details below, we will have to add to this structure a (variable) Dirac operator. When we do this, we obtain\footnote{Up to some  technical details into which we do not want to enter at this stage.} a Spectral Triple, or more precisely the semi-Riemannian generalization called Indefinite Spectral Triple (IST) (see \cite{thesenadir}), or Spectral Spacetime (SST), in the anti-Lorentzian case  (see \cite{part1,SST2}). 

\subsection{Pre-Krein spaces}
We now have to generalize canonical backgrounds to the noncommutative setting. As explained above, we cannot use Krein spaces. We will replace them with ``pre-Krein spaces''. So let us now discuss precisely what this means.

A \emph{pre-Krein space} is a vector space $\K$ equipped with a non-degenerate indefinite metric, which is decomposable into the direct sum $\K=\K_-\oplus \K_+$ of a positive and negative definite subspaces. Giving any such decomposition is equivalent to giving a fundamental symmetry $\eta$, which in turns defines a corresponding norm $\|.\|_\eta$. The completion of $\K$ with respect to this norm will be a Hilbert space $\H_\eta$, and the pair $(\H_\eta,\eta)$ will be a full-fledged Krein space. The key point is that $\H_\eta$ need not be unique, i.e. the $\eta$-norms need not be all equivalent.  

While there is not a unique norm on a pre-Krein space, we can still define a notion of \emph{universally bounded operator}.

\begin{definition} Let   $A$ be a linear or anti-linear  operator on $\K$. Let $\|A\|_\eta$ be the operator norm of $A$ subordinated to the $\eta$-norm. We say that $A$ is universally bounded if $\sup_\eta\|A\|_\eta<\infty$.
\end{definition}
We can observe that in the case of a canonical background, the operators $\chi$ and $C$, as well as the elements of the algebra and $\Omega^1$, are all universally bounded. 

Let $\B_u(\K)$ be the space of linear universally bounded operators on $\K$. It is clear that  $\B_u(\K)$ is a unital algebra and that \emph{the universal operator norm} 
\be 
\|A\|_u:=\sup_\eta\|A\|_\eta
\ee
defines a sub-multiplicative norm on it. Moreover, for all fundamental symmetry $\eta$, $(\B_u,*_\eta,\|\ \|_\eta)$  is a pre-$C^*$-algebra. Let us end this section with a lemma which will be useful later on.

\begin{lemma}\label{adub} Let $A\in B_u(\K)$ and $U$ be a  Krein-unitary operator. Then $UAU^{-1}$ is universally bounded and $\|UAU^{-1}\|_u= \|A\|_u$.
\end{lemma}
\begin{demo}
For every $\Psi\in\K$ and every fundamental symmetry $\eta$, we have
\bea
\|UAU^{-1}\Psi\|_\eta^2&=&(UAU^{-1}\Psi,\eta UAU^{-1}\Psi)\cr
&=&(AU^{-1}\Psi,U^{-1}\eta UAU^{-1}\Psi)\cr
&=&\|AU^{-1}\Psi\|_{U^{-1}\eta U}^2,\mbox{ since }U^{-1}\eta U\mbox{ is a fundamental symmetry}\cr
&\le&\|A\|_u^2\|U^{-1}\Psi\|_{U^{-1}\eta U}^2\cr
&\le&\|A\|_u^2(U^{-1}\Psi,U^{-1}\eta UU^{-1}\Psi)\cr
&\le&\|A\|_u^2\|\Psi\|_\eta^2
\eea
The converse inequality is shown by replacing $A$ with $U^{-1}AU$.
\end{demo}

{\small We will not use this result, but one can also show from the equality $\|A^{*_\eta}\eta\Psi\|_\eta=\|A^\times \Psi\|_\eta$ that $\|A^\times\|_\eta=\|A^{*_\eta}\|_\eta=\|A\|_\eta$. Thus $A^\times$ is universally bounded if $A$ is, with the same universal norm. 
}

We will need two equip our pre-Krein spaces with more structures: a chirality and a real structure. This yields the following definition.

\begin{definition} A $\ZZ_2$-graded real pre-Krein space is a pre-Krein space $\K$ equipped with a linear operator $\chi$ (chirality) and an antilinear operator $C$ (real structure) such that
\be 
\chi^2=1,\quad C^2=  \epsilon,\quad C\chi=  \epsilon''\chi C,\quad C^\times  =  \kappa C,\quad \chi^\times =  \epsilon''  \kappa''\chi\label{kosigns}
\ee
where $\epsilon,\kappa,\epsilon'',\kappa''$ are signs (``KO-metric signs''). A fundamental symmetry $\eta$ is said to be compatible with $\chi$ and $C$ iff 
\be
\chi\eta=\epsilon''\kappa''\eta\chi\mbox{ and }C\eta=\epsilon\kappa \eta C\label{compsigns}
\ee
\end{definition}
The Krein space $\K$ can be decomposed into even and odd subspaces, $\K=\K_0\oplus \K_1$, which are the eigenspaces of $\chi$. An operator $A$ which commutes with $\chi$ will respect this decomposition and will be called \emph{even}. If $A$  anticommutes with $\chi$ it will exchange $\K_0$ and $\K_1$ and be called \emph{odd}. We also say that $A$ is  \emph{$C$-real} if it commutes with $C$, and \emph{$C$-imaginary} if it anticommutes with it. Note that if $\epsilon''\kappa''=1$ then $\chi^\times=\chi$ and this implies that $\K_0$ and $\K_1$ are orthogonal with respect to $(.,.)$. In this case we will say that the Krein product is even. On the contrary if $\epsilon''\kappa''=-1$, $\K_0$ and $\K_1$ are self-orthogonal ($\K_i=\K_i^\perp$) and we say that the Krein product is odd.


Notice that given a general fundamental symmetry, the operators $C^{*_\eta}C=\eta C^\times \eta C$ and $\chi^{*_\eta}\chi=\eta\chi^\times \eta \chi$ must be positive, so that if $\eta$ satisfies commutation relations with $C$ and $\chi$, they are necessarily given by \eqref{compsigns}. For an example of non-compatible fundamental symmetry, consider $K=\CC^4$ equipped with the Krein product $H_0$, chirality $\chi_0$ and real structure $C_0$. Then for all $\lambda\in\RR$, $\eta_\lambda:=(1+{\lambda^2\over 2})\gamma_0+{\lambda^2\over 2}\gamma_0\chi_0+i\lambda\gamma_0\gamma_1$ can be checked to be a fundamental symmetry which has no particular commutation relation with either $\chi_0$ or $C_0$ when $\lambda\not=0$. However it has an odd and real part $(1+{\lambda^2\over 2})\gamma_0$, which can be rescaled to a compatible fundamental symmetry, and this is a general phenomemon. Indeed, let $\eta_\pm={1\over 2}(\eta\pm \chi\eta\chi)$. Then it is immediate to check that $\bra .,.\ket_{\eta_{\pm}}$ is a scalar product when $\chi^\times=\pm\chi$. We can then similarly decompose $\eta_\pm$ into the sum of a part commuting and a part anti-commuting with $C$, one of which yielding a scalar product.

\begin{rem}  In the case of the canonical background of a four dimensional anti-Lorentzian spacetime, the compatible fundamental symmetries are exactly given by the future-directed timelike vector fields, or equivalently by congruences of timelike observers. Two such symmetries yield equivalent $\eta$-norms iff the Doppler shift factor between the vector fields is bounded on the manifold. This result is proved and generalized to other signatures in  \cite{doppler}.
\end{rem}

On a $\ZZ_2$-graded real pre-Krein space, it is natural to define the universal norm by taking the supremum over compatible fundamental symmetries only. It follows from this definition that $\chi$ and $C$ are universally bounded. Lemma \ref{adub} still holds when $U$ is a Krein-unitary operator which commute with $C$ and $\chi$.

\subsection{Algebraic backgrounds: general definition}
We can now come to the main definition of this section. We emphasize that it is a working definition. We keep the number of axioms to a minimum, adding hypotheses when needed (in particular C0 and C1, see below).

\begin{definition}\label{algbgd} An \emph{algebraic background} (AB) is a tuple ${\cal B}=(\A, \K, (.,.),\pi,\chi,C,\Omega^1)$  where:
\begin{enumerate}
\item$(\K,(.,.),\chi,C)$ is a $\ZZ_2$-graded real pre-Krein space, 
\item $\A$ is an algebra and $\pi$ is  a  faithful  representation of it on $\K$ by universally bounded operators,
\item the   chirality operator  $\chi$ commutes with $\pi(a)$ for all $a\in \A$,
\item the ``bimodule of 1-forms'' $\Omega^1$ is an  ${\cal A}$-bimodule of universally bounded operators on $\K$ such that for any $\omega\in \Omega^1$, $\omega\chi=-\chi\omega$.
\end{enumerate}
\end{definition}
\begin{rem}[1] The real structure which is most commonly used in NCG is written $J$ and is related to $C$ by $J=\chi C$. We will call it the ``graded real structure''. 
\end{rem}
\begin{rem}[2] The reader might wonder why we do not mention any $C^*$-structure. We will come back to this issue in a little while.
\end{rem}

Apart from the motivating example of canonical background, one obtains an AB from a spectral triple, an IST, or a spectral spacetime by defining $\Omega^1$ to be the module of noncommutative 1-forms (see  definition \label{defdir} below), and forgetting about the Dirac operator.

The definition of algebraic backgrounds is supposed to generalize canonical backgrounds associated with semi-Riemannian spin manifolds. However, there is nothing in it from which we could retrieve the signature of the metric. The only thing we can do is to define the KO-metric dimension pair $(\mu,\nu)$ defined by table \ref{kometricsigns} below. In the case of a manifold with metric of signature $(p,q)$, one has $\mu=-p-q\ [8]$ and $\nu=q-p\ [8]$. For more on this subject see \cite{bbb2}. (Using the ungraded real structure as we do here is equivalent to using the  ``South-Coast convention'' of \cite{bbb2}, while the graded one corresponds to the West Coast convention. In this latter case, the definition of $(\mu,\nu)$ is changed to $\mu=p+q\ [8]$, $\nu=p-q\ [8]$.)
\begin{table}[hbtp]
\begin{center}
\begin{tabular}{|c||c|c|c|c|}
\hline
$\mu$,$\nu$&0&2&4&6\\
\hline
$ \kappa, \epsilon$&1&-1&-1&1\\
\hline
$ \kappa'', \epsilon''$&1&-1&1&-1\\
\hline
\end{tabular}
\end{center}     
\caption{KO-metric signs in terms of $\mu,\nu$.}\label{kometricsigns}
\end{table}
%

However, in the anti-Lorentzian case,   one can hope to do better at extracting the signature. Indeed, in this case fundamental symmetries are 1-forms. Generalizing to the noncommutative setting yields the following notion:

\begin{definition}
An algebraic background is of anti-Lorentz type if the KO signs are correct and if there exists an element $\beta\in \Omega^1$, called a \emph{time-orientation form}, which satisfies the following properties:
\begin{enumerate}
\item it is Krein self-adjoint,
\item it is real ( $C\beta C^{-1}=\beta$),
\item the hermitian form    $\bra .,.\ket_\beta:=( .,\beta^{-1}.)$ is   definite.
\end{enumerate}
\end{definition}

Let us go back to a general AB, and discuss several important hypothesis we can wish to add. The Krein adjunction and graded real structure  permit to define a crucial linear anti-automorphism of $\End(\K)$ which we denote $A\mapsto A^o$, where $o$ means ``opposite'', and is defined by:
\be 
A^o=JA^\times J^{-1}
\ee
The presence of the chirality here ensures that this definition matches the usual one in Noncommutative Geometry. There is thus a right representation of $\A$ on $\K$, defined by
\be 
\pi^o(a):=\pi(a)^o=C\pi(a)^\times C^{-1}\label{opprep}
\ee
An AB will be said to satisfy the \emph{order $0$ condition} (C0) if for all $a,b\in\A$ one has
\be 
[\pi(a)^o,\pi(b)]=0\label{C0}
\ee 
It will be said to satisfy the \emph{order $1$ condition} (C1) if for all $a\in \A$ and $\omega\in \Omega^1$, one has
\be 
[\pi(a)^o,\omega]=0\label{C1}
\ee 
One can also define a more complex order $2$ condition, but we won't need it in this article (see \cite{BF1,BBB}). To our knowledge, all interesting spectral triples considered so far satisfy the order $0$ condition. The canonical spectral triples attached to a manifold as well the spectral triple of Standard Model also  satisfy the order $1$ and $2$ conditions\footnote{The order $1$ condition on a spectral triple is $[\pi(a)^0,[D,\pi(b)]]=0$, which is    slightly weaker than the order 1 condition on the corresponding algebraic background. However, the combination C0+C1 is the same in both contexts.}. However, the spectral triple of the Pati-Salam model does not satisfy C1 \cite{ccvs}. In the anti-Lorentzian case, one can find natural (and commutative) examples attached to finite graphs which do not satisfy C1 either \cite{SST2}. 

We now come to Dirac operators.

\begin{definition}\label{defdirac} Let  $\B=(\A, \K, (.,.),\pi,\chi,C,\Omega^1)$ be an algebraic background.
\begin{enumerate}
\item A \emph{Dirac operator} on $\B$ is  a symmetric operator on $\K$ which anti-commutes with $\chi$ and $C$.
\item Given a Dirac operator $D$, the \emph{bimodule of noncommutative 1-forms for $D$} is denoted by $\Omega^1_D$ and defined to be the $\A$-sub-bimodule of $\End(\K)$ generated by commutators of $D$ with elements of $\pi(\A)$.
\item A Dirac operator $D$ on $\B$ is said to be \emph{compatible} if $\Omega^1_D\subset \Omega^1$. It will said to be \emph{regular} if $\Omega^1_D= \Omega^1$. The space of all compatible Dirac operators is called the \emph{configuration space} of $\B$, and written ${\cal D}_\B$.
\end{enumerate}
\end{definition}

\begin{rem} We will generally want the configuration space to be non-empty. In this case, any compatible Dirac $D$ provides a first-order differential calculus $(\Omega_D^1,D)$ over $\A$ in the sense of \cite{woronFODC}.  
\end{rem}

According to the second point in the above definition, $\Omega^1_D$ contains finite sums of products of the form $\pi(a)[D,\pi(b)]\pi(c)$, $a,b,c\in\A$. However, using $[D,\pi(b)\pi(c)]-\pi(b)[D,\pi(c)]=[D,\pi(b)]\pi(c)$ we see that it is enough to consider elements of the form $\pi(a)[D,\pi(b)]$. 

\begin{rem} Note that $\D_\B$ is a real vector space which contains the Krein self-adjoint 1-forms anticommuting with $C$.
\end{rem}

Though the notion of signature is still elusive in the general semi-Riemannian case, it will be useful to define the following terminology.

\begin{definition} Let $\B$ be an algebraic background and $D$ a compatible Dirac for $\B$.  Then the pair $(\B,D)$ is called an \emph{indefinite spectral triple} (IST). If $\B$ is of anti-Lorentz type, $(\B,D)$ is called a \emph{spectral spacetime} (SST). Moreover, if $\B$ is the canonical AB associated to $(M,g,or,t-or,\ss)$ and $D^\ss$ is the canonical Dirac operator, then $(\B,D^\ss)$ will be called the \emph{canonical IST} associated to $(M,g,or,t-or,\ss)$.
\end{definition}

These definitions are essentially equivalent to those given in \cite{thesenadir} (for IST) and \cite{SST2} (for SST), to which we refer for more details and examples.

Let us finally discuss the possibility of turning $\A$ into a  $C^*$, or pre-$C^*$-algebra. It is certainly a necessary hypothesis in order to hope for a reconstruction theorem in the commutative case. The reason why we have left the $C^*$-structure out of the general definition is that it is yet unclear how we should implement it for the different signatures. Let us be more specific. There is a Krein adjunction $\times$ on $\End(\K)$, and we can ask $\pi(\A)$ to be stable under $\times$. When it is the case, we say that $\pi(\A)$ is a \emph{$\times$-subalgebra}. Since $\pi$ is injective, we can transport $\times$ to $\A$, which becomes a $*$-algebra, and $\pi$ becomes a $*$-representation by definition. In the Euclidean case, $\times=*$ is the  Hilbert adjoint, hence the $C^*$-identity is satisfied on $\A$, which becomes a pre-$C^*$-algebra. In the anti-Lorentzian case, there are several assumptions to consider. We can ask:
\begin{enumerate}
\item  the SST to be \emph{reconstructible}, which means that there exists a time-orientation form $\beta$ such that $\pi(\A)$ is stable under $\pi(a)\mapsto \beta\pi(a)^\times \beta^{-1}$, or
\item\label{a2} $\pi(\A)$ to be a $\times$-subalgebra, or
\item\label{a3} $\A$ to a be a pre-$C^*$-algebra and $\pi$ to satisfy $\pi(a^*)=\pi(a)^\times$ for all $a\in\A$.
\end{enumerate}  
One can show that if  the SST satisfies C1,  these three assumptions are equivalent. Most authors in semi-Riemannian NCG have put assumption \ref{a3} in the axioms, but most of them also assume C1. However,   the SST attached to a finite graph is reconstructible   but it does not does not satisty \ref{a2} or \ref{a3}. This is why we consider reconstructibility as the priviledge ``$C^*$-assumption'' on spectral spacetimes (see \cite{SST2} for more on this question).  The more complex nature of noncommutative $p$-forms for $p>1$ (the existence of ``junk forms''), forbids any immediate generalization of reconstructibility to other signatures, however.

\subsection{Equivalence of algebraic backgrounds}
We now define the notion of equivalence   for algebraic backgrounds.

\begin{definition}
An equivalence between the algebraic backgrounds $\B=({\cal A},{\cal K},\ldots)$ and  ${\cal B}'=({\cal A}',{\cal K}',\ldots)$ is a Krein-unitary transformation $U$ such that $U\pi(\A)U^{-1}=\pi'(\A')$,  $UCU^{-1}=C'$, $U\chi U^{-1}=\chi'$ and $U\Omega^1U^{-1}=(\Omega^1)'$.
\end{definition}
%
An equivalence of $\B$ with itself will be called an automorphism, and the group of such automorphisms will be denoted by $\Aut(\B)$.

\begin{rem} We make no continuity hypothesis on $U$. We could think of requiring $U$ to be universally bounded, but it would be a bad idea.  As we will see below, a spinomorphism $\Sigma$ defines an important example of  AB automorphism $\Psi\mapsto (x\mapsto \Sigma_x\Psi_x)$ of the canonical background of a manifold, but it is not universally bounded. There is nonetheless some form of continuity coming from the requirement $U\pi(\A)U^{-1}=\pi'(\A')$ and $U\Omega^1U^{-1}=(\Omega^1)'$.
\end{rem}

\begin{rem}
It is worth noticing that an equivalence of algebraic backgrounds automatically stabilizes ${\mathcal C}\ell(A)_o$ and  ${\mathcal C}\ell(\A)_e$ as defined in \cite{dabrodandrea} or equivalently ${\mathcal C_D}(A)$ defined in \cite{lordrennievar}.
\end{rem}

Let $U$ be an AB equivalence as in the definition. For all $a\in \A$ there exists $a'\in\A'$ such that $U\pi(a)U^{-1}=\pi'(a')$. Since $\pi$ and $\pi'$ are faithful, the map $\alpha_U : a\mapsto a'$ is a well-defined isomorphism of algebra which makes the  diagram  

$$\xymatrix{{\cal A}\ar[r]^\alpha_U\ar[d]^\pi& {\cal A}'\ar[d]^{\pi'}\cr
B_u(\K)\ar[r]^{\Ad_U}& B_u(\K')
}$$
commute. In particular, if $\B=\B'$, we have an homomorphism $\alpha : U\mapsto \alpha_U$ from $\Aut(\B)$ to $\Aut(\A)$. We call $\Vert(\B)$ the kernel of $\alpha$. Its elements will be called \emph{vertical automorphisms of }$\B$. We thus have the exact sequence
$$\xymatrix{ 1\ar[r]&\Vert(\B)\ar[r]&\Aut(\B)\ar[r]^\alpha &\Aut(\A)
}$$
A recurring question in what follows will be whether $\alpha$ is surjective and has a section. When it is case, $\Aut(\B)$ will be the semidirect product $\Vert(\B)\rtimes \Aut(\A)$.

We now turn to the action of AB equivalences on Dirac operators.

\begin{propo} Let $U$ be as above. Then $D$ is a Dirac operator for $\B$ iff $UDU^{-1}$ is a Dirac operator for $\B'$. Moreover $D$ is compatible (resp. regular) iff $UDU^{-1}$ is. As a result, $\D_{\B'}=U\D_\B U^{-1}$. In particular,  ${\cal D}_{\B}$ is stable by automorphisms of ${\B}$.
\end{propo}
\begin{demo}
Clearly $UDU^{-1}$ is symmetric since $D$ is and $U$ is Krein unitary. The anticommutation relations with $UCU^{-1}$ and $U\chi U^{-1}$ are obvious.   Now an element of $\Omega^1_{UDU^{-1}}$ is a finite sum of terms like $U\pi(a)U^{-1}[UDU^{-1},U\pi(b)U^{-1}]=U\pi(a)[D,\pi(b)]U^{-1}$, with $a,b\in \A$. Thus $\Omega^1_{UDU^{-1}}=U\Omega^1_D U^{-1}$. Hence if $D$ is compatible, then $\Omega^1_{UDU^{-1}}\subset U\Omega^1 U^{-1}=(\Omega^1)'$ and $UDU^{-1}$ is also compatible. The converse to the previous statements are immediately obtained by replacing $U$ with $U^{-1}$.
\end{demo}
 

\begin{definition} The IST's or SST's $(\B,D)$ and $(\B',D')$ are equivalent iff there exists an AB equivalence $U$ from $\B$ to $B'$ such that $D'=UDU^{-1}$.
\end{definition}
Note that in the case of SST's, $U$ transports time-orientation forms to time-orientation forms.

\section{The case of a canonical background}
In this section we fix a space and time oriented semi-Riemannian manifold $(M,g,or,t-or)$ and  spin structure $\ss$. We will now compute the group of AB automorphisms of $\B_{can}$, as well as its configuration space.

\subsection{The automorphism group}
We need to translate  the  constructions of section \ref{mfss}  to AB equivalences.

\begin{propo}\label{prop7} Let $r\in{\Gamma({\rm Gl}}(M))$. Let $S_r\in\End(\Gamma^\infty_c(\S))$ be defined by
\begin{equation}
\forall x\in M, (S_r\Psi)_x=|\det r_x|^{-1/2}\Psi_x\label{defV}
\end{equation}
Then $S_r$ is an AB equivalence from $\B:=\B_{can}(M,g,or,t-or,\ss)$ to $\B':=\B_{can}(M,r\cdot g,r\cdot or,r\cdot t-or,r\cdot\ss)$.
\end{propo}
\begin{demo}
{\small 
Let $\B'=(\tilde {\cal C}^\infty_c(M),{\cal K}',\pi',\chi',C',\Omega_1')$. From proposition \ref{vertactss} we see that $\K'$ is the space of sections of the same bundle  $\S$ as $\K$, but equipped with the Krein product $(\Psi,\Phi)'=\int_M H_x(\Psi_x,\Phi_x){\rm vol}_{r\cdot g}$. Hence we obviously have $\pi'=\pi$, and moreover we have $\chi'=\chi$ by construction, and $\Omega_1'=i\rho\circ r^{-1}(\Gamma^\infty_c(TM))=i\rho(\Gamma^\infty_c(TM))=\Omega_1$. Moreover, let $R$ and $G$ be the matrices of $r_x$ and $g_x$ respectively, in some local coordinates. Then the matrix of $r\cdot g$ in the same coordinates is $\transp{R}^{-1}GR^{-1}$. Hence, $\sqrt{{\rm vol}_{r\cdot g}\over {\rm vol}_g}(x)=\left({\det R^{-2}|\det G|\over |\det G|}\right)^{1/4}=|\det r_x|^{-1/2}$. It follows easily from this that  $S_r$ is a Krein-unitary transformation from ${\cal K}'$ to ${\cal K}$. The other verifications are immediate.}
\end{demo}

We recover as a particular case of this proposition that if $r$ is unimodular, then $\B=\B'$. 

\begin{propo}\label{prop8} Let $\ss=(\S,\ldots)$ and $\ss'=(\S',\ldots)$ be two $(g,or,t-or)$-spin structures and let $\Sigma :  \S\rightarrow  \S'$ be an isomorphism between them. Define $U_\Sigma : \Gamma^\infty_c(\S)\rightarrow\Gamma^\infty_c(\S')$ by 
\be 
(U_\Sigma\Psi)_x=\Sigma_x\Psi_x\label{defusigma}
\ee
Then $U_\Sigma$ is an AB equivalence between $\B_{can}(\ss)$ and $\B_{can}(\ss')$.
\end{propo}
\begin{demo}
{\small 
The pre-Krein space ${\cal K}$ is $\Gamma^\infty_c(\S)$ with   product $(\Psi,\Phi)=\int_{M}H_x(\Psi_x,\Phi_x){\rm vol}_g$, whereas ${\cal K}'$ is $\Gamma^\infty_c(\S')$ with $(\Psi',\Phi')=\int_M H_x'(\Psi_x',\Phi_x'){\rm vol}_{g}$. Then  
\bea
(U_\Sigma\Psi,U_\Sigma\Psi)'&=&\int_{M}H_x'(\Sigma_x\Psi_x,\Sigma_x\Psi_x){\rm vol}_{g}(x)\cr
&=&\int_{M}H_x(\Psi_x,\Psi_x){\rm vol}_g(x)=(\Psi,\Psi)\nonumber
\eea
Hence $U_\Sigma$ is a unitary transformation. Now we immediately have by the definition of spin structure equivalences that $\rho'=\Ad_{U_\Sigma}(\rho)$, which yields $U_\Sigma\Omega^1U_\Sigma^{-1}=(\Omega^1)'$, $C'=\Ad_{U_\Sigma}C$, $\chi'=\Ad_{U_\Sigma}(\chi)$, $\pi'=\Ad_{U_\Sigma}(\pi)$.}
\end{demo}

In particular, if $\ss$ and $\ss'$ are tetradic, then $\Sigma$ is a spinomorphism, $\B_{can}(\ss)=\B_{can}(\ss')$ and $U_\Sigma$ is an automorphism of $\B_{can}(\ss)$ which induces the identity on the algebra of functions. It is is easy to see that every such automorphism arises in this way. We promote this observation to a theorem.

\begin{theorem}\label{uniquespino}
Let $\ss_e$ be a tetradic spin structure. Then $\Sigma\mapsto U_\Sigma$ is an isomorphism from the spinomorphism group $\Gamma(\Spin(p,q)^0)$ to $\Vert(\B_{can}(e))$.
\end{theorem}
\begin{demo}
{\small 
Let us show the surjectivity. Let $U$ be an automorphism of $\B_{can}(e)$ which commutes with $\A=\tilde{\cal C}_c^\infty(M)$. Then $U$ is of the form $(U\Psi)_x=\Sigma_x\Psi_x$ where the local operator $\Sigma_x$ belong to $\End(\S_x)$. Clearly $\Sigma_x$ must be unitary for $H_x$ in order for $U$ to be Krein unitary. Moreover, from the definitions of AB equivalence, we see that $\Sigma_x$ must stabilize $\Omega^1$ (which in particular implies that $x\mapsto\Sigma_x$ is smooth), commute with $C$ and $\chi$. This means by definition that $\Sigma$ is a spinomorphism. All the other verifications are trivial.}
\end{demo}

We now turn our attention to diffeomorphisms.

\begin{propo}\label{prop9} Let $\theta\in \Diff(M)$. Let  $U_\theta: \Gamma^\infty(\S)\rightarrow\Gamma^\infty(\theta_*\S)$ be defined  by
\be
(U_\theta\Psi)_y:=\Psi_{\theta^{-1}(y)}\label{pfspinorfield}
\ee
Then $U_\theta$ is an AB equivalence between $\B:=\B_{can}(M,g,or,t-or,\ss)$ and  $\B':=\B_{can}(M,\theta_*g,\theta_*or,\theta_*t-or,\theta_*\ss)$. Moreover, $U_\theta D^\ss(g) U_{\theta}^{-1}=D^{\theta_*\ss}(\theta_*g)$.
\end{propo}
\begin{demo}
First we have, from proposition \ref{pfss} and \eqref{pfspinorfield}:
\bea
(U_\theta \chi U_\theta^{-1}\Psi)_y&=&(\chi U_\theta^{-1}\Psi)_{\theta^{-1}(y)}\cr
&=&\chi_{\theta^{-1}(y)}(U_\theta^{-1}\Psi)_{\theta^{-1}(y)}\cr
&=&\theta_*\chi_y\Psi_y=(\theta_*\chi\Psi)_y\nonumber
\eea
Thus $U_\theta \chi U_\theta^{-1}\Psi=\theta_*\chi$. A similar calculation yields $U_\theta CU_{\theta}^{-1}=\theta_*C$. We also immediately obtain from diagram \eqref{pfspinc} that $U_\theta \Omega^1 U_\theta^{-1}=(\Omega^1)'$, where $\Omega^1=i\Gamma^\infty_c(\rho(TM))$ and $(\Omega^1)'=i\Gamma^\infty_c(\theta_*\rho(TM))$.

Let us  check that $U_\theta$ is a unitary transformation from $\K$ to  $\K'$ with the adequate Krein products:
\bea
( U_\theta\Psi,U_\theta\Psi')&=&\int_{M}\theta_*H_y( U_\theta\Psi_y,U_\theta\Psi_y') {\rm vol}_{\theta_*g}(y)\cr
&=&\int_{M}H_{\theta^{-1}y}( \Psi_{\theta^{-1}(y)},\Psi_{\theta^{-1}(y)}')\theta_*{\rm vol}_g (y)\cr
&=&\int_{\theta^{-1}(M)}H_x( \Psi_x,\Psi_x'){\rm vol}_{g}(x)\cr
&=&(\Psi,\Psi')
\eea
%

%
To finish showing that $U_\theta$ is an AB equivalence, we also need to check the intertwining property of $U_\theta$. For all $f\in {\cal C}^\infty(M)$, one has
\bea
(U_\theta\pi(f)U_\theta^{-1})\Psi_y&=&(\pi(f)U_\theta^{-1}\Psi)_{\theta^{-1}(y)}\cr
&=&f(\theta^{-1}(y))(U_\theta^{-1}\Psi)_  {\theta^{-1}(y)}\cr
&=&f(\theta^{-1}(y))\Psi_y\cr
&=&\pi'(\theta_*f)\Psi_y
\eea
%
We now need to check that $D(\theta_*g)=U_\theta D(g) U_\theta^{-1}$. This is done in \cite{dabdoss}, but here  are some indications for the reader's convenience. Let $(e_a)$ be a $(g,or,t-or)$-frame. Then $(\theta_*e_a)$ is a $(\theta_*g,\theta_*or,\theta_*t-or)$-frame. Hence we just need to prove that for all $a=1,\ldots,n$, we have $U_\theta \rho(e_a) U_\theta^{-1}=\theta_*\rho(\theta_*e_a)$, which is trivial, and that $\nabla^{\theta_*\ss,\theta_*g}_{\theta_*e_a}=U_\theta \nabla^{\ss,g}_{e_a}U_\theta^{-1}$. For this, it suffices to prove that for all vector field $Y$, one has

$$(\nabla^{\theta_*\ss,\theta_*g}_{Y}\Psi)_y=(U_\theta \nabla^ {\ss,g}_{(T_x\theta)^{-1}(Y)}U_{\theta}^{-1}\Psi)_y$$
For this we can define a connection $\nabla'$ on $ {\cal S}'$ by the RHS of the above equation, and use the uniquess property of the spin connection. That is, we just have to check that $\nabla'$ so defined is metric, preserves $U_\theta CU_{\theta}^{-1}$ and is Clifford. This is all easy abstract non-sense. The Clifford property of $\nabla'$ follows from the identity
$$T_x\theta(\nabla_X^g v)=\nabla_{T_x\theta(X)}^{\theta_*g}T_x\theta(v)$$
for the Levi-Civita connections of $g$ and $\theta_*g$ respectively. This identity is maybe most clearly seen to hold by comparing the parallel transport of both connections along a curve and its image by $\theta$.
\end{demo}

Now, let us look at the proof of proposition \ref{prop9} and look for the circumstances under which a diffeomorphism can be represented by an \emph{automorphism} of a canonical background. Clearly, a first condition is that the spinor bundle be trivial, so that $\K=\K'$. Moreover, in order for the transformation  to be Krein unitary, we must compensate for the scale factor and define
\begin{equation}
(V_\theta\Psi)_y=\sqrt{ 
{
{\rm vol}_{\theta_*g}\over {\rm vol}_g
}(y)}
\Psi_{\theta^{-1}(y)}\label{defvtheta}
\end{equation}
instead of $U_\theta$ (this formula can also be found, with a typo, in \cite{fgr}, p. 7). Note that this formula is meaningful even if $\theta$ reverses the orientation, since the semi-Riemannian volume form of any metric is a positive multiple of $dx_1\wedge \ldots\wedge dx_n$ on any positively oriented coordinate system, by definition. We clearly also have $V_\theta \chi V_\theta^{-1}=\theta_*\chi$, but this must be equal to $\chi$. Similarly we must have $\theta_*C=C$ and $\theta_*H=H$. The only possibility which works for every $\theta$ is to assume $\chi,H,C$ to be constant. The same is true of the space  $\rho_x(T_xM)$. But this is precisely the conditions which characterize tetradic spin structures, by proposition \ref{tetradsurj}. We thus obtain the following result.
\begin{theorem}\label{lemdiff} Let $M$ be a metric-parallelizable manifold and $\ss_e$ a tetradic spin structure on it. Let $\theta$ be a diffeomorphism of $M$, and $V_\theta : \K\rightarrow \K$ be defined by \eqref{defvtheta}. Then $V_\theta$ is an automorphism of $\B_{can}(e)$. Conversely, if \eqref{defvtheta} defines au automorphism of $\B_{can}(g,or,t-or,\ss)$ for every diffeomorphism $\theta$, then $(M,g)$ is metric-parallelizable and $\ss$ is tetradic.
\end{theorem}

Let us compute the scale factor $s(y)=\sqrt{ 
{
{\rm vol}_{\theta_*g}\over {\rm vol}_g
}(y)} $ in \eqref{defvtheta} using the $g$-frame $e$ as a basis of the tangent space at each point, and using local coordinates around $y$ such that $\partial_i=e_i(y)$ (normal coordinates).  We have ${\rm vol}_{\theta_*g}(\theta_*e_1,\ldots,\theta_*e_n)=\pm 1$ according to the orientation of $\theta_*e$, by definition of ${\rm vol}_{\theta_*g}$. Thus:
\bea
s(y)&=&(\pm 1{\rm vol_g}(\theta_*e_1,\ldots,\theta_*e_n)(y))^{-1/2}\cr
&=&(\pm {\rm vol_{g_y}}(T_x\theta(e_1(x)),\ldots,T_x\theta(e_n(x)))^{-1/2},\mbox{ with }x=\theta^{-1}(y)\cr
&=&|\det({\rm Mat}(T_x\theta))|^{-1/2}\cr
&=&|\det({\rm Mat}(T_x\theta)^{-1})|^{1/2}\cr
&=&|\det({\rm Mat}(T_y\theta^{-1})|^{1/2}
\eea
where ${\rm Mat}(T_x\theta))$ and ${\rm Mat}(T_y\theta^{-1})$ are written in the bases $e(x)$ and $e(y)$ of the respective tangent spaces. Hence, $V_\theta$ can also be written
\be 
(V_\theta\Psi)_y= 
|\det(T_y\theta^{-1})|^{1/2}
\Psi_{\theta^{-1}(y)},\label{vtheta}
\ee
and we immediately obtain that $\theta\mapsto V_\theta$ is a group homomorphism from $\Diff(M)$ to $\Aut(\B_{can}(e))$ by applying the chain rule.  


We can now obtain the structure of the automorphism group of a tetradic canonical background. Remember that we have an exact sequence
$$\xymatrix{ 1\ar[r]&\Vert(\B_{can}(e))\ar[r]&\Aut(\B_{can}(e))\ar[r]^\alpha &\Aut(\A)
}$$
We already know that $\Vert(\B_{can}(e))\simeq \Gamma(\Spin(p,q)^0)$, and it is easy to see that $\Aut(\A)\simeq \Diff(M)$. Indeed, let $\varphi$ be an automorphism of $\A=\tilde {\cal C}^\infty_c(M)$. By lemma \ref{adub}, $\varphi$ is continuous (and has in fact norm $1$). Thus $\varphi$ extends uniquely to the completion $\bar \A$. Now $\|f\|_u=\|f\|_\infty$, and the completion of $\tilde {\cal C}^\infty_c(M)$ for the supremum norm is $\tilde {\cal C}_0(M)$.  The automorphisms of this algebra are exactly the pullbacks by homeomorphisms, thus there exists an homeomorphism $\theta : M\rightarrow M$ such that $\varphi(f)=f\circ \theta^{-1}$, for all $f\in \tilde {\cal C}_0(M)$. Now $\varphi$ sends smooth functions to smooth functions, thus $\theta$ is a  diffeomorphism. We will continue to call $\alpha$ the map $\Aut(\B_{can}(e))\rightarrow \Diff(M)$ obtained by identifying $\Diff(M)$ with $\Aut(\A)$. Since $V : \Diff(M)\rightarrow \Aut(\B_{can}(e))$ is a group homomorphism and satisfies $\alpha\circ V=\id$, we obtain:

\begin{theorem}
Let $\ss_e$ be a tetradic spin structure.  The sequence 
$$\xymatrix{ 1\ar[r]&\Gamma(\Spin(p,q)^0)\ar[r]&\Aut(\B_{can}(e))\ar[r]^(0.55)\alpha &\Diff(M)\ar[r]&1
}$$
is exact, and $\alpha$ has a section. Thus  $\Aut(\B_{can}(e))\simeq \Gamma(\Spin(p,q)^0)\rtimes \Diff(M)$.
\end{theorem}
 

Note that $\Gamma(\Spin(p,q)^0)\rtimes \Diff(M)$   is exactly the group of symmetry of tetradic GR.

\subsection{Transformation of $k$-forms and $k$-vector fields under AB automorphisms}\label{transform}
Let us consider the canonical AB $\B_{\rm can}(e)$ associated with a moving frame. The vector space isomorphism $\Theta$ from $\Lambda TM$ to $\CC l(TM)$ allows to identify the sections of $\CC l(TM)$ with multivector fields, or, if we  compose with the musical isomorphism (defined with $g_e$), with differential forms. Either way, when we  compose with $\rho$, we obtain fields with values in $\End(\S)$. Let us now see how these fields transform under the AB automorphisms. We can work on the example of a simple $k$-vector field $u_1\wedge \ldots\wedge u_k$. It is readily checked that
\bea
U_\Sigma\rho(\Theta(u_1\wedge\ldots\wedge u_k))U_\Sigma^{-1}&=&\rho(\Theta(\Lambda u_1\wedge\ldots \wedge\Lambda u_k))\cr
V_\theta\rho(\Theta(u_1\wedge\ldots\wedge u_k))V_\theta^{-1}&=&\rho(\Theta(u_1\wedge\ldots\wedge u_k))\circ\theta^{-1}\label{transfield}
\eea
where in the first equation, $\Lambda$ is the element of $SO(TM)^0$ on which $\Sigma$ projects. Of course there is a natural action $\theta_*$ of the diffeomorphism $\theta$ on the $k$-vector fields, but we see that this natural action is not the one induced by $V_\theta$ (which exists only because the spinor bundle is trivial). Thus when we speak later of vector fields or forms, one should bear in mind that they are transformed according to \eqref{transfield}. This situation might seem strange. However, in QFT on curved spacetimes it is well-known that ``spinor fields transforms as spinors under local Lorentz transformations, and as scalar under diffeomorphisms''. Here the situation is the same, apart from the scale factor which cancels in the adjoint action, and the  equations \eqref{transfield} are the  consequence of the transformation law of spinors. Note that a virtue of \eqref{transfield} is to make quantities which appear in the Fermionic Lagrangian such as $(\Psi,\rho(v)\Psi)$, where $\Psi$ and $v$ are spinor and vector fields, respectively, manifestly invariant under diffeomorphisms.

\begin{rem}
In fact, if we wanted to keep the natural pushforward operation of vector fields, we would have to find a similar transformation for spinors, i.e. for every diffeomorphism $\theta : M\rightarrow M$, we would have to find a lift $\Sigma\theta : \S\rightarrow \S$ to the spinor bundle, which would allow to define $\theta_*\Psi$ by $(\theta_*\Psi)_y=\Sigma_x\theta(\Psi_x)$ ($y=\theta(x)$). This is a tricky business: see for instance \cite{dabdoss}, \cite{swift}, \cite{bourguignongauduchon}. While there is no such lift which is defined for all diffeomorphism and respects the chain rule, satisfactory solutions can be found, at least for diffeomorphisms close to the identity, and for the Riemannian signature. Anyway, we would certainly want our lift to make the following diagram commute for all $v\in T_xM$:
$$
\xymatrix{
\S_x\ar[r]^{\rho(v)}\ar[d]^{\Sigma_x\theta}& \S_x\ar[d]^{\Sigma_x\theta}\\
\S_x\ar[r]^{\rho(T_x\theta(v))}& \S_x
}
$$
Under a diffeomorphism, the quantity $H(\Psi,\rho(v)\Psi)_y$ would thus be transformed to $H(\Sigma_x\theta \Psi_x,\rho(T_x\theta(v_x))\Sigma_x\theta \Psi_x)$ ($y=\theta(x)$), which by the above diagram is exactly $H(\Psi_x,\rho(v_x)\Psi_x)$, i.e. $H(V_\theta\Psi, (V_\theta\rho(v)V_\theta^{-1})V_\theta\Psi)_y$. Thus the transformation law $(\Psi,v)\rightarrow (\theta_*\Psi,\theta_*v)$ yields the same result as the (simpler) rule $(\Psi,\rho(v))\rightarrow (V_\theta\Psi, V_\theta\rho(v)V_\theta^{-1})$. A similar situation arises in Quantum Field Theory, where the transformation law of a spinor field $\Psi$ under an element of the Lorentz group $\Lambda$ is generally written like  (see for instance \cite{peskinschroeder}, equation 3.8):
\be
\Psi(x)\rightarrow \tilde \Lambda \Psi(\Lambda^{-1}x)\label{recipe}
\ee
where $\tilde \Lambda$ is one of the two lifts of $\Lambda$ to the spin group. In our notations this can also be written $\Psi\rightarrow U_{\tilde \Lambda}V_\Lambda \Psi$.  If we consider $\Lambda$ as a diffeomorphism, our  transformation law for $\Psi$ is  $\Psi\rightarrow V_\Lambda\Psi$. The two points of view are reconciled when scalar products, such as $\bra \Psi,\Psi'\ket_{\gamma_0}:=(\Psi,\gamma_0\Psi')$, are computed.  In QFT,  $\gamma_0=\rho(e_0)$ is considered to be fixed. Thus the transformation of the scalar product $\bra \Psi,\Psi'\ket_{\gamma_0}$ is
\bea
\bra \Psi,\Psi'\ket_{\rho(e_0)}&\rightarrow &\bra U_{\tilde \Lambda}V_\Lambda\Psi,  U_{\tilde \Lambda}V_\Lambda\Psi'\ket_{\rho(e_0)}\cr
&=&( U_{\tilde \Lambda}V_\Lambda\Psi,\rho(e_0)  U_{\tilde \Lambda}V_\Lambda\Psi')\cr
&=&(V_\Lambda \Psi,U_{\tilde\Lambda}^{-1}\rho(e_0)U_{\tilde\Lambda}V_\Lambda\Psi')\cr
&=&(V_\Lambda \Psi,\rho(\Lambda^{-1}e_0)V_\Lambda\Psi')\cr
&=&\bra V_\Lambda\Psi,V_\Lambda\Psi\ket_{\rho(\Lambda^{-1}e_0)}
\eea
We see that if we take the fundamental symmetry into account, the transformation law of QFT is really $(\Psi,\gamma_0)\rightarrow (U_{\tilde \Lambda}V_\Lambda\Psi,\gamma_0)$, and is  equivalent to  $(\Psi,e_0)\rightarrow (V_\Lambda\Psi,\Lambda^{-1}e_0)$.
\end{rem}

\subsection{The configuration space}
First, let us introduce some notations. Let $e$ be a moving frame on $M$. Then we write $D(e)$ for the canonical Dirac operator associated with the metric $g_e$ and spin structure $\ss_e$. Let $\omega$ be a section of $\Lambda TM$ and $\zeta_\omega$ be the multiplication operator by $\rho(\Theta(\omega))$. We then have:


\begin{theorem}\label{configcan} Let $\B=\B_{can}(e)$ and $D$ be a regular element of $\D_\B$. There exist a unique element $r\in{\Gamma({\rm Gl}}(M))$, and a unique element $\omega\in\Gamma(\bigoplus_{m\ge 0}\Lambda^{3+4m}TM)$   such that
\be
D=\delta_r+i\zeta_\omega\label{decdir}
\ee
where $$\delta_r=S_r^{-1} D(r\cdot e)S_r.$$
Conversely, the operators of the above form are all regular elements of  $\D_\B$.
\end{theorem}
\begin{demo}
First, it is easily verified that \eqref{decdir} defines an element of $\D_\B$. Conversely, since $[D,.]$ is a derivation of $\A=\tilde{\cal C}^\infty_c(M)$ into $\Omega^1$, we see that we have
$$[D,f]=i\sum_a\delta_a(f)\gamma^a$$
for all $f$, where $\delta_a$, $a=0,\ldots,n-1$, are derivations of the algebra $\A$. There thus exist vector fields $X_a$ such that
$$[D,f]=i\sum_a (X_a\cdot f)\gamma^a$$
Now on any coordinate chart, we can expand $X_a$ on the local basis $\partial_\mu$, and we see (introducing a minus sign for convenience) that
$$[D,f]=-i\sum_\mu \rho_e(v_\mu)\partial_\mu f$$
where $v_\mu$ is a local section of $TM$. Hence $D$ must have the form
\begin{equation}
D=-i\sum_\mu \rho_e(v_\mu)\partial_\mu+\zeta_1\label{decompdir}
\end{equation}
where $\zeta_1$ is an operator which commutes with every function and is thus of order $0$. Note that the family $(v_\mu)$ is linearly independent since $D$ is regular. We now look for a vertical automorphism $r$ of the tangent bundle, defined locally for the moment, such that 
\begin{equation}
\rho_e(v_\mu)=\rho_{r\cdot e}((dx^\mu)^{\sharp})\label{meteq}
\end{equation}
where $\sharp$ is the musical isomorphism relative to $r\cdot g_e$. In the local basis $(\partial_\mu)$, $(dx^\mu)^{\sharp}=(G')^{-1}\partial_\mu$, where $G'$ is the matrix of $r\cdot g_e$. Now if $G$ is the matrix of $g_e$, we have $G'=\transp{R}^{-1}GR^{-1}$, with $R$ the matrix of $r$. Moreover, $\rho_{r\cdot e}=\rho_e\circ r^{-1}$. Thus \eqref{meteq} translates matricially as $V_\mu=R^{-1}RG^{-1}\transp{R}\partial_\mu$ for all $\mu$, i.e. $V=G^{-1}\transp{R}$, where $V$ is the matrix whose columns are $V_\mu$. We thus find a unique solution $R$, which is invertible since $V$ is. Since on any local chart, we have a unique $r$ which satisfies \eqref{meteq}, we see that it can in fact be defined globally.

Now let us write \eqref{decompdir} in the form $D=\tilde D+\zeta_1$. We have just proven that $D(r\cdot e)$ has the same first order part as $D$, so we can write $D (r\cdot e)=\tilde D+\zeta_2$, where $\zeta_2$ is of order 0. Neither $D(r\cdot e)$ nor $\zeta_2$ is Krein selfadjoint, but using proposition \ref{prop7}, we know that $S_r^{-1}D(r\cdot e)S_r$ is a Dirac operator for $\B$. Now $S_r^{-1}D(r\cdot e)S_r$ differs from  $D(r\cdot e)$ only by a zero-order term $\zeta_3$. Thus we can write $D=S_r^{-1}D(r\cdot e)S_r+\zeta$, where $\zeta=\zeta_1-\zeta_2-\zeta_3$ is a zero-order term.  Since $\End(\S_x)\simeq \CC l(T_xM)$ for all $x$, there exists $\omega\in\Gamma(\Lambda TM\otimes \CC)$ such that $\zeta=\zeta_\omega$. Now both $D$ and $S_r^{-1}D(r\cdot e)S_r$ belong to $\D_\B$, thus $\zeta_\omega$  also does, and consequently anticommutes with $\chi$, $C$ and is  Krein-selfadjoint. Let $\omega=\sum_k\omega_k$ with $\omega_k=\lambda_ke_{i_1}\wedge \ldots \wedge e_{i_k}$. The anticommutation with $\chi$ forces $k$ to be odd, and then the anticommutation with $C$ forces $\lambda_k$ to be pure imaginary. Finally, Krein-selfadjointness is satisfied iff $\rho(e_{i_1}\ldots e_{i_k})^\times=(-1)^{k-1\over 2}\rho(e_{i_1}\ldots e_{i_k})=-\rho(e_{i_1}\ldots e_{i_k})$, which yields $k=3+4m$.

To prove the uniqueness of the decomposition, suppose $\delta_r+\zeta_\omega=\delta_{r'}+\zeta_{\omega'}$. Looking at the first-order part we obtain $r=r'$, and the result follows.
\end{demo}
We   call $\delta_r$ a \emph{frame Dirac}, and $\zeta_\omega$ a \emph{centralizing field}. It is an important fact that the AB automorphisms act separately on these two parts (they preserve the decomposition). Let us first look at the action of diffeomorphisms on the frame part. We have the commutation rule:
\be 
V_\theta S_rV_\theta^{-1}=S_{\theta_*r}
\ee
where $(\theta_*r)_y=T_x\theta r_xT_y\theta^{-1}$, where $y=\theta(x)$. Moreover, $\theta_*(r\cdot e)=\theta_*r\cdot e$ and $\theta_*(r\cdot \ss_e)=\theta_*\ss_{r\cdot e}=\ss_{\theta_*r\cdot e}$. 

Thus 
\bea
V_\theta\delta_r V_{\theta}^{-1}&=&V_\theta S_r^{-1} D(r\cdot e)S_rV_{\theta}^{-1}\cr
&=&S_{\theta_*r}^{-1}V_\theta D(r\cdot e)V_\theta^{-1}S_{\theta_*r}\cr
&=&S_{\theta_*r}^{-1}D(\theta_*r\cdot e)S_{\theta_*r}\cr
&=&\delta_{\theta_*r}
\eea
Let us now look at the action  of a spinomorphism $\Sigma$. Since $\Sigma_x$ belongs to $\Spin(p,q)^0$, there is a unique element $r'\in\Gamma({\rm SO}(p,q)^0)$ such that $\Sigma \rho\Sigma^{-1}=\rho\circ (r')^{-1}$. The image of the spin structure $r\cdot\ss_e$ by $\Sigma$ is  $(rr')\cdot \ss_e$ (since $\rho\circ (r'r)^{-1}= \Sigma \rho\circ r^{-1}\Sigma^{-1}=\rho\circ (r')^{-1}\circ r^{-1}$). Recall that $r'\in {\rm SO}(p,q)^0$ means that $r'$ is an isometry for $g_e$. Thus s$(rr')\cdot g_e(v,w)=g_e((r')^{-1}r^{-1}v, (r')^{-1}r^{-1}w)=r\cdot g_e(v,w)$. Moreover, it is immediate that $U_\Sigma$ commutes with $S_r$, and that $S_{rr'}=S_r$ since $\det(r_x')=1$ at each $x$. Thus we can write the commutation relation
\be
U_\Sigma S_r U_\Sigma^{-1}=S_{rr'} 
\ee 
Finally, we already know from proposition \ref{spinequiv} that $U_\Sigma D(r\cdot g)^{r\cdot \ss_e}U_\Sigma^{-1}=D(r\cdot g)^{(rr')\cdot \ss_e}$, writing spin structures in exponents. But $D(r\cdot g)^{(rr')\cdot \ss_e}=D(rr'\cdot g)^{(rr')\cdot \ss_e}=D((rr')\cdot g)$ since $r'$ is an isometry for $g_e$. From all this, we obtain
\bea
U_\Sigma \delta_rU_\Sigma^{-1}&=&U_\Sigma S_r^{-1} D(r\cdot g) S_r  U_\Sigma^{-1}\cr
&=&S_{rr'}^{-1} D(rr'\cdot g) S_{rr'}\cr
&=&\delta_{rr'}
\eea

As for the centralizing fields, their transformation rules under AB automorphisms were already given in section \ref{transform}. We can either use them to show that the space of centralizing Dirac operators is stable under AB automorphisms, or simply observe that if $\zeta$ is centralizing, then $U\zeta U^{-1}$ also is since ${\rm Ad}_U$ preserves the algebra.



\subsection{Projecting out the centralizing fields}
We can summarize the results of the previous section by the direct sum:
\be 
D_\B=\Delta_\B\oplus Z_\B:\label{eq51}
\ee
where $\Delta_\B$ is the space of frame Dirac operators, and $Z_\B$ the space of centralizing fields. Moreover, $\Delta_\B$ and $Z_\B$ are stable under the adjoint action of $\Aut(\B)$. Thus, the projection on $\Delta_\B$:
\bea
P_\Delta\qquad D_\B  &\longrightarrow & D_\B\cr
 \delta+\zeta &\longmapsto&\delta\label{eq52}
\eea
satisfies $P_\Delta(UDU^{-1})=UP_\Delta(D)U^{-1}$ for any $U\in \Aut(\B)$ and $D\in \D_\B$.

Let us now consider the physical interpretation of these results. First we note that although we had to restrict the configuration space of tetradic GR to some open set of the space of moving frames, the configuration space of the corresponding AB is much larger: it  not only allows for Dirac operators associated with any moving frame, but also for centralizing fields. Our goal was to recast tetradic GR in the language of algebraic backgrounds. Since we have more degrees of freedom than we need, we have either  to get rid of them or explain why they exist and  haven't yet been seen. One way to do the former is to write down an action in which the elements of $Z_\B$ appear as auxiliary fields. Let $f$ be the bijection which sends a moving frame $r\cdot e$ to $\delta_r$, $S_{\rm EH}$ be the  Einstein-Hilbert action, and $P_Z=\id_{\D_\B}-P_\Delta$. Then let 
\be 
S(D):=S_{\rm EH}(g_{f^{-1}(P_\Delta(D))})+\int_M\tr_{S_0}(P_Z(D)^2){\rm vol}_{g_e}(x)
\ee
It is easy to check that it is invariant under $\Aut(\B)$. Since the ``Krein-Schmidt product'' $(A,B)\mapsto\tr_{S_0}(A^\times B)$ is non-degenerate, $D$ is a solution of $\delta S=0$ iff $P_Z(D)=0$ and $P_\Delta(D)=\delta_r$, with $r\cdot g_e$ a solution of Einstein's equations.  Of course, this formula is of little help in practice since $f$ is analytically very involved. We would rather have $S$ be defined by a simple function of $D$, such as the Spectral Action, but unfortunately this is not available in the semi-Riemannian case. Nonetheless, we conclude from this study of canonical backgrounds, that it is possible to remove the centralizing fields without spoiling the symmetry of the theory, and that it is a necessary thing to do if we want to have a theory equivalent to GR. We will keep this principle in mind when we study the Standard Model.

\begin{rem}
Let us consider the possibility of generalizing \eqref{eq51} and \eqref{eq52} to the noncommutative case. For this, we need an algebraic characterization of the summands. In the case of $Z_\B$, this is immediate: the elements of $Z_\B$ are the Dirac operators which commute with $\A$, by definition. Moreover the invariance of $Z_\B$ under  $\Aut(\B)$ follows from this definition. The space $\Delta_\B$ could also be characterized algebraically if \eqref{eq51} were an orthogonal decomposition with respect to some non-degenerate symmetric bilinear form $\varphi$ on the real vector space $\D_\B$. The invariance of $\Delta_\B$ under $\Aut(\B)$ is then automatic if $\varphi$ is $\Aut(\B)$ invariant in the following sense:
\be 
\varphi(UDU^{-1},UD'U^{-1})=\varphi(D,D'),\forall D,D'\in\D_\B,\forall U\in \Aut(\B)
\ee
The restriction of the Krein-Schmidt product defined on $\End(\S)$ by 
\be 
(A,B):=\tr(A^\times B)\label{KS}
\ee
to $\D_\B$ is a real $\Aut(\B)$-invariant symmetric bilinear form. The problem is that it is not well-defined on $\D_\B$ in the case of a canonical background, since $D^2$ is not trace-class, not even bounded. But this is the only problem ! By this we mean that \emph{ at  a formal level} \eqref{eq51} is really an orthogonal decomposition for $(.,.)$, provided the trace satisfies the following properties (with obvious notations)
\begin{enumerate}
\item\label{p1} $\tr(\delta\zeta)=\tr(\zeta\delta)$,  
\item\label{p2} $\tr(\zeta\delta+\zeta')=\tr(\zeta\delta)+\tr(\zeta')$,
\item\label{p3} $\tr(\zeta_\omega)=\int_M\tr_{S_0}(\omega_x){\rm vol}_g$.
\end{enumerate}
In the last property, $\tr_{S_0}$ is the trace in $\End(S_0)$ and $\omega_x$ is really the image of $\omega_x'$ in $\End(S_0)$. This is justified by the fact that $\zeta_\omega$ is a multiplication operator\footnote{The whole reasoning is heuristically justified by the fact that the Dirac operator is ``off-diagonal'' and the centralizing field is diagonal.}.

Let us now start the ``proof''. We first observe that if $\omega$ a $k$-vector field, then the commutator $[\delta_r,\zeta_\omega]$ is the sum of a the multiplication by a $k+1$-vector field and a $k-1$-vector field. To show this we can restrict to the case $r=\id$, i.e. to the canonical Dirac operator associated to $e$. Indeed, when we change the frame $e$  into $r\cdot e$,  $\rho$ is replaced with $\rho\circ r^{-1}$, but the image of the space of $k$-vector fields does not change (the degree  stays the same). We can also restrict to the case of an elementary $k$-vector fields $e_{i_1}\wedge\ldots\wedge e_{i_k}$, which is sent to the element $e_{i_1}\ldots e_{i_k}\in \Gamma(\CC l(TM))$. Then, by the Leibniz rule \eqref{Leibnizrule}, we have 
\bea
[\rho(e_a) \nabla_{e_a}^{\ss_e},\rho(e_{i_1}\ldots e_{i_k})]\Psi&=&\sum_j\rho(e_a e_{i_1}\ldots e_{i_{j-1}}(\nabla_{e_a}^{LC}e_i)e_{i+1}\ldots e_{i_k})\Psi\nonumber
\eea
where the argument of $\rho$ is, by the Clifford commutation relations,  a sum of $k-1$-vectors and $k+1$-vectors. The result follows.

Now we compute $(\delta,\zeta_\omega)$, with $\delta\in \Delta_\B$ and $\omega$ is a $k$-vector:
\bea
(\delta,\zeta_\omega)&=&\tr(\delta\zeta_\omega)\cr
&=&\tr(\delta\chi^2\zeta_\omega)\cr
&=&-\tr(\chi\delta\chi\zeta_\omega)\cr
&=&-\tr(\chi\delta \zeta_{\omega'}),\mbox{ where }\omega'\in \Lambda^{n-k}M\cr
&=&-\tr(\chi \zeta_{\omega'}\delta+\chi\zeta_{\omega''}),\mbox{ where }\omega''\in \Lambda^{n-k-1}TM\oplus \Lambda^{n-k+1}TM\cr
&=&-\tr(\zeta_\omega\delta)+\tr(\zeta_{\omega'''}),\mbox{ where }\omega'''\in \Lambda^{k+1}TM\oplus \Lambda^{k-1}TM\mbox{ and using }\eqref{p2}\cr
&=&-(\delta,\zeta_\omega)+\tr(\zeta_{\omega'''}),\mbox{ using }\eqref{p1}
\eea
Now using \eqref{p3}, $\tr(\zeta_{\omega'''})=\int_M\tr_x(\omega_x'''){\rm vol}_g$. Since $k=3+4m$, $k-1\ge 2$ and $k+1\le n$, thus $\omega_x'''$ is a sum of non-empty products of gamma matrices, and thus has zero trace. It follows that $(\delta,\zeta_\omega)=0$.

We have just seen that $\Delta_\B\subset Z_\B^\perp$, but since $(.,.)$ is not definite, there could be also elements in $Z_\B\cap Z_\B^\perp$. To see that it is not so, let us first consider the case $n=4$. Then $(\zeta_\omega,\zeta_{\omega'})=\tr(\zeta_{\chi v},\zeta_{\chi v'})$, where $v$ and $v'$ are vector fields. If $\zeta_\omega$ is orhogonal to $\zeta_{\omega'}$, we obtain that $\int_M g(v,v'){\rm vol}_g=0$. If this is so for all $\omega'$, then $\omega=0$ since $g$ is not degenerate. The general case is similar, using the canonical extension of $g$ to the Clifford algebra, which is also non-degenerate.

\end{rem}

\section{Almost-commutative algebraic backgrounds}
Let us start with a notation. Let $\K$ be a $\ZZ_2$-graded real pre-Krein space. A vector $\psi\in \K$ or an operator $A$ on $\K$ which  is homogenous (even or odd) is given a degree denoted by $|\psi|$ or $|A|$ respectively, which is an element of $\ZZ_2$.  

Given two algebraic backgrounds $\B_1=(\A_1,\ldots)$ and $\B_2=(\A_2,\ldots,)$, one can define their \emph{graded} tensor product $\B=\B_1\hat\otimes \B_2=(\A,\ldots)$. Let us first define the pre-Krein space. It is the graded tensor product
\be
\K=\K_1\hat\otimes \K_2\label{grtens}
\ee
This means that $\K$ is the tensor product $\K_1\otimes \K_2$, with the grading such that  $|\phi_1\hat\otimes\phi_2|=|\phi_1|+|\phi_2|$, for homogenous vectors. For any two homogenous operators $A_i$ on $\K_i$, $i=1,2$, one defines the graded tensor product
\be 
A_1\hat\otimes A_2(\phi_1\hat\otimes\phi_2)=(-1)^{|A_2||\phi_1|}A_1\phi_1\hat\otimes A_2\phi_2
\ee
Since the algebras are represented by even operators, we define
\be 
\A=\A_1\otimes \A_2\label{prodalg}
\ee
and $\pi=\pi_1\hat\otimes \pi_2(=\pi_1\otimes\pi_2)$. With infinite dimensional spaces and algebras, \eqref{grtens} as well as \eqref{prodalg}  would  require the specification of a particular topological tensor product, but we are only interested in the case where $\A_1$ is $\tilde{\cal C}_c^\infty(M)$  and $\A_2$ finite-dimesional, so we can just take the algebraic tensor product.  The chirality is defined in accordance with \eqref{grtens} by   $\chi=\chi_1\hat\otimes\chi_2(=\chi_1\otimes\chi_2)$.  The real structure is 
\be 
C=C_1\chi_1^{|C_2|}\hat\otimes C_2\chi_2^{|C_1|}\label{prodc}
\ee
The rule is exactly the same for the graded real structure: $J=J_1\chi_1^{|J_2|}\hat\otimes J_2\chi_2^{|J_1|}$. The Krein product needs special care. We have:
\be 
(\phi_1\hat\otimes \phi_2,\psi_1\hat\otimes\psi_2)=(\phi_1,\psi_1)_1(\phi_2,\beta\psi_2)_2\label{kprule}
\ee
where
\be 
\beta=\left\{\matrix{1,\mbox{ if }(.,.)_1\mbox{ is even},\cr
\chi_2,\mbox{ if }(.,.)_1\mbox{ is odd and }(.,.)_2\mbox{ is even}\cr
i\chi_2,\mbox{ if }(.,.)_1\mbox{ and }(.,.)_2\mbox{ are both odd}.
}
\right.\label{prodkrein}
\ee
The justification of \eqref{prodc} and \eqref{prodkrein}  is that they yield the correct real structure and spinor metric when one does the tensor product of two canonical backgrounds. It can be shown that with these rules the KO-metric dimension pair is additive \cite{BBB}. Details can also be found in \cite{thesenadir}. The only additional rule with respect to the tensor product of IST found in \cite{thesenadir} is
\be 
\Omega=\Omega_1\hat\otimes \pi(\A_2)+\pi(\A_1)\hat\otimes \Omega_2
\ee
It is such that if $D_1\in \D_{\B_1}$ and $D_2\in \D_{\B_2}$, then $D_1\hat\otimes 1+1\hat\otimes D_2\in \D_{\B}$, though we will see later that not all elements of $\D_\B$ can be decomposed in this way.

Finally, we define an \emph{almost-commutative AB} to be an AB of the form $\B_{M}\hat\otimes \B_F$ where $\B_M=(\A_M,\ldots)$ is the canonical background of a manifold $M$ equipped with all the required structures, and $\B_F=(\A_F,\ldots)$ is a finite AB (i.e. both the algebra and Krein spaces are finite-dimensional). We will study specific examples later. For now we state some useful general results.

\begin{lemma}\label{commutJ} Let $A\in {\rm End}(\K_M)$ and $B\in{\rm End}(\K_F)$. Then:
\begin{enumerate}
\item $J(A\hat\otimes B)J^{-1}=J_MAJ_M^{-1}\hat\otimes J_F BJ_F^{-1}$,
\item $(A\hat\otimes B)^\times=(-1)^{|A||B|}A^\times\hat\otimes B^\times$,
\item   $(A\hat\otimes B)^o=(-1)^{|A||B|}A^o\hat\otimes B^o$. 
\end{enumerate} 
\end{lemma}

For the proof of the two first properties, we refer to \cite{thesenadir}, lemma 5.1 and 5.2. The third follows easily from the two first.

%

%
%

In the next proposition, we use two hypotheses which will appear again when we compute the automorphisms of the Standard Model AB. We thus feel the need to introduce the following notations, one for the elements of the center of $\A_F$ which are $J_F$-real and Krein selfadjoint, the other for the element of $\Omega^1_F$ which commute with $\pi_F(\A_F)$, with $J_F$, and are Krein anti-selfadjoint:
\bea
Z_J^*(\A_F)&:=&\{a\in \A_F|a\in Z(\A_F), J_F\pi_F(a)=\pi_F(a)J_F, \pi_F(a)^\times=\pi_F(a)\}\cr
C_J^*(\Omega^1_F)&=&\{\omega\in \Omega^1_F|[\omega,\pi_F(\A_F)]=0, \omega J_F=J_F\omega, \omega^\times=-\omega\}\nonumber
\eea
Note that $Z_J^*(\A_F)$ is a sub-$\RR$-algebra of $\A_F$,  $C_J^*(\Omega^1_F)$ is a sub-$Z_J^*(\A_F)$-bimodule of $\Omega^1_F$, and that they are both preserved by AB automorphisms.

\begin{propo}\label{ACconfig}
Let $\B=\B_M\hat\otimes \B_F$ be an almost-commutative AB as above. If $\B_F$ has the following properties:
\begin{enumerate}
\item $Z_J^*(A_F)=\RR$,
\item $C_J^*(\Omega^1_F)=0$,
\end{enumerate}
then the regular Dirac operators of $\B$ are of the form
\be 
D=\delta\hat\otimes 1+\zeta
\ee
where $\delta$ is a frame Dirac of $\S_M$ and $\zeta$ is an operator of order $0$ on $\K$.
\end{propo}
\begin{demo}
Let $f\in \A_M$ and $a\in A_F$. We discard the representations $\pi_M$ and $\pi_F$ to simplify the notations. From $[(f\otimes 1),(1\otimes a)]=0$ we obtain $[[D,f\otimes 1],1\otimes a]=[f\otimes 1,[D,1\otimes a]]=0$ since $[D,1\otimes a]\in \Omega^1$ and $f\otimes 1$ commutes with $\Omega^1$. Let us fix $x\in M$ and write the value of $[D,f\otimes 1]$ at $x$ in the form
$$[D,f\otimes 1]_x=i\sum_a \gamma^a\hat\otimes b_a+1\hat\otimes \alpha$$
where $b_a\in \A_F$ and $\alpha\in \Omega^1_F$. Using linear independence of the $\gamma^a$ and $1$ we see that $[D,f\otimes 1]$ commutes with $1\otimes x$ for all $x\in \A_F$ iff both the $b_a$'s and $\alpha$ commute with $\A_F$. Moreover, since $D$ and $f\otimes 1$ commute with $J=J_M\chi_M^{|J_F|}\hat\otimes J_F\chi_F^{|J_M|}$, we see that $b_a$, which commutes with $\chi_F$, must also commute with $J_F$. Similarly, we find that $1\otimes\alpha$ must commute with $J$, and since $\alpha$ anticommutes with $\chi_F$ we obtain thanks to the graded tensor product that $\alpha$ must commute with $J_F$. Finally, since $D$ and $f\otimes 1$ are Krein selfadjoint, we see that $[D,f\otimes 1]$ is Krein anti-selfadjoint, from which we find that $b_a$ is Krein selfadjoint and $\alpha$ is Krein anti-selfadjoint. From the hypotheses we find that $b_a\in \RR 1_{A_F}$  and   $\alpha=0$. Thus there is a real function $\delta(f)$ such that 
$$[D,f\otimes 1]=i\sum_a\delta(f)\gamma^a\otimes 1$$
Moreover, $\delta$ must be a derivation of $\A_M$. Following the same steps as in the beginning of the proof of theorem \ref{configcan}, we see that there exists a frame Dirac $\delta\in \D_{\B_{M}}$ such that $D-\delta\hat\otimes 1$ commutes with $f\otimes 1$ for all $f$. The result follows.
\end{demo}

\section{Algebraic background of the spectral standard model}
\subsection{The finite algebraic background}
We recall here the noncommutative indefinite triple of finite dimension which is  used in the Lorentzian spectral standard model. We refer to  \cite{thesenadir} for details. The  algebra is 
$$\A_F:=\CC\oplus\HH\oplus M_3(\CC),$$
the  Krein space is 
\be 
\K_F=\K_R\oplus \K_{L}\oplus \K_{\bar R}\oplus \K_{\bar L}\label{quadrupling}
\ee
where for each symbol $\sigma=R,L,\bar R,\bar L$ we have
$${\cal K}_\sigma=\CC^2_{\rm weak\  isospin}\otimes(\CC_{\rm leptons}\oplus \CC^3_{\rm color})\otimes \CC^N_{\rm generations}$$
The number of generations will play little role, so we keep $N$ undetermined for the sake of generality, and also to avoid any confusions between the color and generation space.  The grouping of the two middle summands into a common $\CC^4$ as in Pati-Salam theory would suggest itself, but the representation of the algebra is such that it will be on the contrary more practical to split the first tensor product and write instead
\be
{\cal K}_\sigma=(\CC^2\oplus \CC^2\otimes\CC^3)\otimes \CC^N\label{ksigma}
\ee
In this decomposition, we give names to the elements of the canonical basis, identifying them with elementary particle states. The canonical basis of the leptonic $\CC^2$ is $(\nu,e)$, the one of the quark isospin $\CC^2$ is $(u,d)$, the color $\CC^3$ one is $(r,g,b)$. We will not need to name the basis elements of the generation space. If we need to identify in which $\K_\sigma$ a basis element lives, we attach the symbol $\sigma$ to it. For instance, the basis elements identified respectively with a anti-(right neutrino) and with a left down red quark (of an unspecified generation) are $\nu_{\bar R}$ and $(d\otimes r)_L$. Finally, it will be sometimes useful to decompose $\K_\sigma$ into the sum of a lepton and a quark subspace, defined by $\K_\ell=\CC^2\otimes \CC^N$ and $\K_q=\CC^2\otimes \CC^3\otimes \CC^N$.

A matrix $A$ which is block-diagonal in the decomposition \eqref{quadrupling} will be written $A=\diag(A_R,A_L,A_{\bar R},A_{\bar L})$. If it is also block-diagonal with respect to generations, it will be convenient to write it $A=\diag(a_R,a_L,a_{\bar R},a_{\bar L})\otimes 1_N$ where $A_\sigma=a_\sigma\otimes 1_N$ and $1_N$ is the identity matrix. As an important example, the matrix of the Krein product in the canonical basis is by definition:
\be 
\eta_F=\diag(1,-1,s,-s)\otimes 1_N
\ee
where $1$ is the identity matrix on $\K_\sigma$, and $s$ is a sign, which is $+1$ if we use commuting variables for spinor components, and $-1$ if we use anti-commuting ones. These choices are the only ones which allow us to recover the correct terms in the fermionic action, as explained in \cite{thesenadir}. It is worthy of note that the anti-Lorentzian signature of the manifold forces us to use a non-Euclidean finite part as well, and the ultimate reason for this is the rule \eqref{kprule} for the tensor products of Krein structures. Note that $\eta_F$ is the fundamental symmetry associated with the canonical scalar product on $\K_F\simeq \CC^{32N}$, so that we can write the Krein product on $\K_F$ in the following way:
\be 
(\psi,\phi)_F:=\bra \psi,\eta_F\phi\ket_F=\psi^\dagger\eta_F\phi
\ee

The chirality operator is
\be 
\chi_F=\diag(1,-1,-1,1)\otimes 1_N
\ee
For the real structure, we will use $C_F=\chi_F J_F$ where, in the decomposition \eqref{quadrupling}:
\be  
J_F=\pmatrix{0&0&\epsilon_F&0\cr 0&0&0&\epsilon_F\cr 1&0&0&0\cr 0&1&0&0}\circ c.c.
\ee
Here $\epsilon_F$ is a sign yet unspecified and c.c. is the complex conjugation of the vector components. In the most widely used convention, it is $J_F$ which is called the real structure. We will use it more often than $C_F$ to ease comparison with the literature. In order to avoid any confusion, we will call $C_F$ the real structure, in compliance with our convention, and $J_F$ the graded real structure. It will be convenient to use the particle/antiparticle space decomposition $\K=\K_p\oplus \K_a=(\K_R\oplus \K_L)\oplus(\K_{\bar R}\oplus \K_{\bar L})$ in which $J_F$ is more simply written $J_F=\pmatrix{0&\epsilon_F\cr 1&0}\circ c.c$, so that we have the useful formula
\be 
J_F\pmatrix{A&B\cr C&D}J_F^{-1}=\pmatrix{\bar D&\epsilon_F\bar C\cr \epsilon_F\bar B&\bar A} 
\ee
Before defining the representation of $\A$, we need to introduce a convenient notation. Let $A\in M_2(\CC)$. Then we write
\be
\tilde{A}:=(A\oplus A\otimes 1_3)\otimes 1_N,
\ee
acting on $\K_\sigma$ written as in \eqref{ksigma}. We also need a representation of the quaternions as $2\times 2$ complex matrices, and an embedding of $\CC$ into $\HH$. We will identify  the elements of $\HH$ with the matrices $\pmatrix{\alpha&-\bar\beta\cr \beta&\bar\alpha}$, and the complex $\lambda$ with $$q_\lambda:=\pmatrix{\lambda&0\cr 0&\bar \lambda}.$$
With these notations, the representation $\pi_F$ is
\be 
\pi_F(\lambda,q,m)=\diag(\tilde q_\lambda,\tilde{q},\lambda 1_2\oplus 1_2\otimes m, \lambda 1_2\oplus 1_2\otimes m)\otimes 1_3
\ee
Here we see that $\pi_F(\A_F)$ is a $\times$-algebra and that $\times$ and $\dagger$ coincide on it, such that $\pi_F$ is $*$-morphism for the natural $C^*$-structure on $\A_F$. It also follows that the homomorphism $\alpha_F : \Aut(\B_F)\rightarrow \Aut(\A_F)$ takes values in $*$-$\Aut(\A_F)$, the group of $C^*$-algebra automorphisms of $\A_F$.

The finite Dirac $D_F^0$ which is used to recover the Standard Model, is the following:
\be
D_F^0=\pmatrix{0&-Y_0^\dagger&sM_0^\dagger&0\cr 
Y_0&0&0&0\cr
M_0&0&0&-Y_0^T\cr 
0&0&\bar Y_0&0}
\ee
where $Y_0=\pmatrix{Y_\nu&0\cr 0& Y_e}\oplus \pmatrix{ 1_3\otimes Y_u&0\cr 0&1_3\otimes Y_d}$ and $M_0=s\epsilon_F M_0^T=\pmatrix{Y_R&0\cr 0&0}$. This choice is severely constrained by the axioms of Noncommutative Geometry (Krein-selfadjointness, anticommutation with $\chi_F$, commutation with $J_F$, and the first-order condition). If one admits the second order condition, the only arbitrariness in the choice of $Y_0$ is the diagonal form of the matrices $\pmatrix{Y_\nu&0\cr 0&Y_e}$ and $\pmatrix{ 1_3\otimes Y_u&0\cr 0&1_3\otimes Y_d}$ with respect to weak isospin indices. Note that a non-diagonal form would make the photon acquire a mass thanks to the Higgs mechanism\footnote{This can be axiomatized as the \emph{condition of massless photon}, \cite{connesmarcolli}.}. Note also that their is a neutrino mixing term iff $\epsilon_F=-1$, and the see-saw mechanism occur if furthermore $s=-1$. For this, see \cite{thesenadir}, section 8.10.

We take this opportunity to introduce the notations $M_\nu=Y_\nu Y_\nu^\dagger$, $M_e=Y_e Y_e^\dagger$, \ldots,  $M_R=Y_RY_R^\dagger$, which we will need later.

In the algebraic background point of view $D_F^0$ is not fixed. What we need is   a module of 1-forms $\Omega^1_F$. We select it by the following simple requirement: $D_F^0$ must be a regular Dirac operator, i.e. $\Omega^1_F=\Omega^1_{D_F^0}$. After an easy computation we find:
\be
\Omega^1_F=\{\pmatrix{0&Y_0^\dagger \tilde q_1&0&0\cr \tilde q_2 Y_0&0&0&0\cr 0&0&0&0\cr 0&0&0&0
}|q_1,q_2\in\HH\} \label{finite1forms}
\ee
The finite algebraic background of the Standard Model is thus $\B_F:=(\A_F,\K_F,$ $(.,.)_F,\pi_F,\chi_F,C_F,\Omega^1_F)$. Let us now look for the space of compatible Dirac operators. For any quaternion $q$ we write
\be
\Phi(q)=\pmatrix{0&-Y_0^\dagger\tilde q^\dagger&0&0\cr \tilde q Y_0&0&0&0\cr 0&0&0&0\cr 0&0&0&0
}
\ee
We observe that $\Phi(q)$ is a general Krein-self-adjoint 1-form. For any matrix $m$ acting on the generation space $\CC^N$ and satisfying $m^T=s\epsilon_F m$, we write
\be
\sigma(m)=\pmatrix{0&0&s E_\nu\otimes m^\dagger&0\cr 0&0&0&0\cr
E_\nu\otimes m&0&0&0\cr 0&0&0&0}
\ee
where $E_\nu=\pmatrix{1&0\cr 0&0}$ is the projection on the space spanned by $\nu$. (We will similarly use the notations $E_e,E_u$, etc). We observe that $\sigma(m)$ is Krein selfadjoint and commutes with $J_F$, so that $\sigma(m)^o=\sigma(m)$, and that moreover $\sigma(m)$ commutes with $\pi(\A_F)$ and hence also with $\pi(\A_F)^o$. It is now easy to see that the compatible Dirac operators are those of the form
\bea
D(q,m)&=&\Phi(q)+\Phi(q)^o+\sigma(m)\label{eq17}
\eea
The finite configuration space $\D_{\B_F}$ is thus $\{D(q,m)|q\in \HH, m\in M_N(\CC), m^T=s\epsilon_Fm\}$. When we come to the full AB of the Standard Model, the $q$ and $m$ degrees of freedom will be the origin of the Higgs and complex scalar fields respectively.

\subsection{The gauge group ${\cal G}_{A_F}$}
We recall here the definition of the  so-called ``gauge group'' ${\cal G}_{A_F}$ (see \cite{vs} for details). It consists of the elements of the form $\Upsilon(u):=\pi(u)J_F\pi(u)J_F^{-1}$ where $u$ is in $U(\A_F)$, the unitary group of $\A_F$. The importance of the gauge group lies in its relation with the homomorphism $\alpha_F : \Aut(\B_F)\rightarrow *$-$\Aut(\A_F)$. 

\begin{lemma}\label{lemups}
\begin{enumerate}
\item For all $u\in U(\A_F)$, $\Upsilon(u)$ is an AB automorphism of $\B_F$, such that $\alpha_F(\Upsilon(u))={\rm Ad}_u$.
\item $\Upsilon: U(\A_F)\rightarrow \Aut_{\B_F}$ is a group homomorphism,
\end{enumerate}
\end{lemma}
\begin{demo}
{\small The corresponding property for Euclidean Spectral Triples is well-known, and the proof is almost the same, so it is left to the reader. The only difference is that we require the preservation of $\Omega^1_{F}$, and this comes from the order 1 condition and the fact that $\Omega^1_{F}$ is a bimodule. 
}
\end{demo}
\begin{rem} Note that $\Ad_u \mapsto\Upsilon(u)$ is not a well-defined section of $\alpha_F$. For instance if $u=(e^{i\theta},1,1)$ then $\Ad_u=1$ but $\Upsilon(u)\not=1$.
\end{rem}

It will be important to know what is the image of $\alpha_F$.
\begin{lemma}\label{lemimalphaF} The image of $\alpha_F$ is $*$-${\rm Inn}(\A_F)$, the group of inner $*$-automorphisms of $\A_F$.
\end{lemma}
\begin{demo} By lemma \ref{lemups}, we already know that  $*$-${\rm Inn}(\A_F)\subset \im(\alpha_F)$. To prove the converse, first note that $\Aut(\A_F)=\Aut(\CC)\times \Aut(\HH)\times \Aut(M_3(\CC))$. Since we are dealing with real $*$-algebras we see, using the Skolem-Noether theorem,   that $\Aut(\CC)=\{\id,c_1\}$, where $c_1$ means complex conjugation, $\Aut(\HH)={\rm Inn}(\HH)$ and $\Aut(M_3(\CC))={\rm Inn}(M_3(\CC))\ltimes \{\id, c_3\}$, where $c_3$ is the complex conjugation of $3\times 3$-matrices. Now let us consider the real-subalgebra $\CC\subset \A_F$. The representation $\pi_F$ breaks into a sum of copies of the fundamental representation $\pi_f$ of $\CC$ and its conjugate $\pi_f^*$.  Now observe that $c_1$ exchanges $\pi_f$ and $\pi_f^*$, while there is not an equal number of copies of them in $\pi_F$: this shows that $c_1$ cannot be induced by a unitary operator $U$ such that $U\pi_F(\A_F)U^{-1}=\pi_F(\A_F)$, i.e. $c_1\notin \im(\alpha_F)$. The situation is similar for $c_3$, considering the representation of the center $\CC$ of $M_3(\CC)$ induced by $\pi_F$.  
\end{demo}
Putting together the two previous lemmas, we see that for all $\phi\in\Aut(\B_F)$, there exists $u\in U(\A_F)$ such that $\alpha_F(\phi)=\alpha_F(\Upsilon(u))$.

Let us now consider $u=(e^{i\theta},q,m)\in U(\A_F)$, where  $q,m$ are unitary quaternions and $3\times 3$ matrices respectively, and $\theta$ is real. We easily compute that $\Upsilon(u)=\diag(A,B,\bar A,\bar B)$, with 
\be 
A=(\pmatrix{1&0\cr 0&e^{-2i\theta}}\oplus \pmatrix{e^{i\theta}&0\cr 0&e^{-i\theta}}\otimes \bar m)\otimes 1_N,B=(e^{-i\theta}q\oplus q\otimes \bar m)\otimes 1_N\label{gaugegroup}
\ee
In particular we can see on this  formula that  $\ker\Upsilon=\{\pm 1\}$.

Let us now introduce the unimodular gauge group $S{\cal G}_{\A_F}$, containing the elements $\Upsilon(u)$ such that $\det(\pi(u))=1$. They are of the form $U=\diag(A,B,\bar A,\bar B)$, with 
\be
A=(\pmatrix{1&0\cr 0&e^{-2i\theta}}\oplus \pmatrix{e^{4i\theta/3}&0\cr 0&e^{-2i\theta/3}}\otimes \bar g)\otimes 1_N,B=(e^{-i\theta}q\oplus e^{i\theta/3}q\otimes \bar g)\otimes 1_N\label{gaugegroup}
\ee
with $g\in SU(3)$. Finally, we define $U(1)_X=\{\pi(e^{i\xi},1,e^{i\xi})|\xi\in \RR\}$, and $SU_{\A_F}=\pi({U}(\A_F))\cap \ker\det$.

\begin{lemma}\label{decuni}
Up to discrete symmetries, any element $\pi(u)\in\pi({ U}(\A_F))$ can be uniquely decomposed into the commutative product $\pi(u)=sx$, with $s\in SU_{\A_F}$ and $x\in U(1)_X$.
\end{lemma}
\begin{demo}
{\small Let $u=(\lambda,q,m)\in U(\A_F)$ and define $s=\pi(\lambda',q,(\lambda')^{-1/3}g)$, and $x=\pi(e^{i\xi},1,e^{i\xi})$, with $e^{i\xi}=(\lambda\det m)^{1/4}$, $\lambda'=\lambda^{3/4}(\det m)^{-1/4}$, and $g=(\det m)^{-1/3}m$. One readily checks that $\pi(u)=sx$. For the uniqueness part up to discrete symmetries, observe that $\det(\pi(e^{i\xi},1,e^{i\xi}))=e^{4Ni\xi}=1$ has finitely many solutions.}
\end{demo}
Now let us write  $T^1_X=\Upsilon(U(1)_X)$. By the above lemma and the order $0$ condition, we obtain a decomposition of every element $g\in {\cal G}_{A_F}$ into the product 
\be
g=\tilde g g_X,\mbox{ with }\tilde g\in S{\cal G}_{A_F}, g_X\in T^1_X\label{decompgauge}
\ee
Since $\ker(\Upsilon)=\{\pm 1\}$, this decomposition is unique up to discrete symmetries. Moreover, we can further decompose $\tilde g$ into the commutative product $\tilde g=g_Yg_Wg_C$ with $g_Y(\theta)=\Upsilon(e^{i\theta},1,e^{-i\theta/3})$, $g_W(q)=\Upsilon(1,q,1)$ and $g_C(g)=\Upsilon(1,1,g)$ with obvious notations. For the record, let us explicitly write down these elements:
\bea
g_Y(\theta)&=&\diag(\pmatrix{1&0\cr 0&e^{-2i\theta}}\oplus \pmatrix{e^{4i\theta/3}&0\cr 0&e^{-2i\theta/3}}\otimes 1_3,e^{-i\theta}1_2\oplus e^{i\theta/3}1_2\otimes 1_3,\cr
&&c,c)\otimes 1_N\cr
g_W(q)&=&\diag(1_2\oplus 1_2\otimes 1_3,q\oplus q\otimes 1_3,c,c)\otimes 1_N\cr
g_C(g)&=&\diag(1_2\oplus 1_2\otimes \bar g,1_2\oplus 1_2\otimes \bar g,c,c)\otimes 1_N\cr
g_X(\xi)&=&\diag(\pmatrix{1&0\cr 0&e^{-2i\xi}}\oplus \pmatrix{1&0\cr 0&e^{-2i\xi}}\otimes 1_3,e^{-i\xi}1_2\oplus e^{-i\xi}1_2\otimes 1_3,c,c)\otimes 1_N\cr
\label{gxi}
\eea
where the $c$'s in the 3rd and 4th places stand for the conjugate of the elements in the 1st and 2nd places, respectively (a  notation we will use hereafter).

This decomposition will serve us    to distinguish the gauge fields and the anomalous $X$-field. These have values in the Lie algebras $\underline{S{\cal G}_{\A_F}}$ and $\underline{T^1_X}$. To find these algebras, observe that a curve $G_t$ passing through $1$ in  ${\cal G}_{\A_F}$ can be locally decomposed in a unique way as $G_t=\Upsilon(u_t)$, with $u_t\in U(\A_F)$ and $u_0=1$. Using the same notations as in the lemma above, we can further decompose into $G_t=s_tx_t J_Fs_tx_t J_F^{-1}=s_tJ_Fs_tJ_F^{-1}x_tJ_Fx_tJ_F^{-1}$ with $s_0=x_0=1$. Taking derivatives and using the fact that $u_0'$ and $x_0'$ are anti-selfadjoint, we obtain $G_0'=s_0'-(s_0')^o+x_0'-(x_0')^o$. Consequently, an element $B$ of the Lie algebra $\underline{{\cal G}_{A_F}}$ of the gauge group can be uniquely decomposed as a sum of  the form
\be
B=B_{\rm unimod}+B_X
\ee
where $B_{\rm unimod}\in \underline{S{\cal G}_{\A_F}}$ and $B_X\in\underline{T^1_X}$, moreover $B_{\rm unimod}=b-b^o$, and $B_X=a-a^o$, 
where $b\in \underline{SU_{A_F}}$ and $a\in \underline{U(1)_X}$. Using this and \eqref{gxi}, we obtain the following basis of $\underline{\cal G}_{\A_F}$:
\bea
t_Y&=&\diag(\pmatrix{0&0\cr 0& -2i}\oplus \pmatrix{ {4\over 3}i &0\cr 0& -{2\over 3}i}\otimes 1_3,-i1_2\oplus {i\over 3}1_2\otimes 1_3,c,c)\otimes 1_N\cr
t_W^a&=&\diag(0,i\sigma^a\oplus i\sigma^a\otimes 1_3,c,c)\otimes 1_N\cr
t_C^b&=&\diag(0\oplus 1_2\otimes i \bar\lambda^b,0\oplus 1_2\otimes i \bar\lambda^b,c,c)\otimes 1_N\cr
t_X&=&\diag(\pmatrix{0&0\cr 0&-2i}\oplus \pmatrix{0&0\cr 0&-2i}\otimes 1_3,-i1_2\oplus -i1_2\otimes 1_3,c,c)\otimes 1_N\cr
\eea
where $\sigma^a$, $a=1,2,3$ and $\lambda^b$, $b=1,\ldots,8$ are the Pauli and Gell-Mann matrices respectively.

Let us take this opportunity to introduce some more notations, consistent with the above. We write $U(1)_Y=\{\pi(e^{i\theta},1,e^{-i\theta/3})|\theta\in\RR\}$, $SU(2)_W=\{\pi(1,q,1)|q\in SU(2)\}$, $SU(3)_C=\{\pi(1,1,g)|g\in SU(3)\}$, $T_Y^1=\Upsilon(U(1)_Y)$, $T_W^1=\Upsilon(SU(2)_W)$, $T_C^1=\Upsilon(SU(3)_C)$. Note that $\Upsilon$ is an isomorphism in restriction to these groups.
\subsection{The centralizer and unitary  stabilizer of $\pi(\A_F)$}\label{secom}
Let us look for the centralizer of $\pi(\A_F)$. Using simplified notations, a block matrix $(M_{ij})_{1\le i,j\le 4}$  commutes with $\diag(\tilde q_\lambda,\tilde q,\lambda\oplus m,\lambda\oplus m)$ for all $(\lambda,q,m)\in \A_F$ iff
\bea
\tilde q_\lambda M_{11}&=&M_{11}\tilde q_\lambda\cr
\tilde q_\lambda M_{13}&=&M_{13}(\lambda\oplus m)\cr
\tilde q_\lambda M_{14}&=&M_{14}(\lambda\oplus m)\cr
\tilde q M_{22}&=&M_{22}\tilde q\cr
(\lambda\oplus m)M_{31}/M_{41}&=&M_{31}/M_{41}\tilde q_\lambda\cr
[M_{3/4,3/4},\lambda\oplus m]&=&0\cr
M_{12}=M_{21}=M_{23}=M_{24}=M_{32}=M_{42}&=&0\label{cond20}
\eea
\begin{propo}\label{oddcom}
The elements which are odd, commute with $J_F$ and  with $\pi(\A_F)$ are of the form $\pmatrix{0&0&\epsilon_F \bar M&0\cr 0&0&0&0\cr   M&0&0&0\cr 0&0&0&0}$, with $M=E_\nu\otimes m_\nu$.
\end{propo}
\begin{demo}
{\small In addition to the conditions above, oddness requires $M_{11}=M_{22}=M_{33}=M_{44}=M_{23}=M_{32}=0$, and $J_F$-reality additionnaly sets $M_{34}=M_{43}=0$. Thus only the blocks $M_{13}$ and $M_{31}$ are not required to vanish. However, by conditions \eqref{cond20}, we obtain that $M_{13}=\pmatrix{b&c&0&0\cr 0&0&0&0\cr d&e&0&0\cr 0&0&0&0}$, where we have written the matrix in the basis $(\nu,e,u,d)$ (tensor generations, and for quarks, color). Hence each entry is a matrix acting on generations (and possibly on colors, but these are zeros). On the other hand $M_{31}$ must have a structure of zeros which is transposed with respect to that of $M_{13}$. However $J_F$-reality imposes that $M_{13}=\epsilon_F M_{31}$, hence only $b$ remains non-vanishing.}
\end{demo}

We make the following trivial observation, useful later on.
\begin{lemma}\label{lemoddcom} No odd element in the centralizer of $\pi(\A_F)$ is invertible.
\end{lemma}
\begin{demo}
{\small With the above notations, $M_{12}=M_{42}=0$ and since $M$ is odd we also have $M_{22}=M_{32}=0$.}
\end{demo}

\begin{propo}\label{flavsym}
Let $U\in\End({\cal K}_F)$ be even, commute with $J_F$ and $\pi(\A_F)$. Then $U=\diag(A,B,\bar A,\bar B)$, with $A=(E_\nu\otimes g_\nu\oplus E_e\otimes g_e)\oplus (E_u\otimes 1_3\otimes g_u+E_d\otimes 1_3\otimes g_d)$ and $B=1_2\otimes g_\ell\oplus 1_2\otimes 1_3\otimes g_q$. Such a $U$ is Krein-unitary iff the matrices $g_\nu,g_e,g_u,g_d,g_\ell,g_q$ acting on the generations are unitary.
\end{propo}
\begin{demo}
{\small From the first two conditions we find that $U$ is of the form 
$$U=\pmatrix{A&0&0&C\cr 0&B&D&0\cr 0&\epsilon_F\bar C&\bar A&0\cr \epsilon_F\bar D&0&0&\bar B}$$
Now we want such an element to commute with $\pi(\A_F)$, hence with its complexification. From the commutation with  $\pi(M_3(\CC))$ we obtain that $\bar A$ must act trivially on colors, and thus be of the form $\sum_i\bar  a_i\otimes \bar g_\ell^i\oplus \bar a_i'\otimes 1_3\otimes \bar {g_\ell^i}'$. Then from the $(R,R)$-block we learn that $a_i,a_i'$ must be diagonal in order to commute with the complexification of $\pi(\CC)$. Thus we have
\be
A=(E_\nu\otimes g_\nu\oplus E_e\otimes g_e)\oplus (E_u\otimes 1_3\otimes g_u+E_d\otimes 1_3\otimes g_d)\label{formeA}
\ee
From $(\bar L,\bar L)$ we see that $B$ must also act trivially on colors, but from $(L,R)$ we obtain that it must commute with the complexification of $\pi(\HH)$, hence with any matrix on the isospin space, thus it must only act on generations. We thus obtain
\be
B=1_2\otimes g_\ell\oplus 1_2\otimes 1_3\otimes g_q\label{formeB}
\ee
Now from $(\bar R,L)$ we obtain that $(\lambda \otimes 1_N\oplus 1\otimes m\otimes 1_N)\bar C=\bar C (q\otimes 1_N\oplus q\otimes 1_3\otimes 1_N)$ which immediately yields $\bar C=0$. From  $(L,\bar R)$ we similarly obtain $D=0$.

For the second statement, we just observe that $U$ is block-diagonal, hence it commutes with the fundamental symmetry. Thus it is Krein-unitary iff it is unitary.}
\end{demo}

In the sequel we will call a unitary $U$ such as in the proposition a \emph{flavour symmetry}. The group of flavour symmetries will be denoted by ${\cal F}$. A flavour symmetry written in the notations above will be denoted by $F(g_\nu,g_e,g_u,g_d,g_\ell,g_q)$. The arguments of two elements of such form can be directly multiplied, that is, ${\cal F}\simeq U(N)^6$. Comparing \eqref{gaugegroup} and proposition \ref{flavsym} one finds that an element $U$ belongs to ${\cal G}_{A_F}\cap {\cal F}$ iff both the matrix $m$ and the quaternion $q$ are scalars (hence $q=\pm 1$), and   $g_\nu=1$, $g_e=e^{-2i\mu}$, $g_u=e^{i\mu}\bar m$, $g_d=e^{-i\mu}\bar m$,  $g_\ell=e^{-i\mu}q$ and $g_q=q\bar m$. Thus we obtain $U=F(1,e^{-2i\mu},e^{i(\varphi+\mu)}, e^{i(\varphi-\mu)}, \pm e^{-i\mu},\pm e^{i\varphi})=\Upsilon(e^{i\mu},\pm 1,e^{-i\varphi})$, $\mu,\varphi\in\RR$, which can be rewritten in the form $\Upsilon(1,\pm 1,1)g_Y(\theta)g_X(\xi)$, with the notations of the previous paragraph, though not in a unique way, since the equations to solve are $\mu=\theta+\xi\ [2\pi]$ and $\varphi={\theta\over 3}-\xi\ [2\pi]$. Thus  
\be
({\cal G}_{A_F}\cap {\cal F})^0=\{g_Y(\theta)g_X(\xi)|\theta,\xi\in\RR\}
\ee
%
 
%
%
Let us now define the \emph{unitary stabilizer} of $\pi(\A_F)$ to be 
\be
{\rm Stab}(\A_F):=\{U\in\End(\K_F)|[\chi_F,U]=[J_F,U]=0, UU^\times=1,  U\pi(\A_F)U^{-1}=\pi(\A_F)\}
\ee
\begin{propo}\label{unitstab} Let $U\in {\rm Stab}(\A_F)$. Then $U$ can be written (in a non-unique way) as the commutative product
$$U=\tilde g  g_X  f$$
with $\tilde g\in S{\cal G}_{\A_F}$, $g_X\in T^1_X$, and $  f\in {\cal F}$. 
\end{propo}
\begin{demo}
{\small The conjugation $\Ad_U$ induces an automorphism  of the $*$-algebra $\pi(\A_F)$.  Thus by lemmas \ref{lemups} and \ref{lemimalphaF} there exists an element $g$ of the gauge group, such that $\Ad_U=\Ad_{g}$ on $\pi(\A_F)$. Hence $Ug^{-1}$ commutes with $\pi(\A_F)$, is even, $J_F$-real, and Krein-unitary. It it thus a flavour symmetry $f$.  Since $g\in {\cal G}_{\A_F}$ we use \eqref{decompgauge} to write $g=\tilde g g_X$, as in the statement.
}
\end{demo}
  
\subsection{The automorphism group of the finite algebraic background}
Let us first look at the bimodule structure of $\Omega^1_F$. The left and right actions of $a=(\lambda,q,m)$ on $\phi\in\Omega^1_F$ are 
\be
(\lambda,q,m)\triangleright (q_1,q_2)=(q_\lambda q_1,qq_2), (q_1,q_2)\triangleleft (\lambda,q,m)=(q_1q,q_2q_\lambda)\label{modulestruc}
\ee
We will not refer to $m$ anymore in the rest of this section. The bimodule is also endowed with an involution
\be
(q_1,q_2)^\dagger=(-q_2^\dagger,-q_1^\dagger)
\ee
which comes from the Krein adjoint on $\End(\K_F)$ and exchanges the left action of $(\lambda,q)$ and the right action of $(\bar\lambda,q^\dagger)$.


There is also a scalar product (on the underlying real vector space), which is induced by the real part of the Hilbert-Schmidt product (which is the opposite of the Krein-Schmidt product on 1-forms). We have (neglecting the trivial antiparticle parts of the matrices):
\bea
\Re\tr(\pmatrix{0&Y_0^\dagger \tilde q_1\cr \tilde q_2 Y_0&0}^\dagger\pmatrix{0&Y_0^\dagger \tilde q_1'\cr \tilde q_2' Y_0&0})&=&\Re \tr(Y_0^\dagger\tilde q_2^\dagger\tilde q_2' Y_0)+\Re \tr(\tilde q_1^\dagger Y_0Y_0^\dagger\tilde q_1')\cr
&=&K\Re \tr(q_1^\dagger q_1'+q_2^\dagger q_2')
\eea
where we have used the fact that $\Re\tr (\tilde q Y_0Y_0^\dagger)=\Re\tr(q)\tr(Y_0Y_0^\dagger)$ and $K=\tr(Y_0Y_0^\dagger)=\tr(M_v+M_e+3M_u+3M_d)$.

Let us consider a bimodule automorphism $\theta$. Let us write $\theta(1,0)=(\alpha,\alpha')$ and $\theta(0,1)=(\beta',\beta)$. Using $(1,0)\triangleright (1,0)=(1,0)$ we see that $\alpha'=0$ and using $(0,1)\triangleright (0,1)=(0,1)$ we similarly see that $\beta'=0$. Moreover $(z,0)=(z,0)\triangleright (1,0)=(1,0)\triangleleft (0,q_z)$ for any $z\in\CC$ implies that $\alpha\in\CC$, and we similarly find using $(0,z)=(0,q_z)\triangleright (0,1)=(0,1)\triangleleft (z,0)$ that $\beta\in\CC$. Hence a bimodule automorphism has the form:
\be
\theta(q_1,q_2)=(q_\alpha q_1,q_2q_\beta),
\ee
for $\alpha,\beta\in\CC$. Conversely it is immediate to check that such transformations are bimodule automorphisms. Moreover, such an automorphism respects the scalar product iff $|\alpha|=|\beta|=1$, and it respects the involution iff $\beta=\bar\alpha$. We thus obtain that a bimodule automorphism respecting both structures has the form
\be
\theta(q_1,q_2)=(q_{e^{i\phi}}q_1,q_2q_{e^{-i\phi}}), \phi\in\RR.\label{formeV}
\ee
This result will be used shortly. But first we need two definitions.

\begin{definition}
We say that $Y_0$ is \emph{generic} iff $Y_0$ is invertible and if any $N\times N$     matrix commuting with both $M_\nu$ and $M_e$ (resp. $M_u$ and $M_d$) is scalar.
\end{definition}

\begin{rem}
It is remarkable that the genericity condition has also emerged independently in \cite{dabrositarz} in the investigation of the Hodge property for the finite spectral triple of the Standard Model. As remarked there, it can be equivalently rephrased in this way: $Y_0$ is invertible and  $M_\nu$ and $M_e$ on the one hand, $M_u$ and $M_d$ on the other hand, have no common eigenvector.
\end{rem}

\begin{definition} For $\varphi\in\RR$ we define a \emph{B-L symmetry} to be an element of the form
\bea
g_{B-L}(\varphi)&=&\diag(A,A,\bar A,\bar A)\otimes 1_N\cr
\mbox{with }A&=&e^{-i\varphi}1_2\oplus e^{i\varphi/3}1_2\otimes 1_3,
\eea
and we let $T^1_{B-L}=\{g_{B-L}(\varphi)|\varphi\in\RR\}$.
\end{definition}

\begin{propo}\label{vertf} If $Y_0$ is generic, then  $U\in\Vert(\B_F)$ iff $U=g_Y(\theta)g_X(\xi)g_{B-L}(\varphi)$. This decomposition is commutative and unique up to discrete symmetries.
\end{propo}
\begin{demo}
{\small 
First, let us recall that  $U\in\Vert(\B_F)$ iff $U$ is Krein-unitary, commutes with $\chi_F$, $J_F$, $\pi(\A_F)$, and is  such that $U\Omega_F^1U^{-1}=\Omega_F^1$. It is immediate to check that
\bea
U&=&\diag(A,B,\bar A,\bar B)\cr
\mbox{with }A&=&q_{e^{i\beta}}\otimes e^{i\theta_\ell}1_N\oplus q_{e^{i\beta}}\otimes 1_3\otimes e^{i\theta_q}1_N,\cr
\mbox{and }B&=&1_2\otimes e^{i\theta_\ell}1_N\oplus 1_2\otimes 1_3\otimes e^{i\theta_q}1_N\label{form95}
\eea
has all the required properties and induces on $\Omega^1_F$ the automorphism \eqref{formeV}. Conversely, we know that $U$ is a flavour symmetry, and is thus of the form $U=\diag(A,B,\bar A,\bar B)$, with $A$ and $B$ respectively of the form \eqref{formeA} and \eqref{formeB}. Now the stabilization of $\Omega^1_F$ means that 
\bea
AY_0^\dagger \tilde q_1B^{-1}&=&Y_0^\dagger \tilde q_1',\qquad q_1'\in\HH\cr
B\tilde q_2 Y_0A^{-1}&=&\tilde q_2'Y_0,\qquad  q_2'\in\HH\label{stabform}
\eea
for all $q_1,q_2\in\HH$. Using the unitarity of $U$, we see that the second condition is just the conjugate of the first.




In particular, $AY_0^\dagger B^{-1}$ must be of the form $Y_0^\dagger\tilde q_1'$.  This means that
\bea
(E_\nu\otimes g_\nu Y_\nu^\dagger g_\ell^{-1}+E_e\otimes g_e Y_e^\dagger g_\ell^{-1})&=&\pmatrix{\alpha&\beta\cr 0&0}\otimes Y_\nu^\dagger+\pmatrix{0&0\cr-\bar\beta&\bar\alpha}\otimes Y_e^\dagger\cr
(E_u\otimes 1_3\otimes g_u Y_u^\dagger g_q^{-1}+E_d\otimes 1_3\otimes g_d Y_d^\dagger g_q^{-1})&=&\pmatrix{\alpha&\beta\cr 0&0}\otimes 1_3\otimes Y_u^\dagger\cr
&&+\pmatrix{0&0\cr-\bar\beta&\bar\alpha}\otimes 1_3\otimes Y_d^\dagger\cr
\eea
Hence $g_\nu Y_\nu^\dagger g_\ell^{-1}=\alpha Y_\nu^\dagger$, $g_uY_u^\dagger g_q^{-1}=\alpha Y_u^\dagger$,   $g_eY_e^\dagger g_\ell^{-1}=\bar\alpha Y_e^\dagger$, and $g_dY_d^\dagger g_q^{-1}=\bar\alpha Y_d^\dagger$, with $\alpha\in\CC$. In other words we have $A Y_0^\dagger=Y_0^\dagger B \tilde q_\alpha$. Inserting this in \eqref{stabform}, we see that it is satisfied. From this relation we get $AY_0^\dagger Y_0 A^\dagger=Y_0^\dagger B\tilde q_\alpha \tilde q_{\bar\alpha}B^\dagger Y_0=|\alpha|^2Y_0^\dagger Y_0$. Using the trace we obtain $|\alpha|=1$. We then have $[A,Y_0^\dagger Y_0]=0$. Doing the product $Y_0A^\dagger A Y_0^\dagger$ also yields $[B,Y_0Y_0^\dagger]=0$. We thus obtain
\bea
[g_\nu,Y_\nu^\dagger Y_\nu]=0&& [g_e,Y_e^\dagger Y_e]=0\cr
[g_u,Y_u^\dagger Y_u]=0&& [g_d,Y_d^\dagger Y_d]=0\cr
[1_2\otimes g_\ell,E_\nu\otimes Y_\nu Y_\nu^\dagger+E_e\otimes  Y_eY_e^\dagger]=0&&[1_2\otimes g_q, E_u\otimes Y_u Y_u^\dagger+ E_d\otimes Y_dY_d^\dagger]=0\nonumber
\eea
From the last condition and the genericity hypothesis, $g_\ell$ and $g_q$ are scalars. Thus $B$ commutes with $Y_0^\dagger$, and the condition $AY_0^\dagger=Y_0^\dagger B\tilde q_\alpha$ becomes $A=\tilde q_\alpha B$, using genericity again. Now writing $g_\ell=e^{i\theta_\ell}$, $g_q=e^{i\theta_q}$, $\alpha=e^{i\beta}$, we obtain the form \eqref{form95}. Finally,   we compute $g_Y(\theta)g_X(\xi)g_{B-L}(\varphi)$, and find that it is equal to  $U$ iff $\beta\equiv \theta+\xi$, $\theta_\ell\equiv-\varphi-\theta-\xi$, $\theta_q\equiv {\varphi+\theta\over 3}-\xi$. We find a unique solution for $\varphi$ mod $2\pi$, and unique solution for $(\xi,\theta)$ mod $(\pi/2,\pi/2)$. 
}
\end{demo}

We can now determine the finite algebraic background symmetries.

\begin{propo}\label{finiteAB} If $Y_0$ is generic, then the automorphism group of the finite algebraic background is  ${\cal G}_{\A_F}T^1_{B-L}\simeq{U(1)_Y\times SU(2)_W\times SU(3)_C\times U(1)_{B-L}\times U(1)_X\over \ZZ_2\times \ZZ_2\times \ZZ_3}$
\end{propo}
\begin{demo}
{\small 
Before starting the proof, we first need to explain that $U(1)_{B-L}$ is just the same as $T^1_{B-L}$, but seen as an abstract group isomorphic to $U(1)$ and parametrized by $e^{i\varphi/3}$. Note that we can similarly parametrize $U(1)_Y$ by $e^{i\theta/3}$, and $U(1)_X$ by $e^{i\xi}$. By propositions \ref{unitstab} and  \ref{vertf}, we see that $\Aut(\B_F)={\cal G}_{\A_F}T^1_{B-L}$. We then have a surjective homomorphism $U(1)_Y\times SU(2)_W\times  SU(3)_C\times U(1)_{B-L}\times U(1)_X\rightarrow \Aut(\B_F)$ which sends $(e^{i\theta/3}, q,g,e^{i\varphi/3},e^{i\xi})$ to $g_Y(\theta)g_W(q)g_C(g)g_{B-L}(\varphi)g_X(\xi)$. The computation of the kernel gives $\ZZ_2\times \ZZ_2\times \ZZ_3$, yielding the result. Note that $(a,b,c)\in  \ZZ_2\times \ZZ_2\times \ZZ_3$ is identified with $(e^{i(a+b)\pi}, e^{ia\pi}1_2, e^{2ic\pi/3}1_3,  e^{2ic\pi/3},e^{ib\pi})$.}
\end{demo}


In the sequel we always suppose that the generiticity hypothesis holds, and we write ${\cal G}_{\A_F}^{\rm ext}:={\cal G}_{\A_F}T^1_{B-L}$, which we call  the \emph{extended gauge group}. Note that  ${\cal G}_{\A_F}^{\rm ext}\cap {\cal F}=\Vert(\B_F)=T^1_XT^1_YT^1_{B-L}$. For reason which will appear below, we need to extract the Lie algebra of the extended gauge group as a direct summand of the Lie algebra of ${\rm Stab}(\A_F)$. More precisely, let ${\cal F}'$ be a subgroup of ${\cal F}$ such that any $f\in{\cal F}$ can be written $f=vf'$ with $v\in\Vert(\B_F)$ and $f'\in {\cal F}'$ in a unique way up to discrete symmetries. (Such a subgroup exists: for instance the one defined by $\det(g_\ell)=\det(g_q)=\det(g_\nu)=1$.) Then we will have the Lie algebra decomposition 
\be 
\underline{{\rm Stab}(\A_F)}=\underline{S{\cal G}_{\A_F}}\oplus T_Y^1\oplus T_X^1\oplus T_{B-L}^1\oplus \underline{{\cal F}'}\label{eq102}
\ee
An interesting possibility for ${\cal F}'$ is to consider the orthogonal complement of $\underline{\Vert(\B_F)}$ in $\underline{\cal F}$ with respect to the Krein-Schmidt product $(a,b)=\tr(a^\times b)$. Since the elements of $\underline{\Vert(\B_F)}$  commute with the fundamental symmetry, the Krein-Schmidt product is positive-definite on this Lie algebra, and we have $\underline{\cal F}=\underline{\Vert(\B_F)}\oplus  \underline{\Vert(\B_F)}^\perp$. Moreover, $\underline{\Vert(\B_F)}^\perp$ is a Lie algebra thanks to the fact that $[v,f]=0$ for all $v\in \underline{\Vert(\B_F)}$ and $f\in \underline{{\cal F}}$. Using the basis \eqref{gxi}, we can obtain that  $\underline{\Vert(\B_F)}^\perp$ is defined by the equations:
\bea
\tr(-h_e+2h_u-h_d-h_\ell+h_q)&=&0\cr
\tr(h_e+3h_d+h_\ell+3h_q)&=&0\cr
\tr(h_\nu+h_e-h_u-h_d+2h_\ell+2h_q)&=&0\label{eq103}
\eea
which can be rewritten
\bea
\tr(h_u+h_d)&=&-2\tr(h_q)\cr
\tr(h_\nu+h_e)&=&-2\tr(h_\ell)\cr
\tr(h_\nu-h_e+3h_u-3h_d)&=&0
\eea
Thus the Lie algebra $\underline{\Vert(\B_F)}^\perp$ can be exponentiated to the subgroup ${\cal F}'$ of ${\cal F}$ defined by
\be
\det(g_ug_dg_q^{2})=\det(g_\nu g_eg_\ell^{2})=\det(g_\nu g_e^{-1}g_u^3g_d^{-3})=1
\ee
One can compute that $g_X(\xi)g_Y(\theta)g_{B-L}(\varphi)f(g_\nu,\ldots,g_q)\in {\cal F}'$ leads to the invertible system
\be 
4N\pmatrix{2&2&0\cr 
-1&-1&-1\cr
{1\over 3}&-1&{1\over 3}}\pmatrix{\theta \cr \xi \cr \varphi}
\equiv \pmatrix{\alpha\ [2\pi]\cr \beta\  [2\pi]\cr \gamma\  [2\pi]}
\ee








\subsection{The automorphism group of the total algebraic background}
In this section we are going to compute the automorphisms of $\B=\B_M\hat\otimes\B_F$. It will be important to note that $Z^*_F(\A_F)=\RR$, and $C^*_J(\Omega^1_F)=0$ (for this,  use \eqref{finite1forms} and proposition \ref{oddcom}).


Let us consider an automorphism ${\cal U}$  of   $\B$. It induces an automorphism $\alpha({\cal U})$ of $\A_M\hat\otimes\A_F\simeq\tilde {\cal C}^\infty_c(M,\A_F)$, which must preserve $Z^*_J(\tilde {\cal C}^\infty_c(M,\A_F))=\tilde {\cal C}^\infty_c(M,\RR)$. Thus there exists a diffeomorphism $\theta$ such that $\alpha({\cal U})$ and $\theta_*$ have the same action on $\tilde {\cal C}^\infty_c(M,\RR)$. Thus, using \cite{KR}, example d, we see that $\alpha({\cal U})\circ \theta_*^{-1}$ is a ``local'' automorphism, of the form $x\mapsto \phi_x \in{\rm Inn}(\A_F)$ and the question is whether this map can be lifted to $U(\A_F)$. We have an exact sequence $1\rightarrow \ZZ_2\times \ZZ_3\rightarrow SU(2)\times SU(3)\buildrel{p}\over {\longrightarrow} {\rm Inn}(\A_F)\rightarrow 1$, and since $\ZZ_2\times \ZZ_3$ is central discrete, $p$ is a covering map and $\pi_1({\rm Inn}(\A_F))=\ZZ_2\times \ZZ_3=\ZZ_6$. Thus it will be possible to lift  the map $x\mapsto\phi_x$ to $U(\A_F)$ under the hypothesis $\Hom(\pi_1(M),\ZZ_6)=\{0\}$. We assume this from now on and  let  $u\in{\cal C}^\infty(M,{U}(\A_F))$ be such that
$$\alpha({\cal U})={\rm Ad}_u\circ \theta_*$$
where $u\in{\cal C}^\infty(M,{U}(\A_F))$. The map $x\mapsto \Upsilon(u(x))$ is a smooth map with values in ${\cal G}_{\A_F}$: we call it a \emph{local gauge transformation}. We see that it belongs to $\Aut(\B)$, as well as $V_\theta\hat\otimes 1$, and that their product has the same image as ${\cal U}$ under $\alpha$. We can thus focus on the case ${\cal U}\in \Vert(\B)$. Since in this case ${\cal U}$ commutes   with the algebra $\tilde{\cal C}^\infty_c(M,\RR)$, it is of the form:
\be
{\cal U}(\Psi)_x=U_x\Psi_x
\ee
where $\Psi$ is a section of ${\cal S}\hat\otimes {\K}_F$. We know that $\Omega^1$ is stabilized by $\Ad_{\cal U}$. Thus $\Omega^1_x:=i\rho(T_xM)\hat\otimes\pi(\A_F)\oplus 1\hat\otimes\Omega^1_F$ is stabilized by $\Ad_{U(x)}$ for all $x$. An element of $\Omega^1_x$ is of the form
$$\omega=\sum_a i\gamma_a\hat\otimes\pi_F(x^a)+\sum_b 1\hat\otimes w_b$$
Using the fact that $C^*_F(\Omega^1_F)=0$, we see that $\omega$ commutes with $1\hat\otimes\pi_F(\A_F)$, with $J$, and is Krein anti-selfadjoint iff it is belongs to $i\rho(T_xM)$. Since $U_x$ must preserve these elements, we have
\be
U_x\rho(v)\hat\otimes 1 U_x^{-1}=\rho(\Lambda_x v)\hat\otimes 1
\ee
for all $v\in T_xM$, with $\Lambda_x$ a linear map. Let us drop the $x$ to ease the notation for a while. Squaring both sides we see that $\Lambda\in O(TM)$.  There thus exists an element $\Sigma$ of the Pin or Spin group such that $U$ and $\Sigma\hat\otimes 1$ have the same adjoint action on $\rho(TM)\hat\otimes 1$. Hence $U^{-1}\Sigma\hat\otimes 1$ commutes with $\rho(TM)\hat\otimes 1$, and thus with $\rho(Cl(TM))\hat\otimes 1$. Passing to complex combinations we see that $U^{-1}\Sigma\hat\otimes 1$ commutes with $\rho(\CC l(TM)\hat\otimes 1)=\End(S_0)\hat\otimes 1$ and is thus of the form $1\hat\otimes V$, with $V\in \End(\K_F)$. Hence we conclude that $U=\Sigma\hat\otimes V$. Now, $UU^\times=1$ yields $(-1)^{|\Sigma||V|}(\Sigma\hat\otimes V)(\Sigma^\times\hat\otimes V^\times)=\Sigma\Sigma^\times\hat\otimes VV^\times$. We know that $\Sigma\Sigma^\times=\pm 1$, thus $VV^\times=\pm 1$ with the same sign. Moreover, the commutation between $\chi$ and $U$ shows that $\Sigma$ and $V$ have the same parity. However, $V$ must commute with $\pi(A_F)$, and it is impossible if it is odd (lemma \ref{lemoddcom}). Thus they are both even. Hence $\Sigma$ is in the Spin group. Finally we have 
\bea
J\Sigma\hat\otimes V&=&(J_M\chi_M\hat\otimes J_F\chi_F)\Sigma\hat\otimes V\cr
&=&(-1)^{|\Sigma|}J_M\chi_M\Sigma\hat\otimes J_F\chi_FV\cr
&=&(-1)^{|\Sigma|}(-1)^{|V|}\Sigma J_M\chi_M\hat\otimes J_F V\chi_F
\eea
where we used the fact that an element of Spin group commutes with $J_M\chi_M$, and on the other hand
\bea
\Sigma\hat\otimes V J&=&\Sigma\hat\otimes V(J_M\chi_M\hat\otimes J_F\chi_F)\cr
&=&(-1)^{|V|}\Sigma J_M\chi_M\hat\otimes VJ_F\chi_F
\eea
Comparison between the two expressions shows that $J_FV=(-1)^{|\Sigma|}VJ_F$. Now notice that $V$ is even and commutes with $\pi_F(\A_F)$. By the observation at the beggining of section \ref{secom}, we conclude that $V$ has the form $V=\pmatrix{V_{11}&0&0&V_{14}\cr 0&V_{22}&0&0\cr 0&0&V_{33}&0\cr V_{41}&0&0&V_{44}}$, from which we obtain $V^\times=\pmatrix{V_{11}^\dagger&0&0&-V_{41}^\dagger\cr 0&V_{22}^\dagger&0&0\cr 0&0&V_{33}^\dagger&0\cr -V_{14}^\dagger&0&0&V_{44}^\dagger}$. In particular we obtain $V_{22}V_{22}^\dagger$ for the $(2,2)$-element of $VV^\times$, and this exclude the possibility $VV^\times=-1$. We then conclude that both $\Sigma$ and $V$ are unitary, even, and commute with the real structure\footnote{Note the remarkable interplay between the manifold and finite part in this computation !}. Hence the punchline is that $\Sigma$ is  in the neutral component ${\rm Spin(1,3)}^0$ of the spin group, and $V$ is an element of $\Vert(\B_F)$, the form of which we know by proposition \ref{vertf}. Now if we put the $x$ back, we see that smooth maps $x\mapsto\Sigma_x$ and $x\mapsto V_x$ can be defined  locally such that $U_x=\Sigma_x\hat\otimes V_x$. If either $\Sigma_x$ or $V_x$ can be defined globally, then both of them can. In the first case, it means that $x\mapsto\Lambda_x$ can be lifted to $\Spin(p,q)^0$. We know that this can always be 
 done exactly when $\Hom(\pi_1(M),\ZZ_2)=\{0\}$, which is true since we have already supposed that   $\Hom(\pi_1(M),\ZZ_6)=\{0\}$ (a non trivial morphism $\phi$ from a group $G$ to $\ZZ_2$ induces a non-trivial morphism $\phi\times 0$ from $G$ to $\ZZ_2\times \ZZ_3=\ZZ_6$.)
 


We finally obtain the following result:
\begin{theorem}
If $\Hom(\pi_1(M),\ZZ_6)=\{1\}$ (and if $Y_0$ is generic), then $\Aut(\B_M\hat\otimes \B_F)$ is generated by the elements of the form:
\begin{enumerate}
\item\label{typediffeo} $V_\theta\hat\otimes 1$, where $\theta\in\Diff(M)$,
\item\label{typespin} $U_\Sigma\hat\otimes 1$, where $\Sigma$ is a spinomorphism of $M$,
\item local gauge transformations  $1\hat\otimes \Upsilon(u)$ where $u\in{\cal C}^\infty(M,{U}(\A_F))$,
\item local $B-L$ transformations $1\hat\otimes g_{B-L}(\varphi)$, where\footnote{Note that we use the hypothesis on $\pi_1(M)$ to define $\varphi$ globally.} $\varphi\in {\cal C}^\infty(M,\RR)$.
\end{enumerate}
\end{theorem}
 
For a counter-example if the hypothesis on $\pi_1(M)$ is not satisfied, consider the anti-Lorentz cylinder $M=S^1\times \RR^3$, where $S^1$ is timelike. Then let $\Sigma(e^{it},\vec x)=\cos(t/2)+\sin(t/2)\gamma_3\gamma_4$, with hopefully obvious notations. We also define  $V(e^{it},\vec x)$ to be the element of $\Vert(\A_F)$ of the form $\eqref{form95}$ with $\theta_\ell=e^{it/2}$, $\theta_q=\beta=1$. Then neither $\Sigma$ nor $V$ can be defined globally on $M$, but the product $\Sigma\hat\otimes V$ can, since its value at $t=0$ and $t=2\pi$ is the same. 


\subsection{The total configuration space and the consistent submodels}
To determine the configuration space $\D_\B$ of the Standard Model AB, we use proposition \ref{ACconfig}.  We conclude that $D\in \D_\B$ iff it is of the form $D=\delta\hat\otimes 1+\zeta$, where for all $x$, $\zeta_x$ is a Krein selfadjoint, odd, $J$-real operator on $\S_x\otimes \K_F$, which moreover satisfies  $[\zeta_x,1\otimes \pi_F(a)]\in i\rho(T_xM)\hat\otimes \pi_F(\A_F)\oplus 1\hat\otimes \Omega^1_F$. Let us list these operators, restricting to the case of an anti-Lorentzian signature and space-time dimension $4$. Since $\End( \S_x\otimes \K_F)\simeq \End( \S_x)\otimes \End(\K_F)\simeq \CC l(T_xM)\otimes\End(\K_F)$, we can decompose the local operator as
\be
\zeta_x=\sum_{I\subset\{1;\ldots;4\}}i^{|I|}\gamma_I\hat\otimes B^I\label{eqRx}
\ee
where $\gamma_I=\gamma_{i_1}\ldots\gamma_{i_k}$ if $I=\{i_1,\ldots,i_k\}$ written in ascending order. Moreover, from the requirements on $\zeta_x$, the operator $B^I$ must have the opposite parity as that of $\gamma_I$. The factor $i^{|I|}$ ensures that $i^{|I|}\gamma_I$ is $J_M$-real, so from lemma \ref{commutJ} we also see that $B^I$ must commute with $J_F$. Finally, using $(\gamma_I\hat\otimes B^I)^\times=(-1)^{|I||B|}\gamma_I^\times\hat\otimes (B^I)^\times=\gamma_I^\times\hat\otimes (B^I)^\times$, we see that only the  five following  possibilities are allowed in \eqref{eqRx}:
\begin{enumerate}
\item\label{t1} $i\gamma_i\hat\otimes B^i$ where   $B^i$ is even, $J_F$-real, Krein anti-self-adjoint and $[B^i,\pi(\A_F)]\subset\pi(\A_F)$,
\item\label{t2} $i\gamma_i\gamma_j\gamma_k\hat\otimes B^{ijk}$ where $B^{ijk}$,  is even, $J_F$-real, Krein  self-adjoint and $[B^{ijk},\pi(\A_F)]=0$,
\item\label{t3} $\gamma_i\gamma_j\hat\otimes B^{ij}$, where $B^{ij}$ is odd, $J_F$-real Krein anti-self-adjoint and commutes with $\pi(\A_F)$,
\item\label{t4} $\gamma_1\gamma_2\gamma_3\gamma_4\hat\otimes B^{1234}$, where $B^{1234}$,   is odd, $J_F$-real, Krein self-adjoint  and commutes with $\pi(\A_F)$,
\item\label{t5} $1\hat\otimes B^\emptyset$, where $B^\emptyset$  is odd, $J_F$-real, Krein self-adjoint and   $[B^\emptyset,\pi(\A_F)]\subset\Omega^1_F$.
\end{enumerate}
%

Let us look for the form of the fields $B^i$. They are in the normalizer of the  algebra $\pi(\A_F)$, from which it follows that $\exp( B^i)$ is in the normalizer of the Lie group of invertible elements in $\pi(\A_F)$, hence $\Ad_{\exp(B^i)}$ stabilizes $\pi(\A_F)$ (in formulas: $e^{B^i}\pi(a)e^{- B^i}=e^{[B^i,.]}\pi(a)\in\pi(\A_F)$). Moreover, we have $(B^i)^\times=-B^i$, thus $\exp(B^i)$ is an even Krein-unitary which commutes with $J_F$ and stabilizes $\pi(\A_F)$. We have already computed these elements in proposition \ref{unitstab}: $e^{B^i}$ is the commutative product of an element of the gauge group and a flavour symmetry. Hence $B^i$ belongs to the direct sum of  Lie algebras \eqref{eq102}. Let us analyze these elements in more detail. 

\begin{itemize}
\item A vector field with values in $\underline{S{\cal G}_{\A_F}}$ is a  usual (unimodular) gauge field.
\item A vector field with values in $\underline{T^1_X}$ is   the anomalous \emph{$X$-field}.
\item A vector field with values in $\underline{T^1_{B-L}}$ is of the form $\rho(v)\hat\otimes t^1_{B-L}$ with
\be 
t^1_{B-L}=\diag(-1_2\oplus {1\over 3}1_2\otimes 1_3,-1_2\oplus {1\over 3}1_2\otimes 1_3,1_2\oplus {-1\over 3}1_2\otimes 1_3, 1_2\oplus {-1\over 3}1_2\otimes 1_3)\otimes 1_N
\ee
This is the $Z'$-vector boson of $B-L$ theory.
\item Vector fields with values in  $\underline{{\cal F}'}$ are of linear combinations of $i\gamma_i\hat\otimes\diag(A,B,\bar A,\bar B)$, with $A=(E_\nu\otimes h_\nu+E_e\otimes h_e)\oplus (E_u\otimes 1_3\otimes h_u+E_d\otimes 1_3\otimes h_d)$, $B=1_2\otimes h_\ell\oplus 1_2\otimes 1_3\otimes h_q$, with $h_e,h_u,h_d,h_\nu,h_\ell,h_q$ fields of anti-hermitian $N\times N$ matrices, satisfiying trace relations like for instance \eqref{eq103}. We call them \emph{flavour vector fields}, as their presence in the Fermionic action $(\Psi,D\Psi)$ would create flavour changing neutral currents.
\end{itemize}

Let us now look at the elements of type \ref{t2}. From proposition \ref{unitstab} we see that $B^{ijk}$ only acts on the generations, with matrices $h_\nu,\ldots,h_q$   which are self-adjoint. We call them \emph{flavour pseudovector fields}. Note that they include the centralizing fields coming from $M$ (when all the matrices $h$ are equal and scalar).

From proposition \ref{oddcom}, we see that the fields $B^{ij}$ only act on neutrino generations, via a matrix $m^{ij}$ which must be symmetric if $s\epsilon_F=-1$ and antisymmetric if $s\epsilon_F=1$. We call them \emph{Majorana bivector fields}. 

The field $B^{1234}$ is similar, except that $m^{1234}$ is symmetric if $s\epsilon_F=1$ and antisymmetric if $s\epsilon_F=-1$. We call it the \emph{Majorana pseudoscalar field.}

Finally, the elements of type \ref{t5} are just tensor products of the identity by a compatible finite Dirac. We can write them in the form $1\hat\otimes B^\emptyset=1\hat\otimes(\Phi(q)+\Phi(q)^o+\sigma(m))$ as in \eqref{eq17}. They are the Higgs  and   $\sigma$ fields.

Thus we see the appearance of many unexpected fields, as in usual Kaluza-Klein theory. Some (or why not all) of them may be physical and yet unseen (with the exception of the $X$-field which yields anomalies and will be dealt with later), but if we follow an approach which is as conservative as possible, we now have to see what are the fields which can be removed without spoiling the symmetry of the theory. For this, we are now going to study the transformation properties of the different fields. As we have seen in the previous section, there are 4 kinds of transformations: two which comes from the manifold and have an obvious action, local gauge transformations, and local B-L transformations. We note that all centralizing fields, that is, fields of the form $\omega\hat\otimes B$ where $B$ commutes with $\pi(\A_F)$, are invariant under local gauge transformations (for $B^o=\pm B$ and this implies that $B$ also commutes with $\pi(\A_F)^o$).

Let us write a general compatible Dirac in the form
\be
D=\delta_r\hat\otimes 1+\zeta_{g}+\zeta_X+\zeta_{B-L}+\zeta_{H}+\zeta_\sigma+\zeta_{\rm other}\label{decdir}
\ee
where $\zeta_g$ contains the unimodular gauge fields, $\zeta_X$ the $X$-field, $\ldots$ ,  and $\zeta_{\rm other}$ gathers all the other terms. Let us look at the action of a local gauge transformation $U\in{\cal C}^\infty(M,{\cal G}_{\A_F})$.
\bea
UDU^{-1}&=&\delta_r\hat\otimes 1+U[\delta_r\hat\otimes 1,U^{-1}]+U\zeta_gU^{-1}+U\zeta_XU^{-1}+U\zeta_{B-L}U^{-1}+U\zeta_HU^{-1}\cr
&&+U\zeta_\sigma U^{-1}+U\zeta_{\rm other}U^{-1}\cr
&=&\delta_r\hat\otimes 1+U[\delta_r\hat\otimes 1,U^{-1}]+U\zeta_gU^{-1}+\zeta_X+\zeta_{B-L}+U\zeta_HU^{-1}+\zeta_\sigma +\zeta_{\rm other}\cr
&=&\delta_r\hat\otimes 1+\zeta_g^U+\zeta_X^U+U\zeta_HU^{-1}+\zeta_{B-L}+\zeta_\sigma+\zeta_{\rm other}
\eea
The commutator with the frame Dirac makes unimodular gauge and $X$ fields appear, which are collected into the gauge-transformed terms $\zeta_g^U$ and $\zeta_X^U$. Now let's look at  B-L transformations. They commute with every field except  $\delta_r\hat\otimes 1$ and $1\hat\otimes \sigma(m)   $, and we have:
\bea
(1\hat\otimes g_{B-L}(\varphi))(\delta_r\hat\otimes 1)(1\hat\otimes g_{B-L}(\varphi))^{-1}&=&\delta_r\hat\otimes 1 +\rho\circ r^{-1}(\nabla\varphi)\hat\otimes t^1_{B-L}\cr
(1\hat\otimes g_{B-L}(\varphi))(1\hat\otimes \sigma(m))(1\hat\otimes g_{B-L}(\varphi))^{-1}&=&1\hat\otimes\sigma(e^{2i\varphi}m) \label{gtblsigma}
\eea
%



\section{Conclusion and Outlook}

We see that the smallest subspace of $\D_\B$ which contains the metric and the Higgs  and is stable by the symmetries of the theory, the AB automorphisms, is  spanned by frame Dirac operators as well as unimodular gauge fields, $X$, $Z'_{B-L}$, and Higgs fields. We call $\D^{\rm min}_\B$ the configuration space containing these fields. For $D\in \D^{\rm min}$, the Fermionic action $(\Psi,D\Psi)$ does not yield neutrino mixing and has thus little physical interest. To have this feature, we need to include fields of the form $\sigma(m)$. Thanks to  \eqref{gtblsigma}, we see that any vector subspace of $\{m\in M_N(\CC)|m^T=s\epsilon_F m\}$ will be stable by $\Aut(\B)$. The simplest case is to take the complex multiples of a given $m_0$, so that we add to $\D^{\rm min}$ a single complex scalar field. We thus obtain a configuration space we call $\D^{\rm min+\sigma}_\B$.

A first conclusion we can draw is that none of these models contains the exact same bosonic fields as the Standard Model, since there are at least two additional fields: the $X$-field and the $Z_{B-L}'$-boson. In comparison, the traditional approach in  NCG-based particle models, which uses fluctuations of the metric, leads to the unimodular gauge fields and the $X$-field. Thus, in both cases we need to supplement the theory with the  unimodularity condition, which consists in restricting (arbitrarily) the local gauge transformations to unimodular ones, allowing to leave the $X$-field out of the configuration space. At this point there is, we argue, already some  gains in the AB approach: the B-L boson, which is interesting in many respects and is by far the simplest extension of the SM, and a Kaluza-Klein point of view all the way (no fluctuation). Moreover, the   simplest allowed submodel of the theory which is consistent with the observed neutrino oscillations  includes the complex scalar, which we know is needed in the traditional NCG approach, but  cannot be obtained by fluctations of the metric.

Of course we now need an action to define the theory. In the absence of   the Spectral Action principle in Lorentzian signature, it is still possible to fall back to the older Connes-Lott model, which provides an action principle for fields which are noncommutative 1-forms. This model works without any problem in the Lorentzian setting and allows to recover the SM bosonic action \cite{thesenadir}. The $Z_{B-L}'$-boson, though, is not a noncommutative 1-form. It is however possible to extend the finite algebra $\A_F$ by a factor of $\CC$ to $\A_F^{\rm ext}:=\CC\oplus \CC\oplus \HH\oplus M_3(\CC)$, represented on the same Krein space as before by
\be 
\pi(\lambda,\mu,q,m)=\diag(\tilde q_\lambda,\tilde q,\mu 1_2\oplus 1_2\otimes m,\mu 1_2\oplus 1_2\otimes m)\otimes 1_N
\ee
As for the bimodule of 1-forms, $\Omega^1_{\rm ext}$, we obtain it by requiring as before that $D_0$ be a regular element. If one does this simple changes and redo the previous analysis, one obtains an almost-commutative AB whose automorphism group is still the extended automorphism group ${\cal G}_{\A_F}^{\rm ext}$, but what is remarkable, it is also equal to ${\cal G}_{\A_F^{\rm ext}}$. This means that in this case, the traditional   fluctuations of the metric lead precisely to the configuration space $\D^{\rm min+\sigma}$   !  We can thus import the Connes-Lott action. The resulting model will be studied in a forthcoming paper\footnote{Interestingly, this model violates the order 1 condition, and it is the reason why it has not been found in \cite{KrajPris}, but  in a way which is so mild that the whole Connes-Lott theory goes through  without change.}. We end by another interesting aspect of the AB approach. The compatibility condition $[D,\pi(a)]\in \Omega^1$ defining the configuration space has more solutions when $\Omega^1$ grows and less when $\A$ grows. It means that the number of fields do not necessarily rise too much in unification models, and in the case of centralizing fields, it must even decrease. For instance, in the extended model based on $(\A_F^{\rm ext}, \Omega^1_{\rm ext})$, there are actually \emph{less} fields than in the non-extended one. More precisely, we could have as many as $6$ complex scalar fields in the $\sigma$ sector in the case $N=3$, $s\epsilon_F=1$, but the extended model has only one, since the compatibility condition requires $m$ to be colinear with the $m_0$ contained in $D_0$. The Pati-Salam model would also be important to consider from this point of view, in particular because the $Y$-field will cease to be centralizing. The question of which field to consider physical will then probably become simpler. This will be the subject of future research.

\section*{Acknowledgements}
The author would like to thank Nadir Bizi, Christian Brouder,  Fr\'ed\'eric H\'elein, Pierre Martinetti, and  J\'er\'emie Pierard de Maujouy for inspiring discussions.

\appendix
\section{Equivalence of algebraic and topological spin structures}\label{equivss}
Let us now explain how to obtain a topological spin structure from an algebraic one and {\it vice versa}.

This is essentially a frame bundle/associated bundle construction, and has already been described in \cite{schroder} in the Riemannian case.  Let $\ss=(\S,\rho,\chi,H,C)$ be $(g,or,t-or)$-spin structure. Let $S_0$ be the standard spinor space as in section \ref{sec25}, end $\rho_0,\chi_0,H_0,C_0$ the associated objects. The key point is that the model space $\RR^{p,q}$ used above to define the frame bundle can be identified via $\rho_0$ to a subspace of $\End(S_0)$. In   order to avoid notation clutter, we will implicitly use  the Clifford representations $\rho$ and $\rho_0$ to identify vectors with endomorphisms of spinor spaces. We will write vectors in $\RR^{p,q}$ with upper case letters, and tangent vectors with lower case ones. We define  $P_\ss$ to be the sub-bundle of $\Hom(M\times S_0,\S)$ whose fibre at $x$ contains all the invertible linear maps $p: S_0\rightarrow \S_x$ such that:
\begin{enumerate}
\item $T_xM=p\RR^{p,q}p^{-1}$,
\item $\chi_x=p\chi_0p^{-1}$.
\item $p$ is an isometry from $(S_0,H_0)$ to $(\S_x,H_x)$,
\item $C_x=pC_0p^{-1}$.
\end{enumerate}
The right action of $\Spin(p,q)^0\subset \End(S_0)$ on $P_\ss$ is $p\cdot u:=p\circ u$. It is free and transitive because $\Spin(p,q)^0$ is precisely the set of endomorphisms $u$ of $S_0$ which satisfy $uV_0u^{-1}=V_0$, $u\chi_0=\chi_0$, $uu^\times=1$, and  $uC_0=C_0u$.

\begin{rem}
If we consider a pseudo-orthonormal basis of the spinor space $S_0$, it is transported by $p$ to a pseudo-orthonormal basis of $S_x$. Moreover, Majorana and Weyl spinors (eigenvectors of $C_0$ and $\chi_0$, respectively) are also conserved by $p$. Thus a pseudo-orthonormal basis of Majorana-Weyl spinors of $S_0$ would be transported to such a basis of $\S_x$, and it would be tempting to interpret $P$ as the bundle of ``Majorana-Weyl pseudo-orthonormal spinor frames''. However, the first condition does not seem to have a simple interpretation in terms of spinor frame. Moreover, Majorana-Weyl spinors exist only in KO-dimension $0$.
\end{rem}

Now for all $p\in P_\ss$, we define $\Lambda_\ss(p) : \RR^{p,q}\rightarrow T_xM$ by $\Lambda_\ss(p)X:=pXp^{-1}$. It is now easy to show that $\Lambda_\ss(p)$ is an isometry preserving space and time orientations and that $\Lambda_\ss$ satisfies all the properties of definition \ref{topospin}.

Thus we can define a topological spin structure from an algebraic one. Let us now show the converse. We start with the principal bundle $P$. We define the associated vector bundle
\begin{equation}
\S^P:=P\times_{\Spin(p,q)^0}S_0
\end{equation}
as usual as the quotient space of $P\times S_0$ by the right action $(p,s)\cdot u=(p\cdot u,u^{-1}\cdot s)$  for all $p\in P$, $s\in S_0$ and $u\in \Spin(p,q)^0$. The equivalence classes provide a vector bundle $\S^P$  as is well-known, with a well-defined left action of the Clifford bundle (see \cite{LM}, Chap. II,  Prop. 3.8). Note that the Clifford bundle itself can be viewed as the associated bundle 
\begin{equation}
\CC l(TM,g)=P\times_{\Spin(p,q)^0}\End(S_0)
\end{equation}
where $\End(S_0)\simeq\CC l(p,q)$ and the left action of $\Spin(p,q)^0$ on $\End(S_0)$ is $u\cdot a=uau^{-1}$. In particular, $\Lambda: P\rightarrow {\rm Fr}(M)$ is given by 
\bea
\Lambda(p_x):\ \RR^{p,q}&\rightarrow&T_xM \subset \CC l(TM,g)\cr
X&\mapsto &[p_x,X]\label{deflambda}
\eea
so that $\Lambda(p_x\cdot u)=[p_x,u^{-1}Xu]$ for any $u\in\Spin(p,q)^0$. Now let $p_x\in P_x$. The action of  $[p_x,a]\in \CC l(T_xM)$ on $[p_x,s]$ is 
\begin{equation}
[p_x,a]\cdot [p_x,s]=[p_x,as]\label{clifaction}
\end{equation}
Since the spin group acts freely and transitively on $P_x$, \eqref{clifaction} suffices to define the action of $\CC l(T_xM,g_x)$ on $\S^P_x$ even when the elements are not in the same ``gauge'', since for each $p_x'\in P_x$ there is a unique $u\in\Spin(p,q)^0$ such that $p_x'=p_x\cdot u$, hence we can define
\bea\label{clifaction2}
[p_x,a]\cdot [p_x',s]&=&[p_x,a]\cdot[p_x,u^{-1}s]\cr
&=&[p_x,au^{-1}s]
\eea
We let the reader check the consistency of these definitions (or consult \cite{LM}, Chap. II, prop. 3.8). Now let $A$ be a linear or antilinear operator on $S_0$ which commutes with the action of the spin group. Then we can define a bundle operator $\tilde A$ on $\S^P$ by
\be 
\tilde A([p,s]):=[p,As]\label{atilde}
\ee
In this way we can define $\chi^P$ and $C^P$ from $\chi_0$ and $C_0$. In order to define $H^P$ we can directly invoke theorem \ref{existsspinormetric}. However, in order to be self-complete and consistent with the above approach, let us define it using the associated bundle construction. We set
\begin{equation}
H^P_x([p_x,s],[p_x',s'])=H_0(u\cdot s,s'),\mbox{ where }p_x=p_x'\cdot u
\end{equation}
Let us check that this is well-defined. For this, we compute (with $v,w\in \Spin(p,q)^0$):
\begin{eqnarray*}
H^P_x([p_x\cdot v, v^{-1}s],[p_x'\cdot w,w^{-1}s'])&=&H_0(U\cdot v^{-1}s,w^{-1}s'),\mbox{ with }p_x\cdot v=(p_x'\cdot w)\cdot U\\
&=&H_0(w^{-1}uvv^{-1}s,w^{-1}s'),\mbox{ since }U=w^{-1}uv\\
&=&H_0(us,s'),\mbox{ since }w\in\Spin(p,q)^0\\
&=&H^P_x([p_x,s],[p_x',s'])
\end{eqnarray*}
We must also check that the action of tangent vectors is self-adjoint. Let $X\in \RR^{p,q}\subset \End(S_0)$. We have, with the same notations as above:
\bea
H^P_x([p_x,X]\cdot[p_x,s],[p_x',s'])&=&H^P_x([p_x,X s],[p_x',s'])\cr
&=&H_0(uX s,s')\nonumber
\eea
Now, using \eqref{clifaction2}, we also have
\bea
H^P_x([p_x,s],[p_x,X]\cdot[p_x',s'])&=&H^P_x([p_x,s],[p_x,X u^{-1}s'])\cr
&=&H_0(s,X u^{-1}s')\cr
&=&H_0(X s,u^{-1}s'),\mbox{ since }v\mbox{ is self-adjoint}\cr
&=&H_0(uXs,s'),\mbox{ since }u\in\Spin(p,q)^0\nonumber
\eea
We thus have produced an algebraic spin structure $\ss^P=(\S^P,\ldots,H^P)$ from a topological one, and proven the following theorem:
\begin{theorem}  Let $M$ be a space and time oriented semi-riemannian manifold. The existence of an algebraic spin structure and a topological spin structure on $M$ are equivalent.
\end{theorem}
\begin{rem}
Suppressing some of the objects in the definition we also immediately obtain the equivalence of the algebraic and topological definitions of Clifford and spin-c structures, respectively. We refer to \cite{friedtraut} and \cite{fried}, p. 47 for the topological definitions of Clifford and spin-c structures.
\end{rem}
 
Let us now prove that the constructions $\ss\rightarrow (P_\ss,\Lambda_\ss)$ and $(P,\Lambda)\rightarrow \ss^P$ are inverse of each other up to canonical isomorphisms. Let $[p_x,s]\in \S^{P_\ss}$ and define 
$$\theta([p_x,s])=p_x(s)$$
It is immediate to check that this is a well-defined vector bundle isomorphism from $\S^{P_\ss}$ to $\S$. It is compatible with the Clifford action by \cite{LM}, Chap. II, prop. 3.8. The calculation $H_x(\theta([p_x,s]),\theta([p_x,s']))=H_x(p_x(s),p_x(s'))=H_0(s,s')=H_x^{P_\ss}([p_x,s],[p_x,s'])$ shows that $\theta$ sends $H^{P_\ss}$ to $H$, and   it sends $C^{P_\ss}$ to $C$ and $\chi^{P_\ss}$ to $\chi$ by \eqref{atilde}. Conversely, an element $\pi$ of $P_{\ss^P}$ at $x$ is by definition a morphism from $S_0$ to $\S^P_x$ which satisfies properties 1,\ldots, 4 above. Now let $p_x\in P_x$, and define 
$$\Pi(p_x)=(s\mapsto [p_x,s])$$
Let $X\in \RR^{p,q}\subset\CC l(p,q)$. We have $\Pi(p_x)(Xs)=[p_x,Xs]=[p_x,X][p_x,s]=[p_x,X]\Pi(p_x)(s)$, and this shows property 1. Properties 2,3,4 are also proven by easy abstract nonsense. Thus $\Pi$ is a bundle map from $P$ to $P_{\ss^P}$. Now let $\pi \in P_{\ss^P}$. For every $s\in S_0$, there is a unique (by the freeness and transitivity of the action of the spin group) $p_x(s)\in P_x$ such that $\pi(s)=[p_x(s),s]$. We have to show that $p_x(s)$ does not depend on $s$ in order to prove that $\Pi$ is bijective. For this, we use the isometric property of $\pi$. We have, for all $s,s'\in S_0$:
\bea
H_0(s,s')&=&H_x^P(\pi(s),\pi(s')),\mbox{ by definition of }\pi\cr
&=&H_x^P([p_x(s),s],[p_x(s'),s'])\cr
&=&H_0(u\cdot s,s'),\mbox{ where }p_x(s)=p_x(s')\cdot u
\eea
Thus $s=u\cdot s$ for all $s$, from which we infer that $u=1$ and $p_x(s)=p_x(s')$.Now we have $\Pi(p_x\cdot u)=(s\mapsto[p_x\cdot u,s])$ and $\Pi(p)\cdot u=(s\mapsto [p_x,u\cdot s]=[p_x\cdot u,s])$. Thus  $\Pi$ respects the action of the spin group. Finally, if we unpack $\Lambda_{P_{\ss^P}}(\Pi(p_x))(X)$, we find that it is the map from $\S^{P_{\ss^P}}_x$ to itself, which sends $[p_x,s]$ to $[p_x,X\cdots s]=[p_x,X]\cdot [p_x,s]$. Thus $\Lambda_{P_{\ss^P}}(\Pi(p_x))$ is the map from $\RR^{p,q}$ to $T_xM$ which sends $X$ to $[p_x,X]$, thus, it is $\Lambda(p_x)$ by \eqref{deflambda}. We have thus proven that $\Pi$ is an isomorphism of topological spin structures.

Let us now prove using the notations above, that if $(P,\Lambda)$  and $(P',\Lambda')$ are constructed from $\ss=(\S,\rho,\ldots)$ and $\ss'=(\S',\rho',\ldots)$ and if $\Sigma$ is an isomorphism from $\ss$ to $\ss'$, then  $(P,\Lambda)$ and $(P',\Lambda')$ are also isomorphic in the sense of definition \ref{topisom}.

First, the spin group is identified, on the one hand, as the subgroup of elements of $\End(S_0)$ which stabilize $V_0$, are Krein unitary and commute with $C_0$ and $\chi_0$, and on the other hand as  the subgroup of elements of $\End(S_0')$ which stabilize $V_0'$, are Krein unitary and commute with $C_0'$ and $\chi_0'$. These two identifications are sent to one another by $\Sigma_{x_0}$ for obvious reasons.

Let $p : M \times S_0\rightarrow \S$ be an element of $p$, then we set $f(p)_x(\psi)=\Sigma_x p_x(\Sigma^{-1}_{x_0}\psi)$, according to the following commutative diagram:
$$
\xymatrix{M\times S_0\ar[r]^(0.6)p\ar[d]_{\Sigma_{x_0}} &\S\ar[d]^{\Sigma}\\
M\times S_0'\ar[r]^(0.6){f(p)}&\S'\\
}
$$
\bea
f(p\cdot u)_x(\psi)&=&\Sigma_x (p\cdot u)_x(\Sigma^{-1}_{x_0}\psi)\cr
&=&\Sigma_x p_x( u\Sigma^{-1}_{x_0}\psi)
\eea
while
\bea
f(p)\cdot u_x(\psi)&=&f(p)_x(\Sigma_{x_0}u\Sigma_{x_0}^{-1}\psi)\cr
&=&\Sigma_x p_x(\Sigma^{-1}_{x_0}\Sigma_{x_0}u\Sigma_{x_0}^{-1}\psi)\cr
&=&\Sigma_x p_x( u\Sigma^{-1}_{x_0}\psi)
\eea
where in the first line we used the identification of the two spin groups, seen as subgroups of $\End(S_0)$ and $\End(S_0')$ respectively. Moreover, $f(p)$ is an element of $P'$ thanks to the properties of $\Sigma$ (Krein-unitarity, intertwining property, sends $\chi$ to $\chi'$ and $C$ to $C'$). The second check uses the following commutative cube:
$$\xymatrix{
& \S_x'\ar[dl]_{\Sigma_x} \ar[rr]^{\rho'(\Lambda'(f(p))v)}\ar@{<--}[dd]^(0.8){f(p)}
& & \S_x'\ar[dl]_{\Sigma_x} 
\\
\S_x \ar[rr]^(0.6){\rho(\Lambda(p)v)}
& & \S_x 
\\
& S_0'\ar@{-->}[rr]^(0.4){\rho'(v)}
& & S_0'\ar[uu]^(0.2){f(p)}
\\
S_0\ar[uu]^(0.2)p \ar[rr]^{\rho(v)}\ar@{-->}[ur]^{\Sigma_{x_0}}& & S_0\ar[uu]^(0.2)p \ar[ur]^{\Sigma_{x_0}}
}
$$
Every face of this cube but the top commutes either by definition of $\Lambda$, $\Lambda'$, $f$, or by the intertwining property of $\Sigma$. The top face shows the equality $\Lambda'(f(p))v=\Lambda(p)v$, since it must commute, and $\Sigma$ intertwines $\rho(\Lambda(p)v)$ with $\rho'(\Lambda(p)v)$.

Conversely, if $f$ is an isomorphism in the topological sense, we must define $\Sigma$ out of it.  We let $S_0$ be an irreducible spinor module as above, and let $\S=P\times_{\Spin(p,q)^0}S_0$ and $\S'=P'\times_{\Spin(p,q)^0}S_0$. Obviously $\Sigma$ must be defined by $\Sigma_x(p_x,\psi)=(f(p)_x,\psi)$. We just check that this is well-defined, leaving the other verifications to the reader:
\bea
\Sigma_x(p_x\cdot u,u^{-1}\cdot \psi)&=&(f(p\cdot u)_x,u^{-1}\psi)\cr
&=&(f(p)_x\cdot u,u^{-1}\psi)\cr
&=&(f(p)_x,\psi)\cr
&=&\Sigma(p_x,\psi)
\eea
In summary, we have just proven that $\ss\mapsto (P^\ss,\Lambda^\ss)$ is well-defined at the level of equivalence classes, and we have proved above that it is invertible. We thus obtain:
\begin{theorem}
The sets of equivalence classes of algebraic and topological spin structures are in bijective correspondence.
\end{theorem}

\begin{ex} Let us carry on the previous construction on the Euclidean spin structure $\ss_{a_1,a_2}$ on the $2$-torus (see section \ref{algspinstru}). We obtain $P_{a_1,a_2}=\{((\theta_1,\theta_2),p)\in T\times M_2(\CC)| p=\pmatrix{e^{i\alpha_1}&0\cr 0&e^{i\alpha_2}},$ with $\alpha_1+\alpha_2-a_1\theta_1-a_2\theta_2=0\ [2\pi]\}$. Since $(\theta_1,\theta_2)\mapsto \pmatrix{e^{i\theta_1}&0\cr 0&e^{i\theta_2}}$ is a global section, the bundles $P_{a_1,a_2}$, $a_1,a_2\in\{0,1\}$ are all  trivial, hence isomorphic as abstract bundles. However, the covering $\Lambda_{a_1,a_2}$  at the point $(\theta_1,\theta_2)$ of $T$ sends $\pmatrix{e^{i\alpha_1}&0\cr 0&e^{i\alpha_2}}$ to the rotation of angle $\alpha_1-\alpha_2=2\alpha_1-a_1\theta_1-a_2\theta_2$, and this makes the $(P_{a_1,a_2},\Lambda_{a_1,a_2})$ non isomorphic as topological spin structures. 
\end{ex}

\begin{ex} Let us now consider a tetradic spin structure $\ss_e$. The principal bundle is now always $P=M\times \Spin(p,q)^0$. But the covering is not trivial: for $x\in M$, $g\in \Spin(p,q)^0$ and $X\in \RR^{p,q}$, it is given by $\Lambda(x,g)X=\rho_e^{-1}(p\rho_0(X)p^{-1})$.
\end{ex}

\section{Isomorphism classes of spin structures on a Lorentzian or Riemannian metric parallelizable manifold}\label{preuvetetradss}
In this section we prove that the set of isomorphism classes of tetradic spin structures on a   metric parallelizable space and time oriented manifold $(M,g,or,t-or)$ is in bijection with $\Hom(\pi_1(M),\ZZ_2)$, when $(p,q)\not=(2,2)$. Since this is also the case of the set of isomorphism classes of general spin structures on $(M,g,or,t-or)$, this proves that every spin structure is isomorphic to a tetradic one when this set is finite.

First we look for a necessary and sufficient condition for two tetradic spin structures to be isomorphic. By considering a fixed $(g,or,t-or)$-frame $e$, any other such frame $f$ can be identified with a smooth  map $f : M\rightarrow SO(p,q)^0$. Now if $f,f'$ are two frames, the spin structures $\ss_f$ and $\ss_{f'}$ are isomorphic iff there exists a $\Sigma : M\rightarrow \Spin(p,q)^0$ such that $\rho_{f'}=\Sigma\rho_f\Sigma^{-1}$. Now $\rho_f$ is defined by $\rho_f(f(e_a))=\gamma_a$, and similarly for $\rho_{f'}$. Thus the isomorphism of spin structures is translated as
$$\gamma_a=\rho_f(f(e_a))=\Sigma \rho_f(f'(e_a))\Sigma^{-1}$$
for all $a$, which implies by linearity $\rho_f(f(v))=\Sigma \rho_f(f'(v))\Sigma^{-1}$, for all vector $v$. Replacing $v$ with $f^{-1}(v)$, and writing $L=f'f^{-1}$, we obtain that $\ss_f\simeq \ss_{f'}$ iff 
$$\Sigma^{-1}\rho_f(v)\Sigma=\rho_f(L(v))$$
that is to say, iff $L$ can be lifted to $\Spin(p,q)^0$ so that  the following diagram:
$$\xymatrix{
 & & \ZZ_2\ar[d]\\
 & & \Spin(p,q)^0\ar[d]^\Lambda\\
M\ar@{-->}[urr]^\Sigma\ar[rr]^(0.4){L} & &SO(p,q)^0\\
}$$
commute. By the general theory of covering spaces,  we   know that $L$ can be lifted if, and only if, $L_*(\pi_1(M))\subset \Lambda_*(\pi_1(\Spin(p,q)^0))$, where $*$ denotes the induced homomorphism of fundamental groups and base points are not explictly displayed. Suppose first for simplicity's sake that $p+q\ge 4$ and $p$ or $q$ equals $0$ or $1$. In this case we know that $\pi_1(\Spin(p,q)^0)=\{1\}$. Thus we obtain that $L$ can be lifted iff $(f'f^{-1})_*(\pi_1(M))=\{1\}$, i.e. for every loop $\gamma$, $[f'f^{-1}\circ \gamma]=1$. Now by the Eckmann-Hilton principle, for any based loop $\gamma$, $[f'f^{-1}(\gamma)]=[f'(\gamma)][f(\gamma)]^{-1}$, (products in the Lie group $SO(p,q)^0$ and concatenation yield the same homotopy classes) thus we obtain that 
$$\ss_f\simeq \ss_{f'}\Leftrightarrow f_*=f_*'$$
Moreover,  since $SO(p,q)^0$ is a Lie group, $\pi_2(SO(p,q))=0$, thus by \cite{hatcher}, section 4.3,  each element of $\Hom(\pi_1(M),\pi_1(SO(p,q)^0)$ is induced by some $f$. Finally, from the hypothesis made on $p$ and $q$, $\pi_1(SO(p,q)^0)=\ZZ_2$.  This concludes the proof.

In the general case, $\Spin(p,q)^0$ is not simply connected, but $\Lambda_*$ is injective since $\Lambda$ is a covering map (\cite{hatcher}, prop 1.31).  Write $G=\pi_1(SO(p,q)^0)$, $H=\Lambda_*(\pi_1(\Spin(p,q)^0))$ and $p : G\rightarrow G/H$ the projection map. Leaving aside the case $(p,q)=(1,1)$, we have $G/H\simeq \ZZ_2$. The above argument shows that $\ss_f\simeq \ss_{f'}$ iff $p\circ f_*=p\circ f_*'$. Thus we obtain an injection from the set of isomorphism classes of tetradic spin structures to $\Hom(\pi_1(M),\ZZ_2)$ which is a bijection iff every homomorphism from $\pi_1(M)$ to $\ZZ_2$ can be lifted to $G$. If $p: G\rightarrow G/H$ has a section, then since $G$ and $H$ are abelian, $G\simeq H\times (G/H)$, and the lifting property is true. For $p,q>2$, we have $G=\ZZ_2\times \ZZ_2$, $H\simeq \ZZ_2$ and $G/H\simeq \ZZ_2$. Hence these groups are also vector spaces over $\ZZ_2$ and group homomorphisms are $\ZZ_2$-linear. Thus $p$ has a section. For $p>2$ and $q=2$ (or symmetrically $p=2, q>2$), one has $G=\ZZ\times \ZZ_2$, $H=\ZZ$, $G/H=\ZZ_2$, with the embedding of $H$ into $G$ being $n\mapsto (n,n\ [2])$. Thus the quotient map $p$ is $(n,m\ [2])\mapsto n+m\ [2]$, which has the following section : $s(x)=(0,x)$. In all these case we have thus proven the property. This only leaves the exceptional cases $(p,q)=(2,2)$, and $(2,0)$ (and $(0,2)$). Consider the $(2,0)$ and $(0,2)$ cases. If it is noncompact, then its fundamental group is a free group (\cite{stillwell}, p 142), which must be finitely generated for $\Hom(\pi_1(M),\ZZ_2)$ to be finite. If it is compact, then since it is supposed to be parallelizable, it must be a torus. Either way, every morphism of $\pi_1(M)$ to $\ZZ_2$ can be lifted to $\ZZ$ and the property holds.

This only leaves the case $p=q=2$.  Here we have $G=\ZZ\times\ZZ$. It is known that any fundamental group can arise for a $4$-manifold, but we have the additional requirement of metric-parallelizability. In  \cite{johnsonwalton} it is proven that if a group has a finite presentation with stricly less relations than generators, then this group is the fundamental group of a parallelizable $4$-manifold. In particular, $\ZZ*\ZZ_2$, which has two generators, $x,y$, and the one relation $y^2=1$ is such group, and the morphism which sends $x$ to $0\ [2]$ and $y$ to $1\ [2]$ cannot be lifted to $\ZZ\times \ZZ$. If $M$ is a parallelizable $4$-manifold with fundamental group $\ZZ*\ZZ_2$, then $M$ admits a metric of signature $(2,2)$, since we can declare a particular tetrad to be pseudo-orthonormal with the correct signs. This provides a counter-example.

%
%



\section{Topology on the space of tetradic spin structures}\label{topospin}
We first need to introduce  vertical isometries of $(M,g)$. Let  $O(TM)$ denotes the bundle of isometries, then we call 
\begin{equation}
\Gamma(O(TM))=\{r\in{\Gamma({\rm Gl}}(M))|\forall x\in M, r_x\mbox{ is an isometry of }(T_xM,g_x)\}
\end{equation}
the group of vertical isometries. In the context of GR, this is the group of local Lorentz transformations. We will also denote by $\Gamma(SO^0(TM))$ the subgroup whose elements preserve given space and time orientations, and by $\Aut(\ss)$ the space of automorphisms of a $(g,or,t-or)$-spin structure $\ss$. 


Fix $(g,or,t-or)$ and a corresponding frame $(e_a)$. Let us call $\ss$ the spin structure defined by $(e_a)$, and $\mathfrak{S}$ the space of tetradic spin structures. By definition, the map $r\mapsto r\cdot\ss$ from  $\Gamma(SO^0(TM))$ to ${\mathfrak S}$ is a bijective correspondence, and we can use it to topologize $\mathfrak{S}$ by putting the compact-open topology on $\Gamma(SO^0(TM))$ and transport it. Let us recall that since the topology on  $SO^0(TM)$ can be given by a metric, the compact-open topology on $\mathfrak{S}$ is the topology of uniform convergence on compact subsets.

\begin{rem}
In this section, we set ourselves in the continuous category, for simplicity's sake. The more usual topology on spaces of smooth functions is the Whitney topology. There is however a dense embedding of $({\cal C}^\infty(M,N),$ Whitney topology$)$ into $({\cal C}^0(M,N),$ compact-open topology$)$ when $M,N$ are smooth manifolds. For all these matters, see \cite{hirsch}, chap. 2.
\end{rem}

It is immediate to check that the group operations on $\Gamma(SO^0(TM))$ (pointwise multiplication and inversion) are continuous for this topology. Moreover, the action of $\Gamma(SO^0(TM))\subset {\Gamma({\rm Gl}}(M))$, seen as the group of vertical isometries on ${\mathfrak S}=\Gamma(SO^0(TM))$ seen as the space of spin structures is just by left  multiplication and is thus continuous. The space ${\mathfrak S}/\simeq$ of isomorphism classes of spin structures  being finite, we equip it with the discrete topology. The projection map ${\mathfrak S}\rightarrow {\mathfrak S}/\simeq$ will be continuous iff the classes are closed. %

\begin{theorem}
If $H^1(M,\ZZ_2)$ is finite, the isomorphism classes of spin structures are clopen sets  for the compact-open topology.
\end{theorem}
\begin{demo}
Since there is a finite number of classes, we just need to check that they are closed, and since the group acts transitively on classes by homeomorphisms, we just need to check that the class of $\ss$  is closed. Let $(f_k)_{k\in\NN}$ be a sequence of  maps $M\rightarrow SO^0(p,q)$, which   converges towards $f$, and such that $f_k\cdot \ss\simeq \ss$ for all $k$. The last condition exactly means that there   exists a lift $\tilde f_k$ of $f_k$ to $\Gamma(\Spin(p,q)^0):={\cal C}(M,\Spin(p,q)^0)$ for all $k$, and we need to check that there also exists such a lift of $f$. We can topologize $\Gamma(\Spin(p,q)^0))$ by considering it to be a subspace of ${\cal C}(M,M_{2^{n/2}}(\CC))$ and using the topology of uniform convergence on compact subsets, where $M_{2^{n/2}}(\CC)$ is endowed with the operator norm.  We will prove that there exists a subsequence of $(\tilde f_k)$ which converges to some element $\tilde f\in \Gamma(\Spin(p,q)^0)$. That $\tilde f$ lifts $f$ is then immediate since the covering map $\lambda : \Spin(p,q)^0\rightarrow SO^0(p,q)$ is continuous. To prove the existence of the subsequence, we apply the Arzela-Ascoli theorem (\cite{kelley}, p 233), whose hypotheses we now need to check. First, $(\tilde f_k)(x)$ is uniformly bounded for all $x\in M$:

\begin{lemma}  Let $\tilde A$ be a lift in $\Spin(p,q)$ of $A\in SO(p,q)$. Then $\|\tilde A\|\le \|A\|^{\min(p,q)\over 2}$, where the first norm is the operator norm in $M_{2^{n\over 2}}(\CC)$ and the second norm is the operator norm in $M_n(\RR)$.
\end{lemma}

This lemma is proved in \cite{doppler}. (Note that it degenerates into an equality in the Lorentzian and Euclidean cases.) Finally we need to check the equicontinuity of $(\tilde f_k)$. Let $x\in M$ and $\epsilon>0$. We must check that there exists a neighbourhood $U$ of $x$ such that $\forall x'\in U$, $\forall k\in \NN$, $\|\tilde f_k(x)-\tilde f_k(x')\|\le \epsilon$. 

Let $V$ be a neighbourhood of $f(x)$ such that there exists a section $s : V\rightarrow \Spin(p,q)^0$ of the covering map $\lambda : \Spin\rightarrow SO$. We can choose $V=B(f(x),\delta)$, a ball of radius $\delta$ and center $f(x)$, and $\delta$ small enough to have $s(B(f(x),\delta))\subset B(s(f(x)),\epsilon)$. Let $U$ be a neighbourhood of $x$ such that $f(U)\subset B(f(x),\delta/2)$. We can choose $U$ to be compact without loss of generality. Since $(f_k)$ converges uniformly on $U$, there exists $N\in\NN$ such that $\|f_k-f\|_{\infty,U}\le \delta/2$ for all $k\ge N$, and we thus obtain $\|f_k(x')-f(x)\|\le \|f_k(x')-f(x')\|+\|f(x')-f(x)\|\le\delta$ for all $k\ge N$. By continuity of $f_k$ for $k=0,\ldots,N-1$, we obtain $N$ neighbourhoods $U_k$ such that $f_k(U_k)\subset V$. Hence we have $f_k(U')\subset V$ for all $k\in\NN$ where $U'$ is the intersection of $U,U_0,\ldots,U_{N-1}$. Thus $s(f_k(U'))\subset B(s(f(x)),\epsilon)$. In other words, for all $k\in\NN$ and all $x'\in U'$, we have
\be
\|s(f_k(x'))-s(f_k(x))\|\le \epsilon\label{equi}
\ee
Now since the covering map $\lambda$ is two-to-one, there are exactly to lifts of $f_k$, which are $\tilde f_k$ and $-\tilde f_k$. Hence $s\circ f_k$ must be equal to $\pm \tilde f_k$ on $U'$. It thus follows from \eqref{equi} that $(\tilde f_k)$ is equicontinuous at $x$.
\end{demo}

We can now easily deduce the result which was the aim of this appendix.

\begin{propo}
There exists an open neighbourhood ${\cal U}$ of the unit   in ${\Gamma({\rm Gl}}(M))$ such that the following property holds : 
\be 
\forall r,r'\in {\cal U},  (r\cdot g=r'\cdot g, r\cdot or=r'\cdot or, r\cdot t-or=r'\cdot t-or)\Rightarrow r\cdot\ss\simeq r'\cdot\ss
\ee
\end{propo}
\begin{demo} First, note that $\ss$ is the unit in the identification $\SS=\Gamma(SO(p,q)^0)$. Let $[\ss]$ be the isomorphism class of $\ss$. By the theorem, it is an open set in $\SS=\Gamma(SO(p,q)^0)$. There thus exists an open set ${\cal V}$ in ${\Gamma({\rm Gl}}(M))$ such that ${\cal V}\cap \SS=[\ss]$. By the general properties of topological groups (\cite{foll}, p. 32), there  also exists a neigbourhood ${\cal U}$ of $\ss$ in ${\Gamma({\rm Gl}}(M))$  such that ${\cal U}{\cal U}^{-1}\subset {\cal V}$. Supppose $r$ and $r'$ are as in the claim. Then $r^{-1}r'\in \Gamma(SO(p,q)^0)$ and $r^{-1}r'\in {\cal V}$. Hence $r^{-1}r'\in[\ss]$, which means that $r\cdot \ss\simeq r'\cdot\ss$.
\end{demo}

Let us end with the observation that if $pq=0$ (Riemannian/anti-Riemannian cases), then ${\cal U}$ can be taken to be the whole neutral component.

\begin{propo} If $g$ is Riemmanian, the neutral path-component of $\Gamma(SO^0(TM))$ is the intersection of $\Gamma(SO^0(TM))$ with  the neutral path-component of ${\Gamma({\rm Gl}}(M))^+$.
\end{propo}
\begin{demo}
One inclusion being obvious, we only need to prove that if $1$, the unit map and $L\in \Gamma(SO^0(TM))$ can be connected by a path $L^t$ in ${\Gamma({\rm Gl}}(M))^+$, they can be so connected by a path in $\Gamma(SO^0(TM))$. Now it is well-known that $SO(n)$ is a  retraction of $Gl_n(\RR)$, that is, there exists a continous map $\delta : Gl_n(\RR)\rightarrow SO(n)$ such that $\delta$ is the identity on $SO(n)$. Composing $L^t$ with $\delta$ yields the desired path.
\end{demo}

Combining this proposition with theorem \ref{metcomp} and the fact that the space of Riemannian metrics is path-connected, we obtain the following  result, which shows the independence of the isomorphism class of spin structure from the metric.

\begin{propo} Let $g_0$ be a Riemannian metric on $(M,or)$. Let ${\cal M}(g_0)$ be the path-connected component of $g_0$, and let $\ss$ be a spin structure on $(M,g_0,or)$. Then for every $g\in {\cal M}(g_0)$, there exists $r\in {\Gamma({\rm Gl}}(M))$ such that $g=r\cdot g_0$. Moreover, if If $r\cdot g_0=r'\cdot g_0=g$ and $r\cdot or=r'\cdot or$,  then $r\cdot \ss\simeq r'\cdot \ss$.
\end{propo}

\section{Index of notations}

\begin{longtable}{c@{   }l@{ p }l}
$\tilde{A}$ & canonical extension of $A$ to the Clifford algebra & 6\\
$A^o$ & opposite of $A$ & 25 \\
AB & algebraic background & 24 \\
$\alpha$ & morphism from $\Aut(\B)$ to $\Aut(\A)$ & 27\\
$\B$ & an algebraic background  & 24\\
$\B_{\rm can}$ & canonical AB & 22\\ 
${\cal B}_u(\K)$ & algebra of universally bounded operators & 23\\
$c$ & real structure on $\CC l(V,g)$ & 6\\
$C$ & real structure (acting on spinors) & 9\\
$C_0$ & real structure on $S_0$ & 14\\
${\cal C}$, ${\cal C}(g,\ss)$, ${\cal C}(e)$& configuration space & 20, 21\\
$C_J^*$ & & 39\\
$\tilde{\cal C}^\infty_c(M)$ & smooth functions with compact support + constants & 21\\
$\chi$ & chirality (in $\CC l(V,g)$) & 6\\
$\chi$ & chirality (acting on spinors) & 7\\
$\chi_0$ & chirality on $S_0$ & 14\\
$\CC l(TM)$ & complex Clifford bundle & 7\\
$Cl(V,g)$ & Clifford algebra & 5\\
$\CC l(V,g)$ & Complex Clifford algebra & 5\\
${\cal D}_{\cal B}$ & configuration space of ${\cal B}$ &  26\\
$D^\ss(g)$&   Dirac operator associated to the metric $g$ and spin structure $\ss$ &\\
$\delta_r$ & frame Dirac & 33 \\
$e$ & moving frame & 14\\

$(\epsilon_a)$ & the canonical basis of $\RR^{p,q}$ & 14\\
$\epsilon, \epsilon'', \kappa, \kappa''$ & KO-metric signs & 24\\
$\eta$ & a fundamental symmetry & 23\\
${\rm Fr}(M)$ & the frame bundle & 10\\
$g$ & metric & 5\\
$G^0$ & neutral component of $G$ & 6\\
$(\gamma_a)$ & a standard set of gamma matrices & 14\\
$\Gamma_\CC(V,g)$ & complex Clifford group & 6\\
${\Gamma({\rm Gl}}(M))$ & group of vertical automorphisms of $TM$ & 47\\
 
$\Gamma^\infty_c(\K)$ & smooth sections of $\K$ with compact support &21\\
$\Gamma {\rm Sl}(M)$ & group of special vertical automorphisms & 22\\

$\Gamma(\Spin(p,q)^0)$ & spinomorphism group & 15\\

$H$ & spinor metric & 7\\

${\rm Hol}_\lambda$ & parallel transport along $\lambda$ & 12\\
$H_0$ & Robinson form on $S_0$  & 4\\

IST & Indefinite Spectral Triple & 23\\
$J$ & graded real structure & 9, 24\\
$\lambda$ & the covering map $\Spin(p,q)^0\rightarrow SO(p,q)^0$ & 10\\
$\lambda$ & a curve & 12\\

$n$ & manifold dimension & 5\\

$\Omega^1$ & bimodule of noncommutative 1-forms& 21, 24\\
$\Omega^1_D$ & nc 1-forms defined by $D$ &  26\\
$or$ & orientation & 6\\
$(p,q)$ & signature of the metric & 5\\
$\Psi$ & a spinor field & \\
$\psi$ & a spinor & \\
$\RR^{p,q}$ & standard space of signature $(p,q)$ & 6\\
$\rho$ & representation of the Clifford bundle & 6\\
$\rho_e$ & & 14\\
$\rho_0$ & representation of $\CC l(p,q)$ on $S_0$ & 14\\
$\S$ & a spinor bundle & 6\\

$\ss$ & a spin structure & 10\\
$\ss_e$ & tetradic spin structure & 14\\
$S_r$ & AB equivalence defined by $r$ & 28\\
$S_0$ & standard spinor space  & 14\\
${\cal S}_0$  &   $M\times S_0$ & 14\\
$\Sigma$ & isomorphism of spinor bundles, or spinomorphism & 9, 15\\
$\Spin(V,g)$, $\Spin^c(V,g)$ & spin and spin-c groups & 6\\
SST & Spectral Spacetime & 23\\
$T$ & transpose & 5\\
$\Theta$ & vector space isomorphism & 6\\

$\theta$ & a diffeomorphism & 15\\
$\theta_*$ & pushforward by $\theta$ &  15\\
$t-or$ & time-orientation & 8\\
$U_\Sigma$ & AB equivalence defined by a spinomorphism & 28\\
$U_\theta$ & AB equivalence defined by a diffeomorphism & 29\\
$V_0$ & real space spanned by the gamma matrices & 14\\

$V^\CC$ & complexification of $V$ & 6\\
${\rm Vert}(\B)$ & kernel of $\alpha$ & 27\\

$\times$& anti-automorphism of $\CC l(V,g)$ & 6\\
$\times$ & Krein-adjoint of an operator  & 9\\
$\zeta_\omega$ & centralizing field & 34\\
$Z_J^*$ & & 38\\
$\|\ \|_\eta$ & & 23\\
$\|\ \|_u$ & universal operator norm & 23\\

\end{longtable}


\bibliographystyle{unsrt}
\bibliography{../generalbib/SSTbiblio}
\end{document}